\documentclass[a4paper,11pt]{article}
\usepackage{jheppub} 
\usepackage{lineno}
\usepackage{graphicx} 
\usepackage{amsmath}
\usepackage{tikz}
\usepackage{caption}
\usepackage{subcaption}
\usepackage{amssymb}
\usepackage{braket}
\usepackage{enumitem}

\title{\boldmath Holographic Network, Entanglement Wedge and Traversable Parallel Universe}

\author{Yu Guo, Rong-Xin Miao}
\affiliation{School of Physics and Astronomy, Sun Yat-Sen University,\\
2 Daxue Road, Zhuhai 519082, China}

\emailAdd{guoy225@mail2.sysu.edu.cn}
\emailAdd{miaorx@mail.sysu.edu.cn}

\abstract{This paper investigates the holographic network connecting different CFTs, modeled by Gauss-Bonnet gravity with varying couplings across different bulk branches. By applying the holographic Noether's theorem, we prove that the junction condition on the Net-brane leads to conservation laws at network nodes. We analyze the stability of the gravitational KK modes on the Net-brane and derive the constraints on theory parameters. Additionally, we discuss various proposals for network entropy, confirm that the type I and II network entropies obey the holographic g-theorem, and show that the type III network entropy is non-negative. We explore the two-point functions of various NCFTs at different edges, using examples like free scalars and the AdS/NCFT with a tensionless brane. We find that zero tension results in negative reflectivity at the node, indicating that it is a non-unitary parameter.

We study the wedge inclusion condition, which stipulates that the entanglement wedge must encompass the causal wedge. This condition imposes a lower bound on the tension of the Net-brane, which is stronger than the bound derived from the positivity of reflectivity. Furthermore, we conclude that the tension of Net-branes must be positive; the more edges present, the stronger this bound becomes. We then examine the gravitational dual of compact networks, which feature both EOW branes and Net-branes in the bulk. We derive the joint condition for EOW branes at the Net-brane and analyze vacuum solutions in AdS$_3$/NCFT$_2$. Finally, we demonstrate that AdS/NCFT provides a natural way to envision traversable parallel universes that have different geometries and physical laws. Remarkably, unlike traversable wormholes, our model of parallel universes satisfies all the energy conditions.}

\begin{document}

\maketitle

\flushbottom

\section{Introduction}

\begin{figure}[htbp]
  \centering
\includegraphics[width=.3\textwidth]{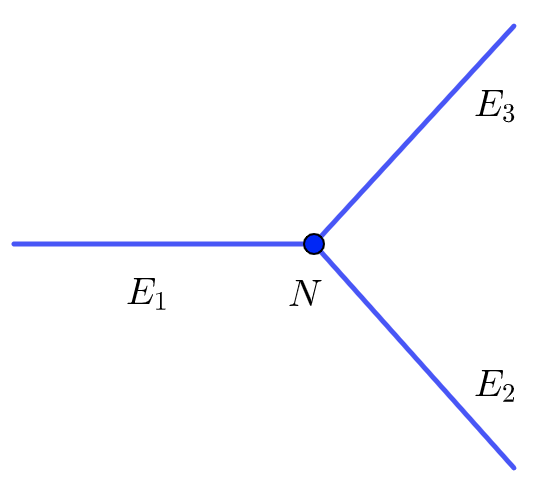}
\qquad \qquad
\includegraphics[width=.4\textwidth]{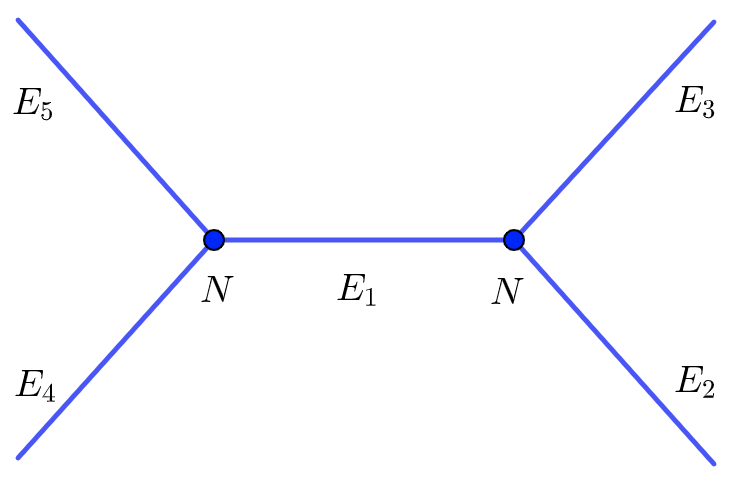}
 \caption{The networks with one node (left) and two nodes (right). The blue lines and points denote the edges ($E_m$) and nodes ($N$) of the networks.  } 
 \label{network}
\end{figure}

\begin{figure}[htbp]
  \centering
\includegraphics[width=.3\textwidth]{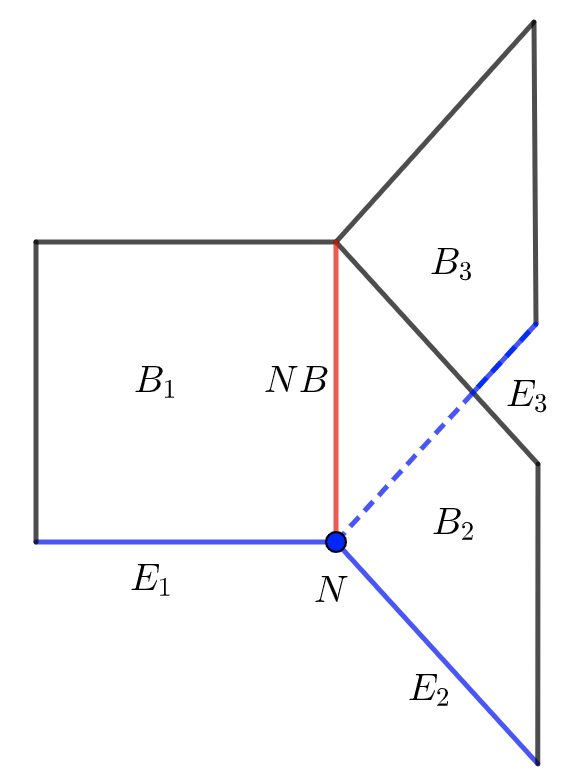}
\qquad \qquad
\includegraphics[width=.4\textwidth]{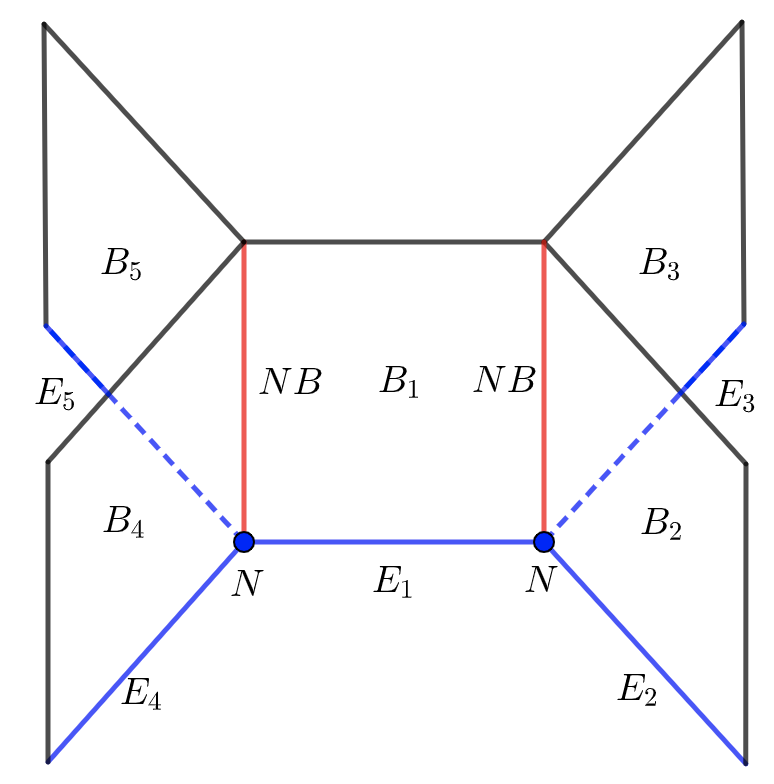}
 \caption{Geometries for holographic networks. The blue lines and points denote the edges ($E_m$) and nodes ($N$) of the networks. The red lines label the Net-branes $NB$, which link the branches $B_m$ (squares) in bulk. The edges ($E_m$) and nodes ($N$) are dual to the branches $B_m$ and Net-branes $NB$ in bulk, respectively. For simplicity, we show only the holographic duals of the networks of Fig. \ref{network}. One can glue above geometries to get the gravity duals of general networks. } \label{holo network}
\end{figure}

A network is a simplified model that describes complex connections and is widely used in fields such as physics, mathematics, and computer science. 
Many physical systems, such as integrated circuit chips, naturally exhibit network structures. Recently, the theory of conformal field theory on networks (NCFT) and its gravity dual (AdS/NCFT) has been established \cite{Guo:2025sbm}. See Fig. \ref{network} and Fig. \ref{holo network} for the geometry of some typical networks and their gravity duals. In theory, NCFT serves as a multi-branch generalization of boundary conformal field theory (BCFT) \cite{Takayanagi:2011zk} and interface conformal field theory (ICFT). In practice, it can describe the physics of phonons and electrons in nanocircuits, offering significant potential applications. The novel network structures give rise to many fascinating quantum effects. For example, unlike BCFTs, the Casimir force of NCFTs can change from attractive to repulsive by adjusting the edge lengths, providing a straightforward method to control the Casimir effect \cite{Zhao:2025npv}. Furthermore, the network entropy, defined as the difference in entanglement entropy between NCFT and BCFT, is always non-negative and effectively captures the network's complexity \cite{Guo:2025sbm}. 
It is important to emphasize that \cite{Guo:2025sbm} concentrates on the bottom-up holographic model of networks. Interestingly, holographic networks can also be constructed within top-down models of string theory, as demonstrated in \cite{DHoker:2007hhe}. However, due to the limited number of top-down solutions available for junctions, this paper will focus on the bottom-up approach outlined in \cite{Guo:2025sbm}. 
Please refer to \cite{Karch:2024udk,Karch:2026spf,Bachas:2020yxv,Chakraborty:2026wip,Chakraborty:2025jtj,Chakraborty:2025dmc,Banerjee:2025zuw,Banerjee:2024sqq,Shen:2024itl,Shen:2026coi,Liu:2025khw} for some related works on AdS/ICFT, string theory, and multi-junctions.

The original work \cite{Guo:2025sbm} examines identical conformal field theories (CFTs) at all network edges. This paper expands the discussion to consider varying CFTs at different edges. Regardless of the specific details of conformal field theory, the conservation of current and energy necessitates the following junction conditions at the nodes \cite{Guo:2025sbm}: 
\begin{align}\label{NCFT BC} 
\text{NCFT}: \ \sum_{m} \overset{(m)}{J}_n|_{N}=0, \ \  \sum_{m} \overset{(m)}{T}_{na}|_{N}=0, 
\end{align} 
where $n$ and $a$ denote the normal and tangential directions to the node $N$. Here, $\overset{(m)}{J}$ and $\overset{(m)}{T}$ represent the current and stress tensor on the edge $E_m$ linked by the same node. This formulation corresponds to Kirchhoff's current law in circuit theory. In the framework of AdS/NCFT, the edges ($E_m$) and nodes ($N$) are extended to bulk branches $B_m$ and Net-branes $NB$. Refer to Fig. \ref{holo network} for the associated geometry. For Einstein gravity, \cite{Guo:2025sbm} proposes imposing the multi-junction condition (JC) on the Net-brane $NB$ \footnote{The multi-junction condition was first introduced in the study of multi-boundary wormholes \cite{Shen:2024itl}. Here, we have different physical motivations in AdS/NCFT.}:
 \begin{align}\label{JC Einstein gravity} 
\text{AdS/NCFT}:\ \sum_m \frac{1}{8\pi G_N}\Big(\overset{(m)}{K}_{ij}-\overset{(m)}{K} h_{ij}\Big)|_{NB}=\mathcal{T}_{ij}, 
\end{align} 
where $\mathcal{T}_{ij}$ is the matter stress tensor on the Net-brane, and $\overset{(m)}{K}_{ij}$ denotes the extrinsic curvatures from bulk branches $B_m$ to the Net-brane. When there is only a tension term on the brane, we have $\mathcal{T}_{ij}=-T h_{ij}/(8\pi G_N)$. The authors of \cite{Guo:2025sbm} demonstrate that the JC (\ref{JC Einstein gravity}) on the Net-brane results in the conservation law (\ref{NCFT BC}) at the node, 
thereby providing validation of the self-consistency for the AdS/NCFT theory.
For general gravity theories, the junction condition takes the form: 
\begin{align}\label{JC general gravity} 
\text{AdS/NCFT}:\ \sum_m \overset{(m)}{T}_{\text{BY}\ ij}|_{NB}=\mathcal{T}_{ij}, 
\end{align} 
where $\overset{(m)}{T}_{\text{BY}\ ij}$ represents the Brown-York (BY) stress tensors from bulk branches $B_m$ to the Net-brane, which may differ across various bulk branches. In Einstein gravity, $\overset{(m)}{T}_{\text{BY}\ ij}$ is given by the left-hand side of (\ref{JC Einstein gravity}), while in Gauss-Bonnet gravity, it is described by (\ref{sect2: TijGB}). In this paper, we prove that the general JC (\ref{JC general gravity}) also leads to the conservation law (\ref{NCFT BC}) at the node. We develop a new alternative proof based on holographic Noether's theorem, which is much simpler than that of  \cite{Guo:2025sbm}.

We focus on Gauss-Bonnet (GB) gravity with varying parameters across different bulk branches. We analyze the gravitational Kaluza-Klein (KK) modes on the Net-brane and derive the ghost/tachyon-free constraints (\ref{sect2: ghost-free condition}) for the GB couplings. These constraints are stronger than those found in AdS/CFT \cite{Buchel:2009sk}. We discuss various proposals for network entropy and confirm that both type I and type II network entropies comply with the holographic g-theorem in general dimensions. Drawing inspiration from AdS/BCFT \cite{Hu:2022ymx}, we express the g-function in higher dimensions as the entanglement entropy of hemispheres connected at the node. We also verify that type III network entropy, defined as the difference in entropy between NCFTs and BCFTs, is non-negative. Additionally, we study the two-point functions of distinct NCFTs at different edges, including free scalar fields and the AdS/NCFT with a tensionless brane. The number of scalars varies at different edges, allowing us to model NCFTs with differing total central charges. 
The tensionless Net-brane results in negative reflectivity at the node, indicating that $T=0$ is a non-unitary parameter.

We investigate the wedge inclusion condition in AdS$_3$/NCFT$_2$, which stipulates that the entanglement wedge must encompass the causal wedge. This requirement necessitates a positive tension on the Net-brane. Furthermore, as the number of edges increases, the lower bound on this tension also rises. Importantly, this condition is stronger than the unitary bound derived from the positivity of reflectivity \cite{Liu:2025khw}.
We then examine the gravity dual of compact networks, which include both end-of-the-world (EOW) branes and Net-branes in the bulk. For example, see Fig. \ref{fig: geometry p=3}. Inspired by the connected Ryu-Takayanagi (RT) surfaces \cite{Ryu:2006bv} in AdS/NCFT \cite{Guo:2025sbm}, we propose that the EOW branes intersect at the same joint point on the Net-brane. By varying the action including the Hayward terms, we derive the joint condition for the EOW branes at the joint point. 
Due to the non-trivial Casimir effect \cite{Zhao:2025npv}, the vacuum state of a compact network is not dual to Poincaré AdS; rather, it is dual to suitably glued AdS solitons. We provide examples in AdS$_3$/NCFT$_2$ for the simplest network with only one node. In general, there are multiple configurations of the Net-brane and EOW branes that satisfy the joint condition, and we choose the configuration with the smallest free energy as the vacuum solution.

Finally, we demonstrate that AdS/NCFT provides a natural way to envision traversable parallel universes. Each branch in the bulk can be viewed as its own universe, potentially exhibiting different geometries, field types, and physical laws than other universes. While causality and unitarity impose certain constraints on the parameters of these universes, they still permit a considerable degree of freedom to connect different universes consistently. Notably, it is possible to travel from one universe to another in a probabilistic sense. For example, when light propagates from one universe to the Net-brane, it has a specific probability of being reflected and a certain probability of transmitting to other universes. We want to emphasize that our model of parallel universes differs from both the many-worlds interpretation in quantum mechanics \cite{Everett, Tegmark:2007wh} and the multiverse proposed in eternal inflation \cite{Eternalinflation}, as these are typically non-traversable between different worlds. Furthermore, unlike traversable wormholes, our model of parallel universes adheres to the null energy condition. We construct two exact models of traversable parallel universes: one is the so-called threefold universe, which connects flat, de Sitter (dS), and anti-de Sitter (AdS) universes; the other glues together universes with and without gravity. In particular, we can couple massless gravity consistently with a non-gravitational bath, which has potential applications in addressing the black hole information paradox \cite{Penington:2019npb, Almheiri:2019psf, Almheiri:2020cfm}.

The paper is structured as follows. In Section \ref{Holographic Network}
, we study the holographic network for Gauss-Bonnet gravity with varying parameters in different bulk branches. We derive the conservation law at nodes using holographic Noether's theorem, analyze the stability of gravitational KK modes on the Net-brane, and examine holographic entanglement entropy along with various aspects of network entropy. 
Additionally, we discuss correlation functions of different Network Conformal Field Theories (NCFTs) at various edges, 
including examples involving free scalars and the AdS/NCFT scenario with a tensionless Net-brane. Section \ref{Wedge inclusion condition}
focuses on the wedge inclusion condition in AdS$_3$/NCFT$_2$, demonstrating that it imposes a strong lower bound on the tension of the Net-brane. 
In Section \ref{Holographic compact network},
we investigate the holographic compact network and establish the joint condition for EOW branes on the Net-brane. Section \ref{Traversable Parallel Universe}
presents the AdS/NCFT as a natural model for traversable parallel universes, demonstrating how universes with distinct physical laws and geometries can be connected through junction conditions on the Net-brane. Finally, we conclude with a discussion of open issues in Section \ref{Conclusions and Discussions}. 
Appendix \ref{app A} looks into the stability of gravitational KK modes on the Net-brane, Appendix \ref{app B}
derives the junction conditions of RT surfaces on the Net-branes for Gauss-Bonnet gravity, and Appendix \ref{app C} 
discusses the joint condition of EOW branes at the joint on the Net-brane. 
The notations used in this paper are outlined in Table \ref{sect2:conenotation}.

\begin{table}[ht]
\caption{Notations of AdS/NCFT}
\begin{center}
    \begin{tabular}{| c | c | c | c |  c | c | c | c| c| c|c| }
    \hline
    & edge $E_m$ & node $N$  & bulk branch $B_m$ & Net-brane $NB$  \\ \hline
 coordinate & $y^i=(\overset{(m)}{x}, y^{a})$ & $y^{a}$ & $x^{\mu}=(z,y^i)=(z,\overset{(m)}{x}, y^{a})$ & $y^i=(z, y^{a})$ \\ \hline
 metric& $h_{ij},\ h_{E\ ij}$ & $\sigma_{ab}$ & $g_{\mu\nu}$ &$h_{ij},\ h_{NB\ ij}$\\ \hline
  normal vector& $\hat{n}$ & $n$ & $\varnothing$ & $\hat{n}$ \\ \hline
 \end{tabular}
\end{center}
\label{sect2:conenotation}
\end{table}

\section{Holographic Network}
\label{Holographic Network}
This section explores the holographic network with different CFTs at its edges, where the corresponding gravity theories across the different bulk branches are generally distinct. For simplicity, we focus on Gauss-Bonnet (GB) gravity in this paper. We derive the junction condition on the Net-brane and demonstrate that it leads to a conservation law at the network node. Additionally, we analyze the spectrum of Kaluza-Klein (KK) modes on the Net-brane and identify ghost-free constraints on the theory's parameters. We also discuss holographic entanglement entropy and network entropy. For clarity, we concentrate on the basic holographic network with $p$ branches depicted in Fig. \ref{holo network} (left), noting that generalizing to more complex cases is straightforward.

Let us start with the GB gravity in bulk
\begin{eqnarray}\label{sect2: GBactionbulk}
  I_{\text{GB\ bulk}}=\sum_{m}^p\frac{1}{16\pi G_{N\ m}}\int_{B_m} d^{d+1}x\sqrt{|g|} \Big(R+\frac{d(d-1)}{L_m^2}+\frac{L_m^2 \lambda_m }{(d-2)(d-3)} \mathcal{L}_{\text{GB}} \Big),
  \end{eqnarray}
 where $\mathcal{L}_{\text{GB}} =R_{\mu\nu\alpha\beta}R^{\mu\nu\alpha\beta}-4R_{\mu\nu}R^{\mu\nu}+R^2$,  \(G_{N\ m}\) and \(\lambda_m\) represent Newton's constant and the GB couplings in the bulk branch \(B_m\). To avoid negative energy fluxes for the dual CFTs, the GB couplings should obey the following constraint \cite{Buchel:2009sk}
  \begin{eqnarray}\label{sect2: GBconstraint}
-\frac{(d-2) (3 d+2)}{4 (d+2)^2}\le \lambda_m \le \frac{(d-3) (d-2) \left(d^2-d+6\right)}{4 \left(d^2-3 d+6\right)^2}. 
  \end{eqnarray} 
As we will discuss in Section \ref{KK modes and stability}
, there are additional constraints on \(\lambda_m\) that arise from the stability of Kaluza-Klein (KK) modes on the Net-brane. It is important to note that \(L_m\) is not the AdS radius due to GB corrections. Instead, the AdS radius \(l_m\) for the branch \(B_m\) is given by \cite{Buchel:2009sk}:
  \begin{eqnarray}\label{sect2: AdS radius}
l_m^2=\frac{2\lambda_m}{1-\sqrt{1-4\lambda_m}}L_m^2,
  \end{eqnarray}  
 where $\lambda_m/(1-\sqrt{1-4\lambda_m})>0$ and $(1-4\lambda_m)>0$ according to the constraint (\ref{sect2: GBconstraint}).

To establish a well-defined action principle, it is necessary to include appropriate Gibbons-Hawking-York (GHY) terms on the Net-brane. For Gauss-Bonnet gravity, we have \cite{Myers:1987yn}:
  \begin{eqnarray}\label{sect2: GBactionbdy}
  I_{\text{GB\ NB}}&=& \sum_m^p\frac{1}{8\pi G_{N\ m}} \int_{NB} d^{d}y\sqrt{|h|} \Big( \overset{(m)}{K}+ \frac{2L_m^2 \lambda_{m}}{(d-2)(d-3)} (\overset{(m)}{J}-2 G^{ij}_{NB} \overset{(m)}{K}_{ij}) \Big) \nonumber\\
&&-\frac{1}{8\pi G_{N}}\int_{NB} d^{d}y\sqrt{|h|} T,
  \end{eqnarray}
where \( T \) represents the brane tension, \( G^{ij}_{NB} = R_{NB}^{ij} - \frac{1}{2} R_{NB} h^{ij} \) is the intrinsic Einstein tensor on the Net-brane \( NB \), \( \overset{(m)}{K}_{ij} \) denotes the extrinsic curvatures from the bulk branch \( B_m \) to the Net-brane, and \( \overset{(m)}{J} \) is the trace of the tensor defined as follows:
 \begin{eqnarray}\label{sect2: Jij}
\overset{(m)}{J}_{ij}=\frac{1}{3}\left(2 \overset{(m)}{K} \overset{(m)}{K}_{ik}\overset{(m)}{K}{}^k_j-2 \overset{(m)}{K}_{ik}\overset{(m)}{K}{}^{kl}\overset{(m)}{K}_{lj}+\overset{(m)}{K}_{ij}\left(\overset{(m)}{K}_{kl}\overset{(m)}{K}{}^{kl}-\overset{(m)}{K}{}^2\right) \right). 
  \end{eqnarray}
  Note that we have defined an effective Newton's constant for the brane tension 
 \begin{align}\label{sect2: effective GN} 
\frac{p}{G_{N}}=\sum_m^p\frac{1}{G_{N\ m}}.
\end{align} 
  Taking the variations of the total action \( I_{\text{GB}} = I_{\text{GB\ bulk}} + I_{\text{GB\ NB}} \), we obtain the following expression on the Net-brane:
  \begin{eqnarray}\label{sect2: variationGB}
\delta I_{\text{GB}}|_{NB}= \frac{1}{2} \int_{NB} d^{d}y\sqrt{|h|}  \Big(-\frac{1}{8\pi G_{N}}T h^{ij}-\sum_m^p \overset{(m)}{T}{}^{ij}_{\text{GB}} \Big)\delta h_{ij}=0,
  \end{eqnarray}
where \( \overset{(m)}{T}{}^{ij}_{\text{GB}} \) denotes the Brown-York stress tensor for Gauss-Bonnet (GB) gravity, expressed as:
 \begin{eqnarray}\label{sect2: TijGB}
\overset{(m)}{T}{}^{ij}_{\text{GB}}=\frac{1}{8\pi G_{N\ m} }\Big(\overset{(m)}{K}{}^{ij}-\overset{(m)}{K} h^{ij}+\frac{2L_m^2 \lambda_m}{(d-2)(d-3)} (\overset{(m)}{Q}{}^{ij}-\frac{1}{3}\overset{(m)}{Q} h^{ij}) \Big),
  \end{eqnarray}
 and $\overset{(m)}{Q}$ is the trace of
    \begin{eqnarray}\label{sect2: Qij}
\overset{(m)}{Q}_{ij}=3\overset{(m)}{J}_{ij}+2 \overset{(m)}{K} R_{NB\ ij}+R_{NB} \overset{(m)}{K}_{ij}-2\overset{(m)}{K}{}^{kl} R_{NB\ kilj}-4R_{NB\ k(i}\overset{(m)}{K}{}^k_{j)}.
  \end{eqnarray}
Above, we assumed that the induced metrics \(\overset{(m)}{h}_{ij}|_{NB} = h_{ij}\) on the Net-brane are continuous. Consequently, the intrinsic curvatures \(R_{NB}\) on the Net-brane do not depend on the branch index \(m\). Since the induced metrics on the Net-brane are dynamical (i.e., \(\delta h_{ij} \neq 0\)), the variational principle (\ref{sect2: variationGB}) leads to the junction condition
  \begin{eqnarray}\label{sect2: GB JC}
\text{JC:}\ \ \  \sum_m^p \overset{(m)}{T}{}^{ij}_{\text{GB}}|_{NB} =\frac{-1}{8\pi G_{N}}T h^{ij}. 
  \end{eqnarray}
  
One can also consider matter fields in bulk. The energy conservation of bulk matter fields leads to the boundary condition on the AdS boundary \cite{Guo:2025sbm}: 
\begin{align}\label{sect2: matter Tij BC} 
\overset{(m)}{T}{}^{\text{matter}}_{\hat{n}j}|_{E_m} = 0, 
\end{align} 
and the junction condition on the Net-brane \cite{Guo:2025sbm}:
\begin{align}\label{sect2: matter Tij JC} 
\sum_{m}^p \overset{(m)}{T}{}^{\text{matter}}_{\hat{n}j}|_{NB} = 0. \end{align} 
These conditions imply that no matter can flow out of the boundary, and that the total energy and tangential momentum fluxes into the Net-brane are zero. Here, $\hat{n}$ and $j$ denote the bulk normal and tangential directions to the AdS boundary and the Net-brane, respectively. 
It is essential to distinguish them from $n$ and $a$, which label the normal and tangential directions to the nodes on either edge $E_m$ or the Net-brane $NB$. The Codazzi equation for Gauss-Bonnet (GB) gravity is given by: 
\begin{align}\label{sect2: C equation} 
D_i \overset{(m)}{T}_{\text{GB}}{}^i_{\ j} = \overset{(m)}{T}{}^{\text{matter}}_{\hat{n}j}. 
\end{align} 
From equations (\ref{sect2: matter Tij BC}, \ref{sect2: matter Tij JC}, \ref{sect2: C equation}), 
we observe that $\overset{(m)}{T}{}^{ij}_{\text{GB}}$ are conserved energy-momentum tensors on both the AdS boundary $E_m$ and the Net-brane $NB$ (in sum sense): 
\begin{align}\label{sect2: C equation Em NB} 
D_i \overset{(m)}{T}{}^{ij}_{\text{GB}}|_{E_m} = 0, \quad \sum_{m}^p D_i \overset{(m)}{T}{}^{ij}_{\text{GB}}|_{NB} = 0. 
\end{align} 
Here, $\overset{(m)}{T}{}^{ij}_{\text{GB}}|_{E_m}$ is defined by (\ref{sect2: TijGB}), with $\overset{(m)}{K}$ representing the extrinsic curvature defined from the bulk to the edge $E_m$. A similar expression holds for $\overset{(m)}{T}{}^{ij}_{\text{GB}}|_{NB}$ on the Net-brane.

\subsection{Conservation law at node}

\begin{figure}[htbp]
  \centering
\includegraphics[width=.4\textwidth]{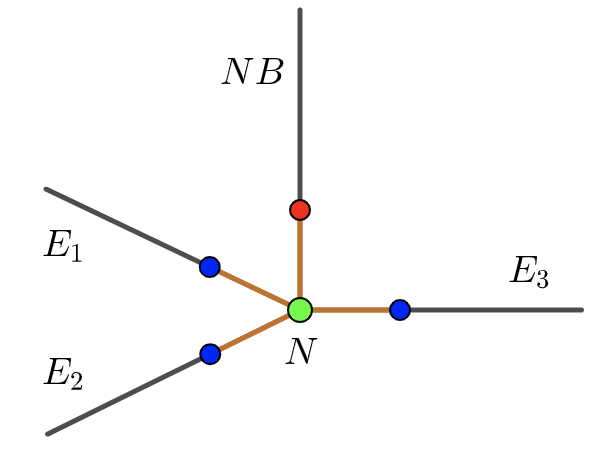}
\qquad 
\includegraphics[width=.5\textwidth]{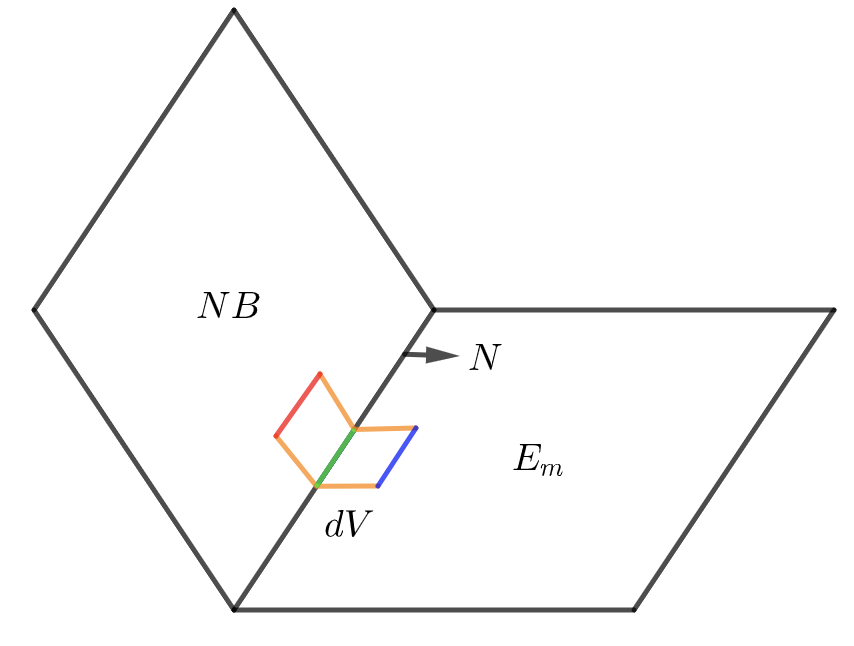}
 \caption{The region $V$, where we apply Gauss's law. This region $V$ is outlined by orange lines with red and blue endpoints in the left figure. The length of each orange line approaches zero ($ dl \to 0$), and the areas at the red, blue, and green points are all $dS$. 
In the right figure, we provide more details about $V$, which is bounded by the red, orange, and blue lines. The red, blue, and green points in the left figure correspond to the lines with the same colors in the right figure. }
  \label{proveNBC}
\end{figure}

Following the approach outlined in \cite{Guo:2025sbm}, we can demonstrate that the GB junction condition (\ref{sect2: GB JC}) on the Net-brane leads to the conservation law (\ref{NCFT BC}) at the network node. The main idea is to select an infinitesimal region $V$ that encompasses small portions of the Net-brane \(NB\), the network node \(N\), and the edges \(E_m\). See Fig. \ref{proveNBC}. We then apply Gauss's law for the conserved stress tensors (\ref{sect2: C equation Em NB}). This leads to the expression 
\begin{align}\label{sect2: physical argument 1} 
\sum_m^p  \Big( \overset{(m)}{T}{}_{\text{GB}}|_{E_m} \Big)_{na} |_N=\sum_m^p  \Big( \overset{(m)}{T}{}_{\text{GB}}|_{NB} \Big)_{na}|_N \sim h_{NB\ na}|_N=0, 
\end{align} 
where \(n\) and \(a\) represent the normal and tangential directions to the node \(N\) on either edge \(E_m\) or the Net-brane \(NB\). The left-hand side of equation (\ref{sect2: physical argument 1}) is associated with the CFT stress tensor on the edges. The third term \(\sim h_{NB\ na} \) is derived from the junction condition (\ref{sect2: GB JC}) on the Net-brane, and the final term \(h_{NB\ na}|_N=0\) vanishes due to the orthogonality between the directions \(n\) and \(a\). We have now outlined the main ideas behind our proof. For further details, please refer to \cite{Guo:2025sbm} for the case of Einstein gravity. 

In this paper, we present a simpler alternative proof. To warm up, we examine the bulk Maxwell theory and demonstrate current conservation at the network node. The action is given by
\begin{align}\label{sect2: matter action} 
I_{\text{Max}} = \sum_{m=1}^p \int_{B_m} d^{d+1}x \sqrt{|g|} \left(-\frac{1}{4 g_m^2} \overset{(m)}{\mathcal{F}}{}^{\mu\nu} \overset{(m)}{\mathcal{F}}_{\mu\nu}\right) + I_{ct},
\end{align}
 where \( \overset{(m)}{\mathcal{F}}= d \overset{(m)}{\mathcal{A}} \), \( g_m \) is the coupling constant on the edge \( E_m \), and \( I_{ct} \) represents the counterterm on the AdS boundary resulting from holographic renormalization \cite{deHaro:2000vlm}. The counterterm is added to ensure a finite action and current for the dual CFTs; its specific expression is not critical to our proof. From the action principle and assuming a continuous induced vector on the Net-brane, we derive the JC for Maxwell's fields \cite{Guo:2025sbm}
 \begin{align}\label{sect2: JC vector} 
 \sum_{m=1}^p\frac{1}{g_m^2} \overset{(m)}{\mathcal{F}}_{\hat{n} j}|_{NB}=0,
\end{align}
where $\hat{n}$ and $j$ denote the bulk normal and tangential directions to the Net-brane. 
 
Since both $\overset{(m)}{\mathcal{F}}$ and $I_{ct}$ are gauge invariant, the total action $I_{\text{Max}}$ is also gauge invariant. Consider an infinitesimal gauge transformation given by $\delta_{\alpha} \overset{(m)}{\mathcal{A}} = d\overset{(m)}{\alpha}$, and assume that the gauge parameter $\overset{(m)}{\alpha}$ is continuous when crossing the Net-brane and the node:
\begin{align}\label{sect2: Noether alpha}
 \overset{(m)}{\alpha}|_{NB}=\alpha|_{NB}, \ \   \overset{(m)}{\alpha}|_{N}=\alpha|_{N}.
\end{align}
We have:
 \begin{align}\label{sect2: gauge transformation} 
0 = \delta_{\alpha} I_{\text{Max}} = & -\sum_{m=1}^p \frac{1}{g_m^2} \int_{NB} d^{d}y \sqrt{|h|} \, \overset{(m)}{\mathcal{F}}{}^{\hat{n} i} \, \delta_{\alpha} \overset{(m)}{A}_i \nonumber \\ & + \sum_{m=1}^p \int_{E_m} d^{d}y \sqrt{|h|} \, \overset{(m)}{J}{}^i \, \delta_{\alpha} \overset{(m)}{A}_i, 
\end{align} 
where $\overset{(m)}{J}$ is the renormalized current for CFTs on the edge $E_m$, $\delta_{\alpha} \overset{(m)}{A}_i = \partial_i \overset{(m)}{\alpha}=\partial_i \alpha$, and 
$\overset{(m)}{A}{}^i = \frac{\partial y^i}{\partial x^{\mu}} \overset{(m)}{\mathcal{A}}{}^{\mu}$ denotes the induced vector on the Net-brane $NB$ and on the edges $E_m$. Due to the JC (\ref{sect2: JC vector}), the first line of (\ref{sect2: gauge transformation}) vanishes, leading us to: 
\begin{align}\label{sect2: gauge transformation 2} 
0 = \delta_{\alpha} I_{\text{Max}} = & \sum_{m=1}^p \int_{E_m} d^{d}y \sqrt{|h|} \, \overset{(m)}{J}{}^i \, D_i \alpha = \sum_{m=1}^p \int_{N} d^{d-1}y \sqrt{|\sigma|} \, \overset{(m)}{J}_n \, \alpha, 
\end{align} 
where we have used integration by parts, $D_i \overset{(m)}{J}{}^i=0$ and $N = \partial E_m$ for the last term of the equation above. For an arbitrary gauge parameter at the node, the vanishing of (\ref{sect2: gauge transformation 2}) results in current conservation (\ref{NCFT BC}) at the network node. We stress that the JC (\ref{sect2: JC vector}) is necessary for the holographic derivation of the current conservation (\ref{NCFT BC}).

Let us now incorporate the theory of gravity. Consider the Lie derivative of the renormalized action under an infinitesimal coordinate transformation defined by \(\delta x^{\mu} =- \xi^{\mu}\). Here, \(\xi\) does not need to be a Killing vector; the only requirements for \(\xi\) are as follows: First, it should not alter the positions of the Net-brane and the node:
\begin{align}\label{sect2: vector condition 1} 
\overset{(m)}{\xi}_{\hat{n}}|_{NB} = \overset{(m)}{\xi}_{n}|_{N} = 0.
\end{align} 
Second, it must be continuous when crossing the Net-brane and the node, indicating that 
\begin{align}\label{sect2: vector condition 2} 
\overset{(m)}{\xi}{}^i|_{NB}=\xi^i|_{NB},\ \ \overset{(m)}{\xi}{}^a|_{N}=\xi^a|_{N}.
\end{align} 
The renormalized action \(I_{\text{ren}}\), or equivalently, the effective action of NCFTs, remains invariant under such an infinitesimal coordinate transformation. Therefore, we have: 
\begin{align}\label{sect2: d Iren} 
0 = \mathcal{L}_{\xi} I_{\text{ren}} &= \int_{NB} d^{d}y \sqrt{|h|} \left[ \left(-\frac{1}{16\pi G_{N}}T h^{ij} -\sum_{m}^p \frac{1}{2} \overset{(m)}{T}{}^{ij}_{\text{GB}}\right) \mathcal{L}_{\xi} h_{ij} + \sum_{m}^p \frac{-1}{g_m^2} \overset{(m)}{\mathcal{F}}{}^{\hat{n}i} \mathcal{L}_{\xi} A_i \right] \nonumber\\
&+ \sum_{m}^p \int_{E_m} d^{d}y \sqrt{|h|} \left( \frac{1}{2} \overset{(m)}{T}{}^{ij} \mathcal{L}_{\xi} \overset{(m)}{h}_{ij} + \overset{(m)}{J}{}^i \mathcal{L}_{\xi} \overset{(m)}{A}_i \right), 
\end{align} 
where \(\overset{(m)}{T}{}^{ij}\) is the renormalized stress tensor of NCFTs. On the Net-brane, the induced metric and vector are independent of the index $m$, while on the edge $E_m$, we have \(\mathcal{L}_{\xi} \overset{(m)}{h}_{ij} = D_i \overset{(m)}{\xi}_j + D_j \overset{(m)}{\xi}_i\), and \(\mathcal{L}_{\xi} \overset{(m)}{A}_i = \overset{(m)}{\xi}{}^j D_j \overset{(m)}{A}_i + \overset{(m)}{A}_j D_i \overset{(m)}{\xi}{}^j\). The first line of (\ref{sect2: d Iren}) vanishes due to the JCs (\ref{sect2: GB JC}, \ref{sect2: JC vector}). 
By integrating by parts and utilizing the relations \(D_i \overset{(m)}{T}{}^{ij} = \overset{(m)}{J}_{i} \overset{(m)}{F}{}^{ji}\) and \(D_i \overset{(m)}{J}{}^{i} = 0\), we derive:
\begin{align}\label{sect2: d Iren 1} 
0=\mathcal{L}_{\xi} I_{\text{ren}}&=\sum_m^p  \int_{N} d^{d-1}y\sqrt{|\sigma|} \Big(  \overset{(m)}{T}{}_{ni}\overset{(m)}{\xi}{}^i + \overset{(m)}{J}_n \overset{(m)}{A}_i \overset{(m)}{\xi}{}^i\Big)\nonumber\\
&=\sum_m^p  \int_{N} d^{d-1}y\sqrt{|\sigma|} \overset{(m)}{T}{}_{na }\xi^a,
\end{align} 
where we have applied the conditions (\ref{sect2: vector condition 1},\ref{sect2: vector condition 2}), $(\overset{(m)}{A}_i \overset{(m)}{\xi}{}^i)|_N=(A_a \xi^a)|_N$ and the conservation of current \(\sum_{m}^p \overset{(m)}{J}_n|_N = 0\). For an arbitrary vector \(\xi^a|_N\) at the node, the above equation implies that the total energy and tangential momentum fluxes into the node are zero:
\begin{align}\label{sect2: sum Tna} 
 \sum_{m}^p \overset{(m)}{T}_{n a}|_N = 0.
\end{align} 
We emphasize that the JCs (\ref{sect2: GB JC}, \ref{sect2: JC vector}) are essential for the holographic discussions presented here. Thus, we have completed the proof of current and energy conservation (\ref{NCFT BC}) at the node based on the junction conditions (\ref{sect2: GB JC}, \ref{sect2: JC vector}) on the Net-brane. The proof presented in this section is simpler than that of \cite{Guo:2025sbm} and does not require handling local Killing vectors in Gauss normal coordinates.

\subsection{KK modes and stability}
\label{KK modes and stability}

This subsection studies the gravitational Kaluza-Klein (KK) modes and its stability on the Net-brane. For simplicity, we focus on the network of $p$ edges $E_{m}$ linked by one node $N$. In each bulk branch $B_{m}$, we take the following ansatz of the perturbative metric
\begin{align}\label{sect2: perturbative metric}
ds^2=dr^2+l_{m}^{2}\cosh^2 \left(\frac{r}{l_{m}}\right) \left( \bar{h}^{(0)}_{ij}(y) + \epsilon \overset{(m)}{H}(r) \bar{h}^{(1)}_{ij}(y)  \right)dy^i dy^j+O(\epsilon^2),
\end{align}
where $\bar{h}^{(0)}_{ij}$ denotes the AdS metric with a unit radius and $\bar{h}^{(1)}_{ij}$ denotes the perturbation obeying the transverse traceless gauge 
 \begin{eqnarray}\label{sect2: hij1gauge}
\bar{D}^i \bar{h}^{(1)}_{ij}=0,\ \ \  \bar{h}^{(0)ij}\bar{h}^{(1)}_{ij}=0,
\end{eqnarray}
where $\bar{D}_i$ is the covariant derivative with respect to $\bar{h}^{(0)}_{ij}$. 
For simplicity, we have suppressed the index \((m)\) for \(\overset{(m)}{r}\). We will reinstate it when necessary. 
By separating variables of the linearized Einstein equations, we obtain
\begin{eqnarray}\label{sect2: EOMMBCmassivehij}
&& \left(\bar{\Box}+2-M^2\right)\bar{h}^{(1)}_{ij}(y)=0,\\
&& l_{m}^{2}\cosh^2\left(\frac{r}{l_{m}}\right) \overset{(m)}{H}{}''(r)+dl_{m} \sinh \left(\frac{r}{l_{m}}\right) \cosh \left(\frac{r}{l_{m}}\right)\overset{(m)}{H}{}'(r) + M^2 \overset{(m)}{H}(r)=0, \label{sect2: EOMMBCmassiveH}
\end{eqnarray}
where $M$ denotes the mass of gravitons on the Net-brane. Solving the radial equation of motion (EOM) (\ref{sect2: EOMMBCmassiveH}) and imposing the Dirichlet boundary condition (DBC) on the AdS boundary $\overset{(m)}{H}(-\infty)=0$, we obtain
\begin{equation}\label{sect2: Htwocase}
\overset{(m)}{H}(r)= \overset{(m)}{c}H_{m}(r)= \overset{(m)}{c} \begin{cases}
 \ \text{sech}^{\frac{d}{2}}(r/l_{m})  \ P_{l_M}^{\frac{d}{2}}(-\tanh (r/l_{m})),&\
\text{even $d$} ,\\
\  \text{sech}^{\frac{d}{2}}(r/l_{m}) \ Q_{l_M}^{\frac{d}{2}}(-\tanh (r/l_{m})),&\
\text{odd $d$}.
\end{cases}
\end{equation}
where $\overset{(m)}{c}$ are constants, $P$ and $ Q$ denote the Legendre polynomials, and $l_M$ is given by
 \begin{eqnarray}\label{sect2: aibia1}
l_M=\frac{1}{2} \left(\sqrt{(d-1)^2+4  M^2}-1\right).
\end{eqnarray}
Here, we ignore the mass subscript $M$ of $\overset{(m)}{H}_{M}$ and $H_{M,m}$.
In each branch $B_{m}$, we assume the invariant location of Net-brane under the linear perturbation:
\begin{align}
    \text{Net-brane}:~\overset{(m)}{r}=\rho_{m}+O(\epsilon^{2}).
\end{align}

Recall that we require an identical induced metric from each branch \( B_m \), which leads to the following conditions:
\begin{align}\label{sect 2: induced metric condition}
    l_{m}\cosh{\left(\frac{\rho_{m}}{l_{m}}\right)}=l_{1}\cosh{\left(\frac{\rho_{1}}{l_{1}}\right)},~\overset{(m)}{H}(\rho_{m})=\overset{(1)}{H}(\rho_{1}).
\end{align}
The leading order of the JC (\ref{sect2: GB JC}) provides the parameterization of brane tension as follows:
\begin{align}\label{sect 2: brane tension GB}
    T=\sum_{m}^{p}\frac{G_{N}}{G_{N~m}}\frac{d-1}{l_{m}}\tanh{\left( \frac{\rho_{m}}{l_{m}} \right)}\left[1-\frac{1-\sqrt{1-4\lambda_{m}}}{6}\left(5+\cosh\left( \frac{2\rho_
    {m}}{l_{m}} \right)\right)\text{sech}^{2}\left( \frac{\rho_{m}}{l_{m}} \right)\right].
\end{align}
Using the relation $\overset{(m)}{H}(\rho_{m})=\overset{(1)}{H}(\rho_{1})$, the linearized order of JC (\ref{sect2: GB JC}) yields a matrix equations: 
$\mathcal{M}\cdot \mathcal{C}=0$, where $\mathcal{C}=(\overset{(1)}{c},\overset{(2)}{c},... )^T$. To obtain non-zero solutions, we require the determinant of the matrix to be zero, i.e., \(|\mathcal{M}| = 0\). This leads to the following constraint on the spectrum of Kaluza-Klein (KK) modes:
\begin{align}\label{sect2: spectrum constraint}
    \sum_{m}^{p}\left[\left(A_{m} {H}_{m}'(\rho_{m}) -C_{m}M^{2}{H}_{m}(\rho_{m})\right)\Pi_{q\neq m}{H}_{q}(\rho_{q})\right]=0. 
\end{align}
Here, we define the coefficients:
\begin{align}\label{sect2: AmBm}
    A_{m}=\sqrt{1-4\lambda_{m}}\frac{G_{N}}{G_{N~m}},~C_{m}=\frac{1-\sqrt{1-4\lambda_{m}}}{d-2}\frac{G_{N}}{G_{N~m}}\frac{2}{l_{m}}\frac{\tanh{\left(\frac{\rho_{m}}{l_{m}}\right)}}{\cosh^{2}{\left(\frac{\rho_{m}
    }{l_{m}}\right)}}.
\end{align}
Note that we have $A_m>0$ from (\ref{sect2: GBconstraint}, \ref{sect2: effective GN}) and $G_{N~m}>0$. Following the approach of \cite{Miao:2023mui}, we derive the ghost-free and tachyon-free condition for the gravitational KK modes:
\begin{align}\label{sect2: ghost-free condition}
    \sum_{m}^{p}C_{m}=\sum_m^p\frac{1-\sqrt{1-4\lambda_{m}}}{d-2}\frac{G_{N}}{G_{N~m}}\frac{2}{l_{m}}\frac{\tanh{\left(\frac{\rho_{m}}{l_{m}}\right)}}{\cosh^{2}{\left(\frac{\rho_{m}
    }{l_{m}}\right)}} \ge 0. 
\end{align}
Please refer to Appendix \ref{app A} for the derivations. We remark that the gravitational KK modes are linearly stable under the above condition. In particular, as shown in Appendix \ref{app A}, we have positive mass spectrum \(M^2 > 0\) under this condition.

It should be noted that, in addition to the constraint (\ref{sect2: ghost-free condition}) on gravitational KK modes, there are further constraints arising from the unitarity and causality of the dual NCFTs. For an example in AdS$_3$/NCFT$_2$, refer to Section \ref{Wedge inclusion condition}. Discussions of higher-dimensional cases are left for future work. A crucial point is that the tension of the Net-brane must be positive and sufficiently large to ensure a unitary and causal theory.

Next, we will analyze the spectrum of KK modes that obey the constraint (\ref{sect2: spectrum constraint}). To warm up, we will first examine the symmetric case, where the theory parameters are uniform across all bulk branches: \( G_{N\ m} = G_N \), \( \lambda_m = \lambda \), \( l_m = l \), and \( \rho_m = \rho \). In other words, we have identical NCFTs at all network edges. Under these conditions, the spectrum described in (\ref{sect2: spectrum constraint}) simplifies to:
\begin{align}\label{sect2: spectrum constraint case I}
 \left(A_1 H'_1(\rho) -C_{1}M^{2} H_1(\rho)\right) H_{1}(\rho)^{p-1}=0,
\end{align}
where the index $1$ can be replaced with any index $m$. Clearly, we obtain \( (p-1) \) classes of degenerate modes that obey Dirichlet boundary conditions (DBC):
\begin{align}\label{sect2: spectrum case I DBC}
\text{DBC}:\  H_{1}(\rho)=0,
\end{align}
and one class of modes that comply with Neumann boundary conditions (NBC):
\begin{align}\label{sect2: spectrum case I NBC}
\text{NBC}:\   \left(A_1 H'_1(\rho) -C_{1}M^{2} H_1(\rho)\right)=0.
\end{align}
We remark that (\ref{sect2: spectrum case I NBC}) is just the NBC for Gauss-Bonnet gravity in AdS/BCFT. Similar to a BCFT, the modes obeying NBC (\ref{sect2: spectrum case I NBC}) are confined to one edge, $\overset{(m)}{T}_{na}|_N=0$. We refer to these modes as isolated modes \cite{Guo:2025sbm}. In contrast, the modes satisfying DBC (\ref{sect2: spectrum case I DBC}) can move freely between edges, and we designate them as transparent modes \cite{Guo:2025sbm}. 

\begin{figure}[!h]
    \centering
    \includegraphics[width=0.5\linewidth]{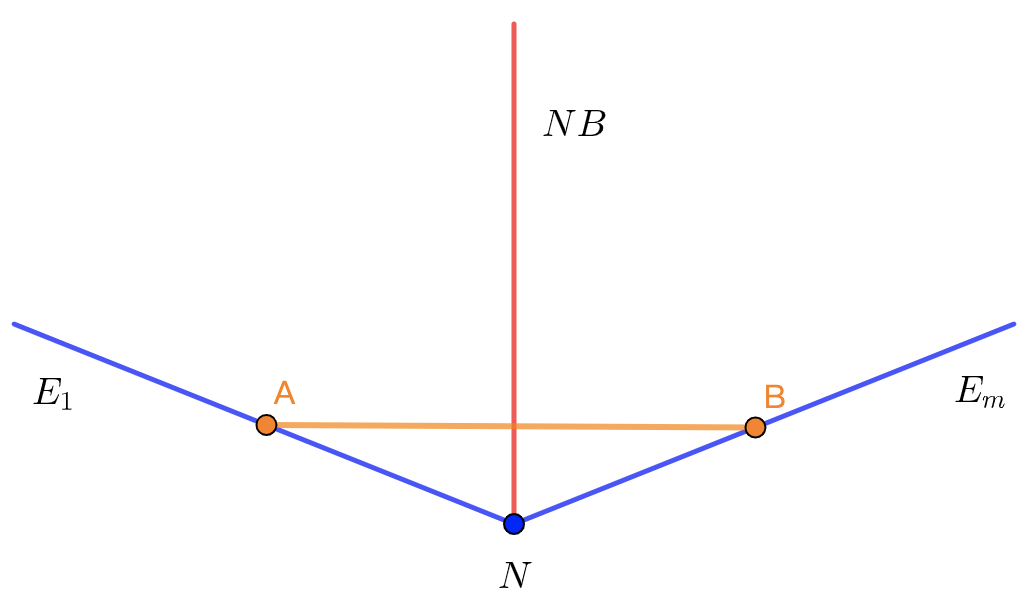}
    \caption{When $\rho<0$, the bulk path connecting two boundary points A and B is shorter than the one on the AdS boundary. It implies that, from the boundary perspective, one can superluminally transmit signals from A to B, which violates the causality of NCFTs.}
    \label{causal GB}
\end{figure}

For the symmetric case, the stability conditions (\ref{sect2: GBconstraint}, \ref{sect2: ghost-free condition}) become:
\begin{align}\label{sect2: stability conditions case I1}
&\text{for}\ \rho\ge 0:\ \ 0\le \lambda \le  \frac{(d-3) (d-2) \left(d^2-d+6\right)}{4 \left(d^2-3 d+6\right)^2},\\ \label{sect2: stability conditions case I2}
&\text{for}\ \rho\le 0:\ \ -\frac{(d-2) (3 d+2)}{4 (d+2)^2}\le \lambda \le 0.
\end{align}
Note that for a causal NCFT, the parameter \(\rho\) must be non-negative.  As illustrated in Fig. \ref{causal GB}, when \(\rho < 0\), the bulk path connecting two boundary points, A and B, is shorter than the path along the AdS boundary. This scenario implies that, from the boundary perspective, it is possible to transmit signals superluminally from A to B, thereby violating the causality of NCFTs. Consequently, we focus on the case where \(\rho \geq 0\) in (\ref{sect2: stability conditions case I1}). Remarkably, this condition results in a positive GB coupling \(\lambda\), which is stronger than the constraint (\ref{sect2: GBconstraint}) in AdS/CFT.

Let us now calculate the spectrum for some typical parameters $(\rho, \lambda)=(1, 9/100)$ and $(1, 0)$ for $l=1$, $d=4$ and $p=3$. See the Table \ref{tab: KK modes 1}, where $M^2>0$ implies stability. Additionally, Table \ref{tab: KK modes 1} reveals that the spectrum includes one NBC mode (\ref{sect2: spectrum case I NBC}) and two degenerate DBC modes (\ref{sect2: spectrum case I DBC}).

\begin{table}[htbp]
    \centering  
    \caption{Spectrum for KK modes for symmetric case}  
    \begin{tabular}{|c|c|c|c|c|c|c|c|c|c|c|}  
        \hline
        $\rho_{1,2,3}$&$\lambda_{1,2,3}$&\multicolumn{9}{|c|}{$M^{2}$}\\ 
        \hline
        1 &9/100 & 0.47 & 4.50 & 4.50 & 5.90 &12.07 &12.07 &14.39&22.91 &22.91 \\  
        \hline
        1&0  &0.64  & 4.50 &4.50 &6.50 &12.07 &12.07 &15.65 &22.91 &22.91 \\  
        \hline
    \end{tabular}
    \label{tab: KK modes 1}  
\end{table}

Let us proceed to discuss the case with varying theory parameters in spacetime branches. In general, the spectrum condition cannot be separated into DBC and NBC. Specifically, the NBC (\ref{sect2: spectrum case I NBC}) contradicts the spectrum constraint generally. As a result, there cannot be any isolated eigenmodes that satisfy the spectrum constraint while also adhering to the isolated condition \(\overset{(m)}{T}_{na}|_N=0\). However, it is possible to find suitable linear combinations of all modes that meet the requirement \(\overset{(m)}{T}_{na}|_N=0\), even though no single mode does. 
In principle, any solution that obeys the conservation law \(\sum_m \overset{(m)}{T}_{na}|_N=0\), including the isolated ones where \(\overset{(m)}{T}_{na}|_N=0\), can be expanded using the complete basis of the spectrum. This issue falls outside the paper's primary focus, and we will address it in future work.

To end this section, we present the spectrum for the first few KK modes for the asymmetric case.
Without loss of generality, we choose the parameters $G_{N\,1}=2G_{N\,2}=3G_{N\,3}$, $l_{1}=1,~l_{2}=0.95,~l_{3}=0.9$
, $\rho_{1,2,3}\approx (1,1.01,1.02)$ 
and $p=3$ for $d=4$. We note that these parameters comply with the stability constraints (\ref{sect2: GBconstraint}, \ref{sect2: ghost-free condition}). As shown in Table \ref{tab: KK modes 2}, we observe a positive spectrum \(M^2>0\), which confirms stability under the conditions specified. Note that the case $\lambda_{1,2,3}=0$ corresponds to Einstein gravity. 

\begin{table}[htbp]
    \centering  
    \caption{Spectrum for KK modes for asymmetric case}  
    \begin{tabular}{|c|c|c|c|c|c|c|c|c|c|c|c|}  
        \hline
        $\lambda_{1}$&$\lambda_{2}$&$\lambda_{3}$&\multicolumn{9}{|c|}{$M^{2}$}\\ 
        \hline
        $0$&$0$&$0$ & 0.55 & 4.38 &4.48 &6.17&11.63&11.99&14.94&21.96&22.73\\  
        \hline
        $0.01$&$0.05$&$0.08$ & 0.46 & 4.37 &4.48 &5.83&11.62&11.98&14.23&21.94&22.70\\  
        \hline
    \end{tabular}
    \label{tab: KK modes 2}  
\end{table}

\subsection{Holographic entanglement entropy}
\label{Holographic entanglement entropy}
\begin{figure}[!h]
    \centering
    \includegraphics[width=0.6\linewidth]{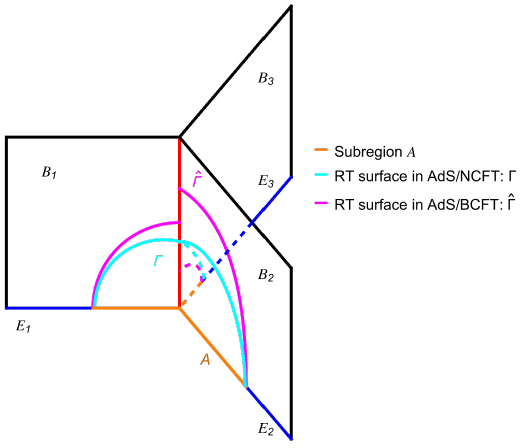}
    \caption{RT surfaces in AdS/NCFT. The orange lines represent the connected subregion $A$, while the cyan-blue curves indicate the dual RT surfaces $\Gamma$, which remain continuous across the Net-brane in AdS/NCFT. In contrast, the magenta curves corresponding to the RT surfaces $\hat{\Gamma}$ in AdS/BCFT for each bulk branch \( B_m \) are usually discontinuous across the Net-brane. }
    \label{GBRT}
\end{figure}

This subsection investigates the holographic entanglement entropy (HEE) for the network with different CFTs at its edges. For simplicity, we focus on a connected subsystem \(A\) within the network, consisting of one node and \(p\) edges. Refer to Fig. \ref{GBRT} for the geometry: the orange lines represent the subsystem \( A \), the cyan-blue curves indicate the Ryu-Takayanagi (RT) surfaces in AdS/NCFT, and the magenta curves denote the RT surfaces in AdS/BCFTs for each bulk branch. Note that the RT surfaces for NCFTs are interconnected at the same intersection on the Net-brane. As argued in \cite{Guo:2025sbm}, it is natural that a connected subsystem of NCFTs corresponds to a connected RT surface in bulk. Unlike the end-of-the-world (EOW) brane in AdS/BCFT, the Net-brane is not a boundary of the spacetime in AdS/NCFT. Consequently, physical continuity requires that RT surfaces be connected across the Net-brane. For a more detailed discussion on the connected RT surfaces in AdS/NCFT, please refer to \cite{Guo:2025sbm}.

Let us now focus on the connected RT surfaces in AdS/NCFT. The HEE for Gauss-Bonnet gravity is given by \cite{Hung:2011xb}
\begin{align} \label{sect2: HEEGB}
S &=\sum_m^p \frac{1}{4 G_{N\ m}} \int_{\Gamma_m} d^{d-1}\xi\sqrt{h_{\Gamma_m}}\left(1+\frac{2L_m^2 \lambda_{m}}{(d-2)(d-3)} \mathcal{R}_m \right) \nonumber\\
&+  \sum_m^p \frac{1}{ G_{N\ m}} \int_{\gamma} d^{d-2}s\sqrt{h_{\gamma}} \frac{L_m^2 \lambda_{m}}{(d-2)(d-3)}\mathcal{K}_m,
\end{align}
where $\Gamma_m$ denotes the RT surface in the branch $B_m$, $\mathcal{R}_m$ is the intrinsic Ricci scalar on $\Gamma_m$, $\gamma=\Gamma_m\cap NB$ represents the intersection of the RT surfaces with the Net-brane $NB$, and $\mathcal{K}_m$ is the extrinsic curvature defined from $\Gamma_m$ to $\gamma$.

Now, let us discuss how the RT surfaces in different branches are interconnected on the Net-brane. Similar to the junction condition for the metric, the connecting condition for RT surfaces can be derived from the variational principle. For detailed derivations, please refer to Appendix \ref{app B}. 
Below, we list the junction condition for RT surfaces:
 \begin{align}\label{sect2: JC RT}
&\sum_{m=1}^p \frac{1}{G_{N\ m}} \Big[\overset{(m)}{n}_{\alpha} \left( (1+\hat{\lambda}_m \mathcal{R}_m )h_{\Gamma_{m}}^{\alpha\beta}-2\hat{\lambda}_m \mathcal{R}_m^{\alpha\beta} \right) \partial_{\beta} \overset{(m)}{x}{}^{\nu}g_{\mu\nu}\nonumber\\
&-2\hat{\lambda}_m \left( \hat{D}_a (\mathcal{K}_m h_{\gamma}^{ab}-\mathcal{K}_m^{ab})\partial_b \overset{(m)}{x}{}^{\nu} g_{\mu\nu} +  (\mathcal{K}_m h_{\gamma}^{ab}-\mathcal{K}_m^{ab})  \overset{(m)}{K}_{\gamma\ \mu ab}\right) \Big] \frac{\partial \overset{(m)}{x}{}^{\mu}}{\partial y^i}|_{\gamma}=0,
\end{align}
where $\hat{\lambda}_m=2L_m^2 \lambda_{m}/(d-2)(d-3)$, $\overset{(m)}{n}{}^{\alpha}$ are the unit vectors normal to the intersection $\gamma=\Gamma_m\cap NB$ directed along the RT surface $\Gamma_m$ approaching the Net-brane $NB$, $x^{\mu}$, $\xi^{\alpha}$, $ s^a$ and $y^i$ are the coordinates of the bulk branch $B_m$, the RT surfaces $\Gamma_m$, the intersection $\gamma$ and the Net-brane $NB$, respectively. The extrinsic curvatures \(\overset{(m)}{K}_{\gamma\ \mu ab}\) are defined from the bulk branch \(B_m\) to the intersection \(\gamma\). See (\ref{app B: Kgamma}) for its explicit expression.

In AdS/BCFT, the holographic boundary entropy \cite{Takayanagi:2011zk}
 \begin{align}\label{sect2: boundary entropy}
S_{\text{bdy}}=S_{\text{BCFT}}-S_{\text{CFT}}=S_{\text{BCFT}}(\rho)-S_{\text{BCFT}}(0),
\end{align}
decreases with the boundary RG flow
 \begin{align}\label{sect2: g theorem}
\partial_{\rho} S_{\text{bdy}}\ge 0,
\end{align}
which is the so-called g-theorem. In this context, $S_{\text{CFT}} = S_{\text{BCFT}}(0)$, and $\rho$ is related to the brane tension $T = (d-1) \tanh(\rho)$, 
which labels the boundary energy scale.  We remark that the EOW brane approaches the AdS boundary for large brane tensions $\rho\to \infty$, while it extends deep into the bulk spacetime for small brane tensions $\rho \to 0$. Therefore, it is natural to associate a large $\rho$ with the ultraviolet (UV) regime and a small $\rho$ with the infrared (IR) regime. Consequently, (\ref{sect2: g theorem}) means 
 \begin{align}\label{sect2: g theorem 1}
 S_{\text{bdy}}|_{\text{UV}} \ge  S_{\text{bdy}}|_{\text{IR}}.
\end{align}

Similar to the boundary entropy, in AdS/NCFT, we can define two types of network entropy as follows \cite{Guo:2025sbm}:
\begin{align}\label{sect2: network entropy I}
S_{\text{I}}=S_{\text{NCFT}}-S_{\text{CFT}},
\end{align}
and 
 \begin{align}\label{sect2: network entropy II}
S_{\text{II}}=S_{\text{NCFT}}(\rho_m)-S_{\text{NCFT}}(0). 
\end{align}
Since $S_{\text{CFT}}=S_{\text{BCFT}}(0) \ne S_{\text{NCFT}}(0)$ generally, $S_{\text{I}}$ and $S_{\text{II}}$ are different. The values of \( S_{\text{BCFT}} \) and \( S_{\text{NCFT}} \) are determined using the disconnected magenta curves and the connected cyan-blue curves in Fig. \ref{GBRT}, respectively.
For identical CFTs on edges, \( S_{\text{I}} \) equals \( S_{\text{II}} \) for symmetric subsystems with the same edge lengths \( \overset{(m)}{L} = L \). However, they differ when \( \overset{(m)}{L} \neq \overset{(n)}{L} \) \cite{Guo:2025sbm}. Furthermore, it has been verified in \cite{Guo:2025sbm} that both \( S_{\text{I}} \) and \( S_{\text{II}} \) satisfy the holographic g-theorem in AdS$_3$/NCFT$_2$:
\begin{align}\label{sect2: network g theorem}
\partial_{\rho} S_{\text{I}}\ge 0,\  \partial_{\rho} S_{\text{II}}\ge 0,
\end{align}
even in cases where \( \overset{(m)}{L} \neq \overset{(n)}{L} \).

\begin{figure}[!h]
    \centering
    \includegraphics[width=0.6\linewidth]{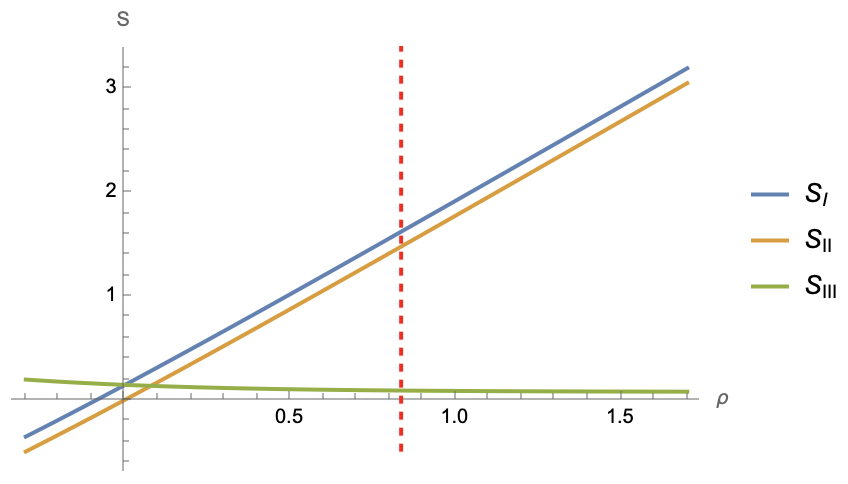}
    \caption{Network entropies. Without loss of generality, we select the parameters \( p = 3 \), \( \overset{(1)}{L} = 1 \), \( \overset{(2)}{L} = 2 \), \( \overset{(3)}{L} = 3 \), \( 4 G_{N\ 1} = 3 \), \( 4 G_{N\ 2} = 2 \), and \( 4 G_{N\ 3} = 1 \). It demonstrates that \( S_{\text{I}} \) and \( S_{\text{II}} \) increase as the tension parameter \( \rho \) rises, while \( S_{\text{III}} \) decreases. Additionally, \( S_{\text{III}} \) remains non-negative (i.e., \( S_{\text{III}} \ge 0 \)), whereas \( S_{\text{I}} \) and \( S_{\text{II}} \) can be either positive or negative. The causal constraint (\ref{sect 3: tension bound final2}) requires $\rho \gtrsim 0.83$, which is labeled by the dotted red line.}
    \label{Netentropy}
\end{figure}

Let us examine whether the holographic g-theorem holds for different CFTs at the network edges. To warm up, we first study the case of AdS$_3$/NCFT$_2$ with the same AdS radius $l_m=l=1$ but different Newton's constant $G_{N\ m}$ at the edges. According to (\ref{sect 2: induced metric condition}), we have the same tension parameters $\rho_m=\rho$, when \( l_m = l \). In the Poincaré  AdS$_3$, the area of RT surface $\Gamma_m$ in each branch $B_{m}$ 
is given by \cite{Guo:2025sbm}
\begin{align}\label{sect2: minimal surface}
\overset{(m)}{A}=\log \left(\frac{(\overset{(m)}{L}-x_b)^2+z_b^2}{z_b \epsilon }\right),
\end{align}
where \( \epsilon \) is the UV cut-off of \( z \), \( x_b = -\sinh(\rho) z_b \), and \((z_b, x_b)\) represents the common intersection point of the RT surfaces on the Net-brane. Note that the Gauss-Bonnet terms vanish in AdS$_3$. Consequently, the HEE (\ref{sect2: HEEGB}) becomes
 \begin{align}\label{sect2: AdS3 SNCFT}
S_{\text{NCFT}}(\rho)=\sum_m^p \frac{1}{4 G_{N\ m}} \log \left(\frac{(\overset{(m)}{L}+\sinh(\rho) z_b)^2+z_b^2}{z_b \epsilon }\right),
\end{align}
where $z_b$ can be determined by minimizing the above entropy. On the other hand, $S_{\text{BCFT}}$ and $S_{\text{CFT}}$ are given by \cite{Guo:2025sbm}
 \begin{align}\label{sect2: AdS3 SBCFT}
&S_{\text{BCFT}}(\rho)=\sum_m^p \frac{1}{4 G_{N\ m}} \left( \log (\frac{2\overset{(m)}{L}}{ \epsilon })+\rho\right),\\
&S_{\text{CFT}}=S_{\text{BCFT}}(0)=\sum_m^p \frac{1}{4 G_{N\ m}} \log (\frac{2\overset{(m)}{L}}{ \epsilon }). \label{sect2: AdS3 SCFT}
\end{align}
In the symmetric case, where \( \overset{(m)}{L} = L \), we get $z_b=L \text{sech}(\rho )$. From the above equations, we can derive the network entropies
\begin{align}\label{sect2: SI}
S_{\text{I}}=S_{\text{II}}=\sum_m^p \frac{\rho}{4 G_{N\ m}}=\frac{p \rho}{4 G_{N}},
\end{align}
which obey the holographic g theorem (\ref{sect2: network g theorem}). In the asymmetric case with $\overset{(m)}{L}\ne \overset{(n)}{L}$, we can calculate the network entropies numerically and verify that they also satisfy the holographic g-theorem. See Fig. \ref{Netentropy} for example. 

In addition to \( S_{\text{I}} \) and \( S_{\text{II}} \), there is a third type of network entropy, defined as the difference in entropy between NCFT and BCFT \cite{Guo:2025sbm}:
 \begin{align}\label{sect2: network entropy II}
S_{\text{III}}=S_{\text{NCFT}}-S_{\text{BCFT}}, 
\end{align}
where \( S_{\text{NCFT}} \) corresponds to the connected cyan-blue curves and \( S_{\text{BCFT}} \) corresponds to the disconnected magenta curves in Fig. \ref{GBRT}. By definition, $S_{\text{BCFT}}$ represents the minimal entropy, with the RT surfaces terminating on the Net-brane without necessarily intersecting at the same point. In contrast, $S_{\text{NCFT}}$ requires the RT surfaces to be interconnected at the same intersection on the Net-brane. This leads to the conclusion that $S_{\text{NCFT}} \ge S_{\text{BCFT}}$, resulting in a non-negative value for $S_{\text{III}}$. In AdS/NCFT, $S_{\text{BCFT}}$ is independent of the internal edges, as the corresponding disconnected RT surfaces can extend to infinity and result in zero areas \cite{Guo:2025sbm}. On the other hand, $S_{\text{NCFT}}$ is dependent on the internal edges because the connected condition prevents the RT surfaces from going to infinity for these internal edges. Therefore, $S_{\text{III}}$ serves as a good measure of the internal edges. On the other hand, $S_{\text{I}}$ and $S_{\text{II}}$ provide good descriptions of the node degrees of freedom, as they obey the holographic g theorem. From (\ref{sect2: AdS3 SNCFT},\ref{sect2: AdS3 SBCFT}), we can derive $S_{\text{III}}$ in AdS$_3$/NCFT$_2$.  As illustrated in Fig. \ref{Netentropy}, $S_{\text{III}}\ge 0$ and disobeys the holographic g theorem (\ref{sect2: network g theorem}).

Let us further discuss the various network entropies illustrated in Fig. \ref{Netentropy}. We select the following parameters: \( p = 3 \), \( \overset{(1)}{L} = 1 \), \( \overset{(2)}{L} = 2 \), \( \overset{(3)}{L} = 3 \), \( 4 G_{N\ 1} = 3 \), \( 4 G_{N\ 2} = 2 \), and \( 4 G_{N\ 3} = 1 \). The results show that \( S_{\text{I}} \) and \( S_{\text{II}} \) increase as the tension parameter \( \rho \) rises, while \( S_{\text{III}} \) decreases. According to Section \ref{Wedge inclusion condition}, 
the causal constraint (\ref{sect 3: tension bound final2}) establishes a lower bound on brane tension, which is \( \rho \gtrsim 0.83 \). This limit is indicated by the dotted red line in Fig. \ref{Netentropy}. For all physical parameters where \( \rho \gtrsim 0.83 \), all network entropies remain positive. However, for non-physical parameters, \( S_{\text{I}} \) and \( S_{\text{II}} \) may become negative, while \( S_{\text{III}} \) remains positive. Besides, we have \( S_{\text{I}} >S_{\text{II}} >S_{\text{III}} \) within the physical parameters.

Next, let us discuss higher-dimensional HEE with non-trivial Gauss-Bonnet terms. We focus on the case with the same AdS radius $l_m=1$ but different Newton's constants $G_{N\ m}$ and GB couplings $\lambda_m$ across various bulk branches. We choose the symmetric subsystem $A$ to be a hemisphere at each edge. The advantage of this choice is that it admits an analytical expression for the RT surface, and the corresponding entanglement entropy can be taken as the g-function in higher dimensions \cite{Hu:2022ymx}. For convenience, we adopt the following form of the AdS metric in each branch \( B_m \):
\begin{eqnarray}\label{sect2: AdSmetricbrane}
 ds^2=d\overset{(m)}{r}{}^2+\cosh^2 (\overset{(m)}{r}{}) \frac{dw^2-dt^2+ dr^2+ r^2 d\Omega^2}{w^2}, \ \  -\rho\le \overset{(m)}{r}<\infty,
\end{eqnarray}
where \( d\Omega \) represents the line element of a unit \((d-3)\)-sphere, 
the Net-brane is located at $\overset{(m)}{r}=-\rho$ and the AdS boundary is at $\overset{(m)}{r}=\infty$. The hemisphere on the AdS boundary is given by \cite{Hu:2022ymx}:
\begin{eqnarray}\label{halfball1}
w^2+ r^2 \le r_0^2, \ w\ge 0,
\end{eqnarray}
with \( r_0 \) being the same at all edges. 
Note that $r$ differs from $\overset{(m)}{r}$.
The dual RT surface in bulk is given by \cite{Akal:2020wfl}
\begin{eqnarray}\label{sec2: RTsurface ball}
w^2+r^2=r_0^2, \ \ t=\text{constant},
\end{eqnarray}
for Einstein gravity. Due to the symmetry of the hemisphere, this expression also serves as the RT surface for Gauss-Bonnet gravity \cite{Hu:2022ymx}. One can check that (\ref{sec2: RTsurface ball}) satisfies the extremal condition (\ref{app B: extremal surface}) in bulk and the junction condition (\ref{app B: JC RT}) on the Net-brane for Gauss-Bonnet gravity. 

From (\ref{sect2: AdSmetricbrane},\ref{sec2: RTsurface ball}), we obtain \cite{Hu:2022ymx}
 \begin{eqnarray}\label{sect2: formulas}
 && \mathcal{R}_m=-(d-1)(d-2),\ \ \  \mathcal{K}_m= (d-2) \tanh (\rho ),\nonumber\\
&& \int_{\Gamma_m} d\xi^{d-1} \sqrt{h_{\Gamma_m}}=\int_{-\rho}^{\infty} \cosh ^{d-2}(\overset{(m)}{r}) d\overset{(m)}{r}  \int r_0 r^{d-3} \left(r_0^2-r^2\right)^{\frac{1}{2}-\frac{d}{2}} dr d\Omega, \\
&& \int_{\gamma} ds^{d-2} \sqrt{h_{\gamma}}= \cosh ^{d-2}(\rho) \int r_0 r^{d-3} \left(r_0^2-r^2\right)^{\frac{1}{2}-\frac{d}{2}} dr d\Omega.\nonumber
 \end{eqnarray}
 Substituting these equations into (\ref{sect2: HEEGB}), we derive the HEE
\begin{align} \label{sect2: HEE hemisphere}
S(\rho)&=\sum_m^p \frac{1}{4 G_{N\ m}} \left(\left(1-\frac{2 (d-1) L_m^2 \lambda _m}{d-3}\right)\int_{-\rho}^{\infty} \cosh ^{d-2}(\overset{(m)}{r}) d\overset{(m)}{r} +  \frac{4 L_m^2 \lambda _m \sinh (\rho )\cosh{(\rho)}^{d-3}}{d-3} \right) \nonumber\\
& \times \int r_0 r^{d-3} \left(r_0^2-r^2\right)^{\frac{1}{2}-\frac{d}{2}} dr d\Omega.
\end{align}
Recall that we focus on a symmetric subsystem \( A \), where the hemisphere has the same radius at each edge. Similar to the case of AdS$_3$/NCFT$_2$, for the symmetric subsystem, we have:
\begin{align} \label{sect2: hemisphere SI SII}
S_{\text{I}}&=S_{\text{II}}=S(\rho)-S(0)\nonumber\\
&=\sum_m^p \frac{1}{4 G_{N\ m}} \left(\left(1-\frac{2 (d-1) L_m^2 \lambda _m}{d-3}\right)\int_{-\rho}^{0} \cosh ^{d-2}(\overset{(m)}{r}) d\overset{(m)}{r} +  \frac{4 L_m^2 \lambda _m \sinh (\rho )\cosh{(\rho)}^{d-3}}{d-3} \right) \nonumber\\
& \times \int r_0 r^{d-3} \left(r_0^2-r^2\right)^{\frac{1}{2}-\frac{d}{2}} dr d\Omega.
\end{align}
Additionally, we have
\begin{align} \label{sect2: hemisphere SIII}
S_{\text{III}}=S_{\text{NCFT}}-S_{\text{BCFT}}=S(\rho)-S(\rho)=0.
\end{align}
Above, we have used the fact $S_{\text{BCFT}}=S_{\text{NCFT}}$ for the symmetric subsystem. 

From (\ref{sect2: hemisphere SI SII}) and $L_m^2$ given by (\ref{sect2: AdS radius}), we derive
\begin{align} \label{sect2: hemisphere SI SII g theorem}
\partial_{\rho}S_{\text{I}}&=\partial_{\rho}S_{\text{II}}\nonumber\\
&=\sum_m^p \frac{1}{4 G_{N\ m}}\cosh ^{d-4}(\rho ) \left(\sqrt{1-4 \lambda_{m}}+\left(2-\sqrt{1-4\lambda_{m}}\right) \sinh ^2(\rho )\right) \nonumber\\
& \times \int r_0 r^{d-3} \left(r_0^2-r^2\right)^{\frac{1}{2}-\frac{d}{2}} dr d\Omega.
\end{align}
For positive values of $\lambda_m$, it is straightforward to see that $\partial_{\rho} S_{\mathrm{I}} = \partial_{\rho} S_{\mathrm{II}} > 0.$ 
For negative values of $\lambda_m$ satisfying the constraint~(\ref{sect2: GBconstraint}), one finds
\begin{align} \label{sect2: hemisphere SI SII g theorem 1}
\left(2-\sqrt{1-4\lambda_{m}}\right)\geq\frac{4}{d+2}>0.
\end{align}
As a consequence, the quantity in~(\ref{sect2: hemisphere SI SII g theorem}) is positive, and the result is therefore consistent with the holographic g-theorem.

In this subsection, we have examined the HEE in AdS/NCFT within the context of Gauss-Bonnet gravity. We derived the junction condition for the RT surfaces on the Net-brane. Additionally, we confirmed through various examples that \(S_{\text{I}}\) and \(S_{\text{II}}\) follow the holographic g-theorem, and that \(S_{\text{III}}\) remains non-negative.

\subsection{Correlation Function}

In this subsection, we examine the two-point function of the stress tensor in both the free theory and the holographic Gauss-Bonnet gravity. We focus on a one-node network with \( p \) edges, where the CFTs on each edge may differ, for example, in their central charges.

In our study of NCFTs, we define coordinates along each edge \( E_{m} \) as $\overset{(m)}{\textbf{y}}:~\overset{(m)}{y}{}^{i}=(\overset{(m)}{x},\overset{(m)}{y}{}^{a})$ with $\overset{(m)}{x}\ge 0$, where $a = 2, \ldots, d$ represents the transverse directions. The node is located at $\overset{(m)}{x}=0$. In our previous work \cite{Guo:2025sbm}, by applying the symmetry of the restricted conformal group $O(d,1)$, we derived the general form of the stress-tensor two-point function as \footnote{See also the related discussions for BCFTs \cite{McAvity:1993ue,Herzog:2017xha}.}
\begin{align}\label{sect3: TTfromHHnew}
 &\langle  \overset{(m)}{T}_{ij}({\bf{y}})\overset{(q)}{T}_{kl}({\bf{y'}})  \rangle \nonumber\\
 =&\frac{1}{s^{2d}} \Big[ \overset{(mq)}{\delta}(v)\,\delta_{ij}\delta_{kl}
 + \overset{(mq)}{\epsilon}(v)\,(I_{ik}I_{jl}+I_{il}I_{jk})
 +\big( \overset{(mq)}{\beta}(v)- \overset{(mq)}{\delta}(v)\big)(\hat{X}_{i}\hat{X}_{j}\delta_{kl}+\hat{X}'_{k}\hat{X}'_{l}\delta_{ij}) \nonumber\\
&\qquad  
 -\big( \overset{(mq)}{\gamma}(v)+ \overset{(mq)}{\epsilon}(v)\big)
 (\hat{X}_{i}\hat{X}'_{k} I_{jl}+\hat{X}_{j}\hat{X}'_{l}I_{ik}
 +\hat{X}_{i}\hat{X}'_{l}I_{jk} +\hat{X}_{j}\hat{X}'_{k}I_{il}  ) \nonumber\\
&\qquad 
+\big( \overset{(mq)}{\alpha}(v)-2 \overset{(mq)}{\beta}(v)
+4 \overset{(mq)}{\gamma}(v)+ \overset{(mq)}{\delta}(v)
+2 \overset{(mq)}{\epsilon}(v) \big) 
\hat{X}_{i}\hat{X}_{j}\hat{X}'_{k}\hat{X}'_{l} \Big].
\end{align}
On the same edge with $q=m$, the relevant quantities are
\begin{eqnarray}\label{sect3: Osbornformula}
\begin{split}
&\mathbf{s}=\mathbf{y}-\mathbf{y'},~s=|\mathbf{s}|, ~\mathbf{\bar{s}}=\mathbf{y}-\mathbf{\bar{y}}',~\bar{s}=|\mathbf{\bar{s}}|\\
&v=\sqrt{\frac{(x-x')^2+(y_a-y'_a)^2}
{(x+x')^2+(y_a-y'_a)^2}}, \\
&I_{ij}=\delta_{ij}-2\frac{(y_i-y'_i)(y_j-y'_j)}{s^2}, \\
&\hat{X}_i=\frac{1}{s\bar{s}}
\left(x^{2}-x'^{2}-(y_a-y'_a)^{2},~ 2x (y_a-y'_a) \right),\\
&\hat{X}'_i=\frac{1}{s\bar{s}}
\left(x'^{2}-x^{2}-(y_a-y'_a)^{2},~ -2x' (y_a-y'_a) \right).
\end{split}
\end{eqnarray}
For mixed edges with $q\neq m$, one replaces $\mathbf{y}'=(x',y'_a)$ in 
$v, I_{ij},\hat{X}_{i},\hat{X}'_{i}$ 
with the reflected coordinate 
$\mathbf{\bar{y}}'=(-x',y'_a)$. Note that we also preserve the following definition:
\begin{align}
    \mathbf{s}:~s^{i}=(\overset{(m)}{x}-\overset{(q)}{x}{}',\overset{(m)}{y}{}_{a}-\overset{(q)}{y}{}'_{a}),
\end{align}
for the mixed edges.
For convenience, we define the following invariants:
\begin{align}\label{sect3: invariant}
    v^{2}:\begin{cases}
        v_{\text{I}}^{2}
&=\frac{(\overset{(m)}{y}_a-\overset{(m)}{y}{}'_a)^{2}
+(\overset{(m)}{x}-\overset{(m)}{x}{}')^{2}}
{(\overset{(m)}{y}_a-\overset{(m)}{y}{}'_a)^{2}
+(\overset{(m)}{x}+\overset{(m)}{x}{}')^{2}},\\
v_{\text{II}}^{2}
&=\frac{(\overset{(m)}{y}_a-\overset{(q)}{y}{}'_a)^{2}
+(\overset{(m)}{x}+\overset{(q)}{x}{}')^{2}}
{(\overset{(m)}{y}_a-\overset{(q)}{y}{}'_a)^{2}
+(\overset{(m)}{x}-\overset{(q)}{x}{}')^{2}},
    \end{cases}
\end{align}
where $v_{\text{I}}$ and $v_{\text{II}}$ represent the invariants on the same and mixed edges, respectively. 
Since $\overset{(m)}{x}\ge 0$, the physical distances in our notation are
 $|\overset{(m)}{\mathbf{y}}-\overset{(m)}{\mathbf{y}}{}'|^{2}
=|\mathbf{s}|^2=(\overset{(m)}{y}_a-\overset{(m)}{y}{}'_a)^{2}
+(\overset{(m)}{x}-\overset{(m)}{x}{}')^{2}$ and $|\overset{(m)}{\mathbf{y}}-\overset{(q)}{\mathbf{y}}{}'|^{2}
=|\mathbf{\bar{s}}|^2=(\overset{(m)}{y}_a-\overset{(q)}{y}{}'_a)^{2}
+(\overset{(m)}{x}+\overset{(q)}{x}{}')^{2}$.

To be consistent with the conservation law (\ref{NCFT BC}), we request 
\begin{eqnarray}\label{sect3: TT conservation}
\sum_{m}^{p} \lim_{\overset{(m)}{x}\to 0}  \langle \overset{(m)}{ T}_{na}({\bf{y}})\overset{(q)}{ T}_{kl}({\bf{y'}}) \rangle =0, 
\end{eqnarray} 
which leads to \cite{Guo:2025sbm}
\begin{align}\label{sect3: TT constraint} 
\overset{(qq)}{\gamma}(v_{\text{I}}=1) - \sum_{m\neq q} \overset{(mq)}{\gamma}(v_{\text{II}}=1) = 0. 
\end{align} 

In the following subsections, we compute the stress-tensor two-point functions for the free scalars and the AdS/NCFT based on Gauss–Bonnet gravity with a tensionless Net-brane, where the CFTs living on different edges may differ. Finally, we verify that they all obey the constraints (\ref{sect3: TT constraint}).

\subsubsection{Free scalars}
\label{Free scalars}
We consider a network consisting of a single node connected to \(p\) edges. To ensure that the total central charges vary from edge to edge, the number of scalar fields can differ from edge to edge. The corresponding action is given by
\begin{align}
    I=\frac{1}{2}\sum_{m}^{p}\sum_{\hat{a}}^{\hat{N}_{m}}\int_{E_{m}} d^{d}y~\partial_{i}\overset{(m)}{\phi_{\hat{a}}}\partial^{i}\overset{(m)}{\phi_{\hat{a}}},
\end{align}
where \( \hat{N}_{m} \) represents the number of free scalar fields residing on edge \( E_{m} \), and \( \hat{a} \) indexes the individual fields \( \overset{(m)}{\phi_{\hat{a}}} \) on that edge. By taking the variation of the action, we derive the following boundary term at the node: 
\begin{align}\label{sect3: scalar variation}
    \delta I=\sum_{m}^{p}\sum_{\hat{a}}^{\hat{N}_{m}}\int_{N}d^{d-1}y~\partial_{n}\overset{(m)}{\phi_{\hat{a}}}\delta \overset{(m)}{\phi_{\hat{a}}},
\end{align}
where \( n \) denotes the out-pointing normal direction from each edge to the node. The total central charges for scalars on edge \( E_{m} \) are given by \( \overset{(m)}{C}{}_{T}^{\phi} = \hat{N}_{m} C_{T}^{\phi} \), where \( C_{T}^{\phi} \) is the central charge defined in the two-point function of stress tensors for a single free scalar.

We assume that all the induced scalar fields on the node are identical. Under this assumption, the variation expressed in equation (\ref{sect3: scalar variation}) leads to the following JC:
\begin{align}\label{sect3: scalar JC}
    \text{scalar JC: }\ \sum_{m}^{p}\sum_{\hat{a}}^{\hat{N}_{m}}\partial_{n}\overset{(m)}{\phi_{\hat{a}}}|_{N}=0,~\overset{(m)}{\phi_{\hat{a}}}|_{N}=\overset{(q)}{\phi_{\hat{b}}}|_{N}.
\end{align}
The above JC (\ref{sect3: scalar JC}) is consistent with the conservation of energy flow on the node:
\begin{align}\label{sect3: conservation Tna}
\sum_{m=1}^p \overset{(m)}{T}_{na}|_N=0,
\end{align}
where the stress tensor is given by
\begin{align}\label{sect3: scalar stress tensor}
    \overset{(m)}{T}_{ij}=\sum_{\hat{a}}^{\hat{N}_{m}}\frac{1}{2(d-1)}\left( d\partial_{i}  \overset{(m)}{\phi_{\hat{a}}}\partial_{j}\overset{(m)}{\phi_{\hat{a}}}-(d-2)\overset{(m)}{\phi_{\hat{a}}}\partial_{i}\partial_{j}\overset{(m)}{\phi_{\hat{a}}}-\delta_{ij}(\partial\overset{(m)}{\phi_{\hat{a}}})^{2} \right).
\end{align}

The scalar two-point functions that satisfy the JC  (\ref{sect3: scalar JC}) can be solved as follows: 
\begin{align}\label{sect3: scalar two point function}
    \langle \overset{(m)}{\phi_{\hat{a}}}(\mathbf{y})\overset{(m)}{\phi_{\hat{a}}}(\mathbf{y'}) \rangle=&\frac{\kappa}{d-2}\frac{1+c_{r_1} v_{I}^{d-2}}{|\overset{(m)}{\mathbf{y}}-\overset{(m)}{\mathbf{y}}{}'|^{d-2}},~
    \langle \overset{(m)}{\phi_{\hat{a}}}(\mathbf{y})\overset{(m)}{\phi_{\hat{b}}}(\mathbf{y}') \rangle=\frac{\kappa}{d-2}\frac{c_{r_2} v_{I}^{d-2}}{|\overset{(m)}{\mathbf{y}}-\overset{(m)}{\mathbf{y}}{}'|^{d-2}},\nonumber\\
    \langle \overset{(m)}{\phi_{\hat{a}}}(\mathbf{y})\overset{(q)}{\phi_{\hat{b}}}(\mathbf{y}') \rangle=&\frac{\kappa}{d-2}\frac{c_{t}}{|\overset{(m)}{\mathbf{y}}-\overset{(q)}{\mathbf{y}}{}'|^{d-2}},
\end{align}
where
\begin{align}
    \kappa=\frac{\Gamma \left(\frac{d}{2}\right)}{2\pi ^{d/2}},~c_{r_1}=\frac{2-\hat{N}}{\hat{N}},~c_{r_2}=c_{t}=\frac{2}{\hat{N}},~\hat{N}=\sum_{m}^{p}\hat{N}_{m}.
\end{align}
From (\ref{sect3: scalar stress tensor},\ref{sect3: scalar two point function}), we derive the two-point functions of the stress tensor as
\begin{align}\label{sect3: scalar TmTm}
    \langle \overset{(m)}{T}{}_{ij}(\mathbf{y})\overset{(m)}{T}{}_{kl}(\mathbf{y}') \rangle=&\overset{(m)}{C}{}_{T}^{\phi}\left( \frac{\mathcal{I}_{ij,kl}(\mathbf{s})}{s^{2d}}+\overset{(m)}{c_{r}}{}^{2}\frac{\bar{\mathcal{I}}_{ij,kl}(\mathbf{\bar{s}})}{\bar{s}^{2d}}+c_{r_{1}}\frac{\#_{ij,kl}}{(s\bar{s})^{2d}} \right),\\
    \langle \overset{(m)}{T}{}_{ij}(\mathbf{y})\overset{(q)}{T}{}_{kl}(\mathbf{y}') \rangle=&\overset{(m)}{C}{}_{T}^{\phi}\left(\overset{(q)}{c_{t}}{}^{2}\frac{\bar{\mathcal{I}}_{ij,kl}(\mathbf{\bar{s}})}{\bar{s}^{2d}}\right)=\overset{(q)}{C}{}_{T}^{\phi}\left(\overset{(m)}{c_{t}}{}^{2}\frac{\bar{\mathcal{I}}_{ij,kl}(\mathbf{\bar{s}})}{\bar{s}^{2d}}\right),\label{sect3: scalar TmTq}
\end{align}
where we have
\begin{align}
    &\overset{(m)}{C}{}_{T}^{\phi}=\hat{N}_{m}C_{T}^{\phi}=\hat{N}_{m}\frac{\kappa^{2}d}{d-1},\\
    &\overset{(m)}{c_{t}}{}^{2}=\hat{N}_{m}c_{t}^{2},~\overset{(m)}{c_{r}}{}^{2}=\left(c_{r_{1}}^{2}+(\hat{N}_{m}-1)c_{r_{2}}^{2}\right), \label{sect2.4.1: cr ct}\\
    &\mathcal{I}_{ij,kl}(s)=\frac{1}{2}\left(I_{ik}(s)I_{jl}(s)+I_{il}(s)I_{jk}(s)\right)-\frac{1}{d}\delta _{ij}\delta_{kl}\\
    &\bar{\mathcal{I}}_{ij,kl}(\bar{s})=\mathcal{I}_{ij,kl}(s)|_{x'\to\bar{x}'}.
\end{align}
Here, the term \(\frac{\#_{ij,kl}}{(s\bar{s})^{2d}}\) arises from the mixing between the contribution in free space \(s\) and its mirror contribution \(\bar{s}\). This component $\#_{na,kl}$ vanishes in the limit as \(\overset{(m)}{x}{}' \rightarrow 0\). 
From the two-point function \eqref{sect3: scalar TmTm}, we can derive the central charge \(\overset{(m)}{C}{}_{T}^{\phi} = \hat{N}_{m} \, C_{T}^{\phi}\) in the limit far aways from the node.

Next, we rewrite the correlators of stress tensors (\ref{sect3: scalar TmTm}, \ref{sect3: scalar TmTq}) into the general form presented in (\ref{sect3: TTfromHHnew}). The corresponding functions are given by
\begin{align}
    \overset{(mm)}{\alpha}=&\hat{N}_{m}\kappa^{2}\left( 1+\overset{(m)}{c_{r}}{}^{2}v_{\text{I}}^{2d}+ \frac{c_{r_{1}}}{4}(d-2)d\frac{d+1}{d-1}v_{\text{I}}^{d-2}(1-v_{\text{I}}^2)^{2} \right),\\
    \overset{(mm)}{\gamma}=&\frac{-\hat{N}_{m}\kappa^{2}d}{2(d-1)}\left( 1-\overset{(m)}{c_{r}}{}^{2}v_{\text{I}}^{2d}+\frac{c_{r_{1}}}{2}(d-2)\frac{d+1}{d-1}v_{\text{I}}^{d-2}(1-v_{\text{I}}^4) \right),\\
    \overset{(mm)}{\epsilon}=&\frac{\hat{N}_{m}\kappa^{2}d}{2(d-1)}\left( 1+\overset{(m)}{c_{r}}{}^{2}v_{\text{I}}^{2d} +\frac{c_{r_{1}}}{2(d-1)}\left( (d-2)(v_{\text{I}}^{d-2}+v_{\text{I}}^{d+2})+2dv_{\text{I}}^{d} \right)\right),
\end{align}
for the same edge and
\begin{align}
    \overset{(m\neq q)}{\alpha}=\overset{(m)}{c_{t}}{}^{2}\hat{N}_{q}\kappa^{2},~\overset{(m\neq q)}{\gamma}=-\overset{(m)}{c_{t}}{}^{2}\frac{\hat{N}_{q}\kappa^{2}d}{2(d-1)},~\overset{(m\neq q)}{\epsilon}=\overset{(m)}{c_{t}}{}^{2}\frac{\hat{N}_{q}\kappa^{2}d}{2(d-1)},
\end{align}
for the mixed edges. The other functions of (\ref{sect3: TTfromHHnew}) can be obtained from the traceless condition of the stress tensor \cite{McAvity:1993ue} as
\begin{align} \label{sect3: tracelessness}
    &\overset{(mq)}{\alpha}+(d-1)\overset{(mq)}{\beta}=0,~\overset{(mq)}{\beta}+(d-1)\overset{(mq)}{\delta}+2\overset{(mq)}{\epsilon}=0.
\end{align}
We verify the constraint (\ref{sect3: TT constraint}) by using the following relation
\begin{align}
    N_{m}(1-\overset{(m)}{c_{r}}{}^{2})=\overset{(m)}{c_{t}}{}^{2}\sum_{q\neq m}\hat{N}_{q}.
\end{align}

\subsubsection{AdS/NCFT}

This subsection examines the two-point function of the stress tensor in holographic GB gravity with a tensionless Net-brane. The scenario involving a tensive Net-brane is too complex, and we will address it in future work. Our findings suggest that zero tension is associated with non-unitary NCFTs. As illustrated in Figure \ref{causal GB}, causal NCFTs require non-negative tension. Consequently, various pieces of evidence indicate that the tension of the Net-brane should be positive in AdS/NCFT. We will further investigate this issue in the next section.

We take the action $I_{\text{GB}}=I_{\text{GB bulk}}+I_{\text{GB NB}}$ (\ref{sect2: GBactionbulk},\ref{sect2: GBactionbdy}) with $T=0$. For simplicity, we consider an identical AdS radius $l_m$ for all spacetime branches and set $l_m=1$. Following the procedure of \cite{Liu:1998bu}, we take the following ansatz for the perturbative metric
\begin{align}\label{sect3: AdSNCFT metric}
    ds^{2}=\frac{dz^{2}+dx^{2}+\delta_{ab}dy^{a}dy^{b}+H_{\mu\nu}dx^{\mu}dx^{\nu}}{z^{2}},
\end{align}
with the gauge
\begin{align}\label{sect3: AdSNCFT gauge}
    H_{zz}(z=0,\mathbf{y})=H_{zi}(z=0,\mathbf{y})=0. 
\end{align}
Imposing the JC (\ref{sect2: GB JC}), the continuous condition $\overset{(m)}{h}_{ij}|_{NB}=h_{ij}$ on the tensionless Net-brane and the DBC
\begin{align}\label{sect3: AdSNCFT DBC}
    H_{ij}(z=0,\mathbf{y})=\hat{H}_{ij}(\mathbf{y}),
\end{align}
on the AdS boundary, we derive the linear metric perturbations 
\begin{align}\label{sect3: metric perturbations}
    \overset{(m)}{H}_{\mu\nu}(z,\mathbf{y})=&\frac{C_{T}}{2d}\left\{ \int_{E_{m}}d^{d}y'\left[ \frac{z^{d}}{S^{2d}}J_{\mu i}J_{\nu j} P_{ijkl}+\overset{(m)}{c_{r}}\frac{z^{d}}{\bar{S}^{2d}}\bar{J}_{\mu i}\bar{J}_{\nu j}P_{ijkl}\right]\overset{(m)}{\hat{H}}_{kl}(\mathbf{y'}) \right.\nonumber\\
    &+\left.\sum_{q\neq m}\overset{(q)}{c_{t}}\int_{E_{q}}d^{d}y' \frac{z^{d}}{\bar{S}^{2d}}\bar{J}_{\mu i}\bar{J}_{\nu j}P_{ijkl}\overset{(q)}{\hat{H}}_{kl}(\mathbf{y'})\right\},
\end{align}
where various notations are given by
\begin{align}
    C_{T}=&\frac{2\Gamma [d+2]}{\pi^{d/2}\Gamma [d/2](d-1)},~\overset{(m)}{c_{r}}=\frac{A_{m}-\bar{A}_{m}}{A_{m}+\bar{A}_{m}},~\overset{(m)}{c_{t}}=\frac{2A_{m}}{A_{m}+\bar{A}_{m}},~\bar{A}_{m}=\sum_{q\neq m}A_{q},
\end{align}
and 
\begin{align}\label{sect3: AdSNCFT metric solution 123}
    S^{2}=&z^{2}+(x-x')^{2}+(y_{a}-y_{a}')^{2},\nonumber\\
    \bar{S}^{2}=&z^{2}+(x+x')^{2}+(y_{a}-y_{a}')^{2},\nonumber\\
    P_{ijkl}=&\frac{1}{2}\left( \delta_{ik}\delta_{jl}+\delta_{il}\delta_{jk} \right)-\frac{1}{d}\delta_{ij}\delta_{kl},\\
    J_{\mu \nu}=&\delta_{\mu\nu}-2\frac{(x_{\mu}-x_{\mu}')(x_{\nu}-x_{\nu}')}{S^{2}},\nonumber\\
    \bar{J}_{\mu\nu}=&J_{\mu\nu}-2X_{\mu}X_{\nu}',\nonumber
\end{align}
and
\begin{align}\label{sect3: AdSNCFT metric solution XbarX}
    X_{\mu}=&\frac{1}{S\bar{S}}\left( 2xz,x^{2}-x'^{2}-(y_{a}-y_{a}')^{2}-z^{2},2x(y_{a}-y_{a}') \right),\\
    \bar{X}_{\mu}=&\frac{1}{S\bar{S}}\left( -2x'z,x'^{2}-x^{2}-(y_{a}-y_{a}')^{2}-z^{2},-2x'(y_{a}-y_{a}') \right). \label{sect3: AdSNCFT metric solution barX}
\end{align}
Here, we recall that $A_{m}$ is given by (\ref{sect2: AmBm}).

From the metric perturbations (\ref{sect3: metric perturbations}), we derive the on-shell quadratic action \cite{Buchel:2009sk}
\begin{align}\label{sect3: AdSNCFT quadratic action}
    I_{2}=&\sum_{m}^{p}\frac{A_{m}}{64\pi G_{N}}\int_{E_{m}} d^{d}y~z^{1-d} \overset{(m)}{H}_{ij}\partial_{z}\overset{(m)}{H}_{ij}\nonumber\\
    =&\frac{C_{T}}{128\pi G_{N}}\left[ \sum_{m}^{p}A_{m}\int_{E_{m}}d^{d}y d^{d}y' \overset{(m)}{\hat{H}}_{ij}(\mathbf{y})\left( \frac{\mathcal{I}_{ij,kl}}{s^{2d}}+\overset{(m)}{c_{r}}\frac{\bar{\mathcal{I}}_{ij,kl}}{\bar{s}^{2d}} \right)\overset{(m)}{\hat{H}}_{kl}(\mathbf{y'}) \right.\nonumber\\
    &~~~~~~~~~~~~~~+\left. \sum_{m}^{p}\sum_{q\neq m}A_{m}\int_{E_{m}}d^{d}y \int_{E_{q}}d^{d}y'\overset{(m)}{\hat{H}}_{ij}(\mathbf{y})\overset{(q)}{c_{t}}\frac{\bar{\mathcal{I}}_{ij,kl}}{\bar{s}^{2d}}\overset{(q)}{\hat{H}}_{kl}(\mathbf{y'}) \right].
\end{align}
where
\begin{align}\label{sect3: AdSNCFT Iijkl}
    \mathcal{I}_{ij,kl}=&\lim_{z\to0}\frac{1}{2}\left( J_{ik}J_{jl}+J_{il}J_{jk} \right)-\frac{1}{d}\delta_{ij}\delta_{kl},\\
    \bar{\mathcal{I}}_{ij,kl}=&\lim_{z\to0}\frac{1}{2}\left( \bar{J}_{ik}\bar{J}_{jl}+\bar{J}_{il}\bar{J}_{jk} \right)-\frac{1}{d}\delta_{ij}\delta_{kl}.
\end{align}
From (\ref{sect3: AdSNCFT quadratic action}), we obtain the holographic two-point function of the stress tensor:
\begin{align}\label{sect4.2: holo TT same}
    \langle \overset{(m)}{T}{}_{ij} (\mathbf{y})\overset{(m)}{T}{}_{kl}(\mathbf{y'})\rangle&=\overset{(m)}{C_{T}}\left( \frac{\mathcal{I}_{ij,kl}(\mathbf{s})}{s^{2d}}+\overset{(m)}{c_{r}}\frac{\bar{\mathcal{I}}_{ij,kl}(\mathbf{\bar{s}})}{\bar{s}^{2d}} \right),\\
    \langle \overset{(m)}{T}{}_{ij} (\mathbf{y})\overset{(q)}{T}{}_{kl}(\mathbf{y'})\rangle&=\overset{(m)}{C_{T}}\overset{(q)}{c_{t}}\frac{\bar{\mathcal{I}}_{ij,kl}(\mathbf{\bar{s}})}{\bar{s}^{2d}}=\overset{(q)}{C_{T}}\overset{(m)}{c_{t}}\frac{\bar{\mathcal{I}}_{ij,kl}(\mathbf{\bar{s}})}{\bar{s}^{2d}}, \label{sect4.2: holo TT mix}
\end{align}
where $\overset{(m)}{C_{T}}=\frac{A_{m}C_{T}}{16\pi G_{N}}$ denotes the central charge of NCFTs on the edge $E_{m}$. Finally, we rewrite the above two-point functions into the general form (\ref{sect3: TTfromHHnew}):
\begin{align}
    \overset{(mm)}{\alpha}=\frac{(d-1)}{d}\overset{(m)}{C_{T}}(1+\overset{(m)}{c_{r}}v_{\text{I}}^{2d}),~\overset{(mm)}{\gamma}=-\frac{\overset{(m)}{C_{T}}}{2} (1-\overset{(m)}{c_{r}}v_{\text{I}}^{2d}),\overset{(mm)}{\epsilon}=\frac{\overset{(m)}{C_{T}}}{2} (1+\overset{(m)}{c_{r}}v_{\text{I}}^{2d})
\end{align}
for the same edge and
\begin{align}
    \overset{(m\neq q)}{\alpha}=\overset{(m)}{c_{t}}\frac{(d-1)\overset{(q)}{C_{T}}}{d},~\overset{(m\neq q)}{\gamma}=-\overset{(m)}{c_{t}}\frac{\overset{(q)}{C_{T}}}{2},~\overset{(m\neq q)}{\epsilon}=\overset{(m)}{c_{t}}\frac{\overset{(q)}{C_{T}}}{2}
\end{align}
for the mixed edges. We can verify the constraint (\ref{sect3: TT constraint}) by using following relation:
\begin{align}
    A_{m}(1-\overset{(m)}{c_{r}})=\overset{(m)}{c_{t}}\sum_{q\neq m}A_{q}.
\end{align}

From the holographic two-point functions of the stress tensors (\ref{sect4.2: holo TT same},\ref{sect4.2: holo TT mix}), we can determine the reflectivity at edge \( E_m \):
\begin{align}\label{sect4.2: holo reflectivity}
  R_m=\overset{(m)}{c_{r}}=\frac{A_{m}-\sum_{q\ne m} A_{q} }{\sum_{n=1}^p A_{n}},
\end{align}
and the transmissivity from edge \( E_m \) to edge \( E_q \):
\begin{align}\label{sect4.2: holo transmissivity}
  T_{mq}=\overset{(q)}{c_{t}}=\frac{2A_{q}}{\sum_{n=1}^p A_{n}},
\end{align}
where $A_m=\sqrt{1-4\lambda_m} G_N/G_{N~m}$. One can verify that these quantities satisfy the probability conservation:
\begin{align}\label{sect4.2: probability conservation}
 R_m+\sum_{q\ne m} T_{mq}=1,
\end{align}
as well as the condition:
\begin{align}\label{sect4.2: Tmq condition}
\overset{(m)}{C_{T}} T_{mq}=\overset{(q)}{C_{T}} T_{qm}. 
\end{align}
In the case of ICFT\(_2\) and NCFT\(_2\), a similar condition holds, with \( \overset{(m)}{C_{T}} \) being replaced by the two-dimensional central charge \( \overset{(m)}{c} \) \cite{Meineri:2019ycm, Liu:2025khw}. 

For a unitary NCFT, the reflectivity must be positive. It is indeed true for the free scalars discussed in Section \ref{Free scalars}. 
From (\ref{sect3: scalar TmTm}), we find that the reflectivity \( R_m = \overset{(m)}{c_{r}}{}^2 > 0 \) for free scalars, where \( \overset{(m)}{c_{r}} \) (\ref{sect2.4.1: cr ct}) differs from the one used in AdS/NCFT. Conversely, in the tensionless Net-brane case, the holographic reflection coefficients in \eqref{sect4.2: holo reflectivity} are typically not all positive; in particular, the coefficients associated with at least \(p-1\) branches are negative. It indicates that \( T = 0 \) is a non-unitary parameter in AdS/NCFT. We will discuss the tension issue more in the next section.

\section{Wedge inclusion condition}
 \label{Wedge inclusion condition}
In this section, we examine the causal constraints on the tension of Net-branes by utilizing the wedge inclusion condition. This condition requires that the entanglement wedge (EW) encompasses the causal wedge (CW), i.e., ``EW$\supseteq$CW." This relationship establishes a lower bound on the tension of Net-branes. For simplicity, we will focus on Einstein gravity in \(\text{AdS}_{3}/\text{NCFT}_{2}\) and set the AdS radius to 1. It is important to note that this causal bound is stronger than the unitary bound, which is derived from the positivity of holographic reflectivity at the network node \cite{Liu:2025khw}.

According to \cite{Liu:2025khw}, the holographic reflectivity at the network node is given by
\begin{align}
    R=\frac{T-p+2}{T+p},
\end{align}
where {we have set the AdS radius to 1} and \(p\) represents the number of edges. For a unitary NCFT, the reflectivity must be non-negative, which establishes a lower bound on the brane tension:
\begin{align}\label{sect 3: reflectivity bound}
{\text{unitary bound}: \ \ T\ge p-2. }
\end{align}
 For AdS/BCFT with \(p = 1\), it leads to \(T \geq -1\) for an end-of-the-world (EOW) brane. Note that the EOW brane acts as a boundary of the bulk spacetime and can have negative brane tension \cite{Miyaji:2021ktr}. 
For AdS/ICFT with \(p = 2\), the normal brane satisfies \(T \geq 0\). This normal brane resembles a thin shell and is expected to have non-negative energy. However, for AdS/NCFT with \(p \geq 3\), we have \(T > 0\).

Remarkably, we find that the requirement ``EW$\supseteq$CW" can provide a stronger constraint on the tension of the Net-brane:
\begin{align}\label{sect 3: EWCW bound}
{\text{causal bound}: \ \ T\ge p \sqrt{\frac{p-2}{p+2}},~~~p\geq 2. }
\end{align}
In this context, ``EW \(\supseteq\) CW'' denotes the causal consistency
requirement that the entanglement wedge should contain the causal wedge in a
well-defined AdS/CFT duality \cite{Headrick:2014cta}. 
We refer to the condition ``EW$\supseteq$CW" as the causal constraint in (\ref{sect 3: EWCW bound}).
Note that ``EW$\supseteq$CW"  always holds in AdS/BCFT with $p=1$. For AdS/ICFT with \(p= 2\), both the unitary and causal bounds coincide, leading to \(T \geq 0\). In contrast, for AdS/NCFT with \(p \geq 3\), the causal bound is stronger than the unitary bound, resulting in \(T \geq p \sqrt{\frac{p - 2}{p + 2}} > p - 2\).

\subsection{The physical picture}

In this subsection, we explain why ``EW$\supseteq$CW" yields a bound on the tension of the Net-brane. We focus on the physical picture in this subsection, and leave the quantitative calculations to the next subsection. For the sake of simplicity, we will assume that \( G_{N~m} = G_{N} \) in this discussion. However, extending the analysis to the case where \( G_{N~m} \neq G_{N} \) is straightforward, and all the arguments presented here will still apply in the general case. Additionally, since our spacetime is static, we will consider the time slices of EW and CW.

\begin{figure}[!h]
  \centering
 \includegraphics[width=0.49\textwidth]{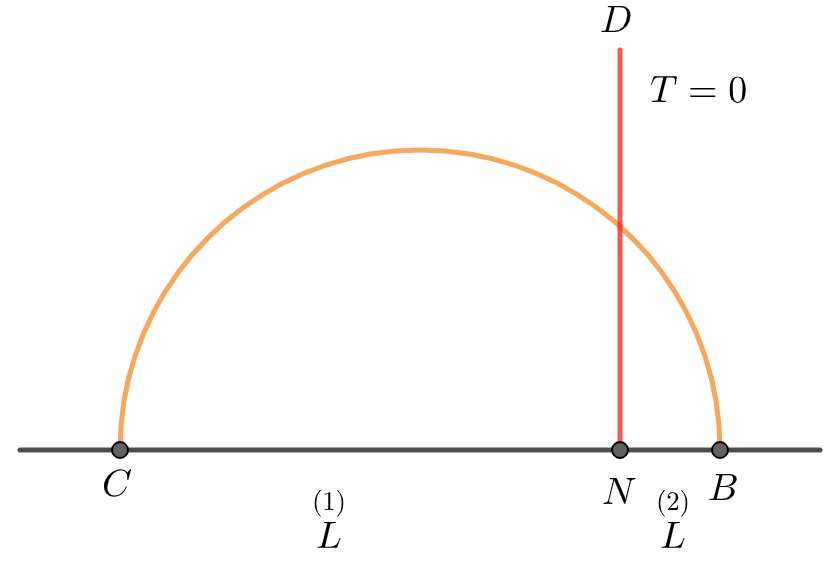} \includegraphics[width=0.49\textwidth]{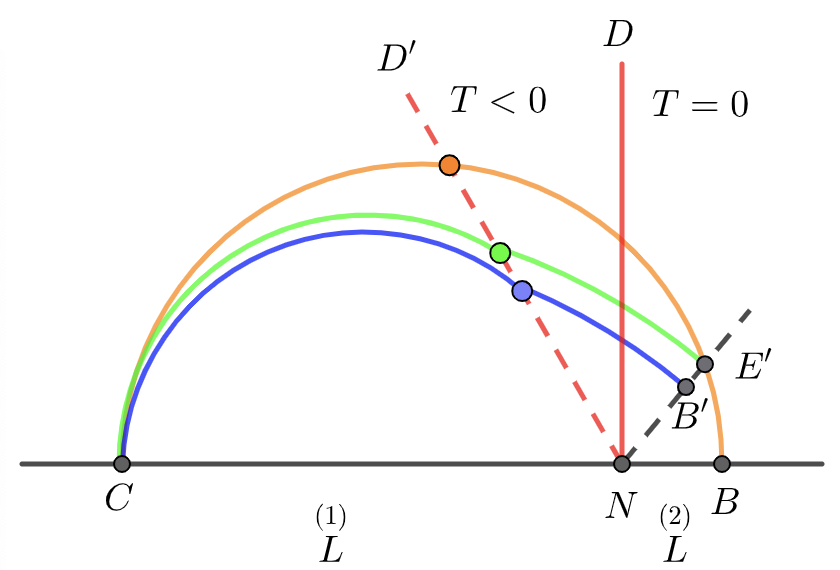}
 \caption{{EW$=$CW for $T=0$ (left);\ \  EW$\subset$CW for $T<0$ (right) in AdS/ICFT}.
 The left figure illustrates the AdS$_3$/ICFT$_2$ with zero brane tension, where the red line $ND$  represents the tensionless brane. The regions \(CN \cup NB\) define the subsystem, with \(CN= \overset{(1)}{L} > NB = \overset{(2)}{L} \). EW and CW are both represented by the area between the orange semicircle and the black line for \(T = 0\).
The right figure depicts the AdS$_3$/ICFT$_2$ with negative brane tension, achieved by rotating the right spacetime branch counterclockwise around node \(N\) by a specific angle. The bulk region is above the black line \(CN\) and dotted black line $NB'E'$, with the dotted red line \(ND'\) marking the brane. The blue curve is the RT surface for the subsystem $CN \cup NB'$. The region between them is the EW. The green curve shows the RT surface for the subsystem \(CN \cup NE'\), serving as an auxiliary line. After the rotation, the CW for the left spacetime branch (the area between \(CN\) and \(ND'\)) remains unchanged, as the subsystem does not alter. However, the CW for the right spacetime branch does change, although this change is irrelevant to our discussion. 
Notably, the EW becomes smaller, leading to EW$\subset$CW for $T<0$ in AdS$_3$/ICFT$_2$. To eliminate this unphysical scenario, we get $T\ge 0$ for AdS$_3$/ICFT$_2$.} 
 \label{EWCW2p}
\end{figure}

At the time slice, EW is given by the region between RT surface and the subsystem on the AdS boundary. Let us quickly review CW in AdS$_{3}$/ICFT$_{2}$ with a tensionless Net-brane (or AdS$_{3}$/CFT$_{2}$). Consider a subsystem $A$ and its domains of dependence $D(A)$ for CFTs. The causal wedge (CW) of $A$ is defined as the intersection of the causal past and causal future of the domain of dependence \( D(A) \) within the bulk spacetime. At the time slice of AdS$_3$, CW is just a half-disk in the bulk. See the region between the orange semicircle and the black line of Fig. \ref{EWCW2p} (left). 

Below, we will argue that $T\ge 0$ for AdS/ICFT and $T>0$ for AdS/NCFT with $p\geq 3$ by using ``EW$\supseteq$CW". 
To warm up, we examine the AdS$_3$/ICFT$_2$ with zero brane tension. See Fig. \ref{EWCW2p} (left), where the red line $ND$ denotes the tensionless brane, and $CN \cup NB$ is the subsystem with $CN=\overset{(1)}{L} >NB=\overset{(2)}{L} $. Similar to the AdS$_3$/CFT$_2$, the EW and CW coincide for $T=0$. Both regions are indicated by the area between the orange semicircle and the black line in Fig. \ref{EWCW2p} (left). Thus, we have EW$=$CW for $T=0$, suggesting that $T = 0$ is a critical point in AdS$_3$/ICFT$_2$. 

Next, let us consider the AdS$_3$/ICFT$_2$ with negative brane tension. As illustrated in Fig. \ref{EWCW2p} (right), this configuration is obtained by rotating the right spacetime branch counterclockwise about point $N$ by a specific angle. In this case, the AdS boundary is represented by the black line $CN$ and the black dotted line $NB'E'$. The brane is depicted by the red dotted line $ND'$, and the subsystem is identified as $CN \cup NB'$ with $CN= \overset{(1)}{L}  > NB' = \overset{(2)}{L} $. After the rotation, the CW for the left spacetime branch (the region between \( CN \) and \( ND' \)) remains unchanged because the subsystem \( A = \overset{(1)}{L}  + \overset{(2)}{L}  \) is the same, as is the domain of dependence \( D(A) \) for the ICFT. Since the bulk region between \( CN \) and \( ND' \) is unchanged, the corresponding CW in this region also remains unchanged. Note that the CW in the right spacetime branch (region between $ND'$ and $NB'$) indeed changed, but is irrelevant to our discussions below. 

Let us go on to discuss the EW for the subsystem $CN \cup NB'$. Our goal is to show that it lies within the CW for $T<0$ in the left spacetime branch. The trick is to use the auxiliary RT surface for a larger subsystem $CN \cup NE'$. The RT surface must be perpendicular to the boundary to minimize its area. Consequently, the orange curve in Fig. \ref{EWCW2p} (right) cannot represent the RT surface, as the angle between the orange curve and the AdS boundary (dotted black line) is obtuse. To form a right angle with the AdS boundary, the RT curve (green line) should bend inward relative to the orange curve. Now, returning to the EW for the subsystem \( CN \cup NB' \): since \( CN \cup NB' \) is contained within the larger subsystem \( CN \cup NE' \), its entanglement wedge must also be contained within that of \( CN \cup NE' \). Therefore, as illustrated in Fig. \ref{EWCW2p} (right), we have
\begin{align}\label{sect 3: EWCW}
EW(CN \cup NB')\subset  EW(CN \cup NE') \subset CW(CN \cup NB'),  \ \text{for} \ T<0,
\end{align}
in the left spacetime branch. To satisfy the condition ``EW \( \supseteq \) CW," we conclude that \( T \) must be greater than or equal to 0 for AdS\(_3\)/ICFT\(_2\).

\begin{figure}[t]
  \centering
  \begin{subfigure}{0.3\linewidth}
    \centering
    \includegraphics[width=\linewidth]{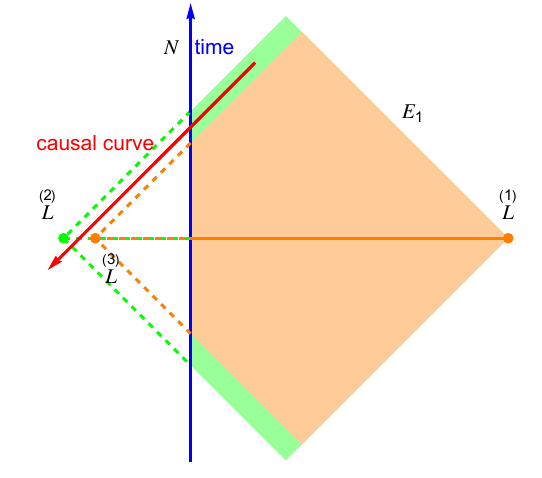}
    \caption{$D(A)$ in $E_{1}$.}
    \label{fig: domain1}
  \end{subfigure}
  \hfill
  \begin{subfigure}{0.3\linewidth}
    \centering
    \includegraphics[width=\linewidth]{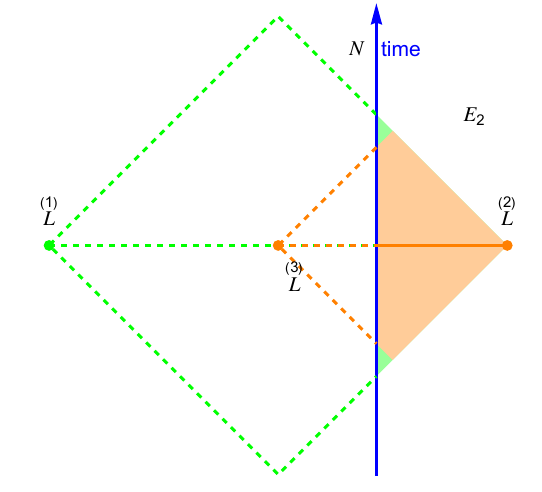}
    \caption{$D(A)$ in $E_{2}$.}
    \label{fig: domain2}
  \end{subfigure}
   \hfill
  \begin{subfigure}{0.3\linewidth}
    \centering
    \includegraphics[width=\linewidth]{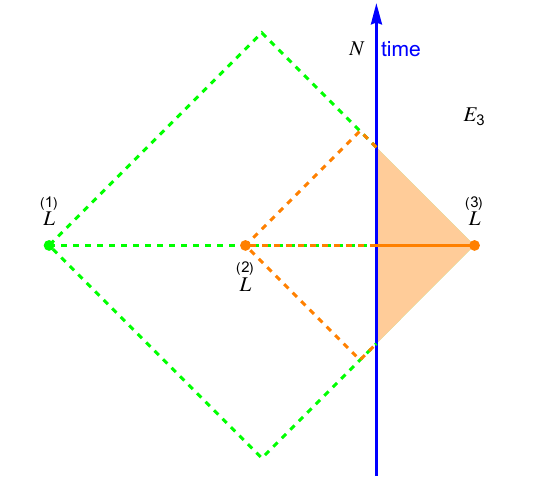}
    \caption{$D(A)$ in $E_{3}$.}
    \label{fig: domain3}
  \end{subfigure}
  \caption{Domain of dependence of the subregion \(A\). 
We consider the case with \(p=3\), \(\overset{(1)}{L}=1\), \(\overset{(2)}{L}=0.4\), and \(\overset{(3)}{L}=0.3\) as an illustrative example. The time direction is represented by the blue line, which runs parallel to node \(N\). The orange solid lines indicate the subregions on the edges \(E_m\) at a fixed time slice. On the left side of the node, the green and orange points represent auxiliary points associated with the subregions on the other two edges. Each of these subregions defines a candidate domain of dependence, as illustrated by the green and orange-shaded regions. Their overlap, specifically the orange shaded region, determines the final domain of dependence. The red arrow indicates a causal curve that begins at a point in the green region, passes through the node, and enters another edge.}
  \label{fig: domain}
\end{figure}

To facilitate a smooth transition into the discussion of AdS/NCFT, we will first provide a brief overview of the domain of dependence in NCFT and its corresponding causal wedge. Let's consider a constant-time slice (Cauchy surface) and choose a boundary subregion \(A\) that includes a specific node. Denote the lengths of subregion \(A\) on the \(p\) edges as \(\overset{(1)}{L} \ge \overset{(2)}{L} \ge \cdots \ge \overset{(p)}{L}\). To determine the domain of dependence on the \(m\)-th edge, we start by pairing this edge with each of the other edges and treating each pair as an independent ICFT. This process yields \(p-1\) candidate domains of dependence. However, for the domain of dependence in NCFT, every inextensible causal curve passing through any point in the domain must intersect the chosen subregion \(A\). Therefore, the actual domain of dependence on the \(m\)-th edge is defined as the intersection of all these candidate domains. For instance, when \(p=3\), as illustrated in Fig. \ref{fig: domain1}, a point on the first edge that lies outside the smaller candidate domain (for example, a point in the green region) can send a causal curve (represented by the red arrow) through the node into \(E_3\) without intersecting subregion \(A\). Consequently, such a point cannot belong to the NCFT domain of dependence. Thus, the domain of dependence on the \(m\)-th edge in NCFT is dictated by the effective ICFT formed by this edge and the edge with the shortest subregion, as indicated by the orange shaded region in Fig. \ref{fig: domain}.

\begin{figure}[!h]
    \centering
    \includegraphics[width=0.9\linewidth]{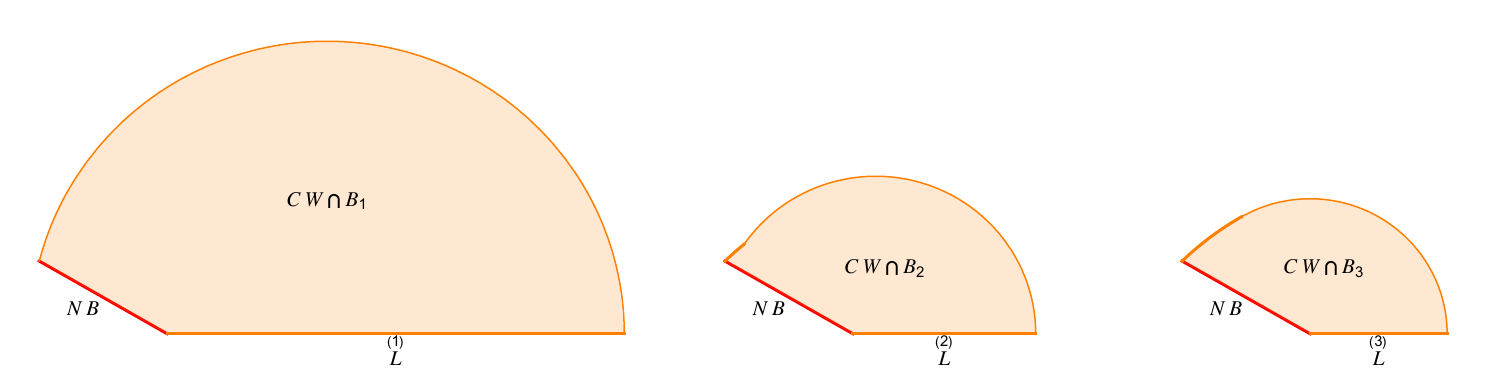}
    \caption{Causal wedge of $A$ in the time slice, where we choose the one-node network with three edges and Net-brane tension $T=3\sin\theta=\frac{3\sqrt{3}}{2}$, and the subregion as $\overset{(1)}{L}=1,~\overset{(2)}{L}=0.4,~\overset{(3)}{L}=0.3$. In the figure, the red and orange lines denote the Net-brane and the subregions on the individual branches, respectively. The orange curves together form the boundaries of the causal wedges in the different branches, with the orange regions representing the corresponding causal wedges in the time slice. The thick orange curves denote the contribution of the causal curve through the Net-brane.}
    \label{fig: CW123}
\end{figure}

According to the domain of dependence, the bulk causal wedge is defined as the intersection of the bulk causal future and past of \(D(A)\). On a static time slice, it can be characterized by the intersection points between this slice and causal curves emerging from \(D(A)\). Thus, the causal wedge on each branch has two types of contributions: the light-cone contribution within the same branch, which forms a semicircle on the time slice, and the contribution from causal curves that reach the Net-brane and then propagate into each branch. Figure \ref{fig: CW123} illustrates the case where \(p=3\) with \(\overset{(1)}{L} =1\), \(\overset{(2)}{L} =0.4\), and \(\overset{(3)}{L} =0.3\). Since the first branch contains the largest subregion, its light-cone contribution includes the portion that passes through the Net-brane; therefore, a single semicircle determines its causal wedge. For the second and third branches, the causal wedge profile consists of two circular arcs: one arc near the Net-brane, which is due to causal curves passing through it, and another arc near the AdS boundary, which arises from the light cone within the same branch.

\begin{figure}[htbp]
  \centering
 \includegraphics[width=0.49\textwidth]{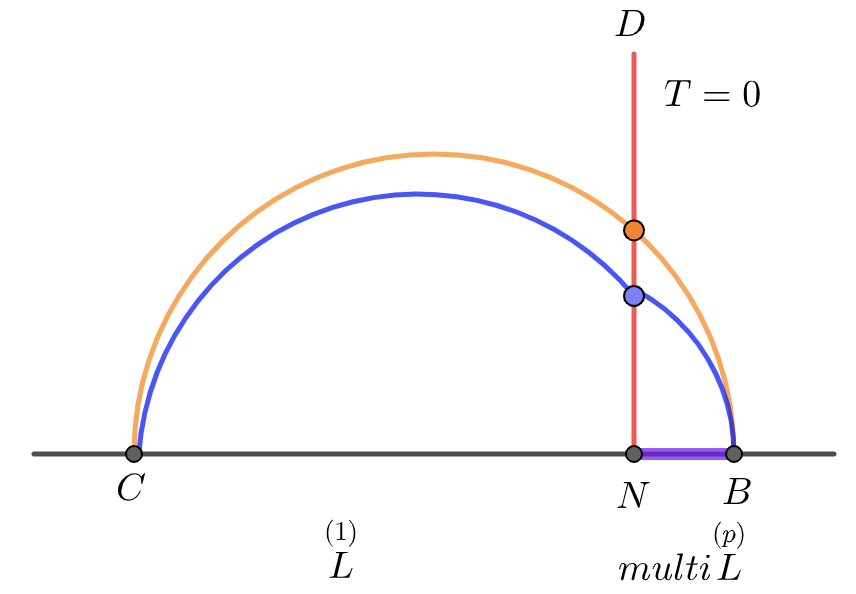}  \includegraphics[width=0.45\textwidth]{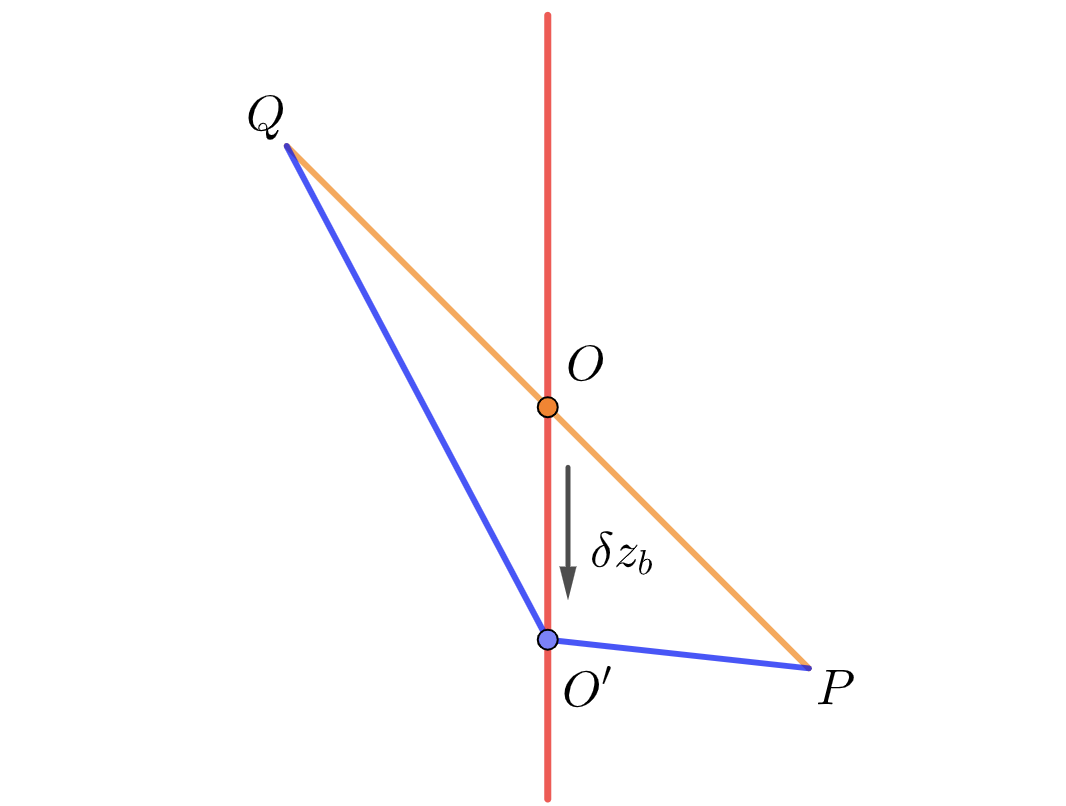} 
 \caption{{EW$ \subset$CW for $T=0$ in AdS/NCFT (left);\ \  Minimizing RT surface  (right)}. The left figure illustrates AdS/NCFT with a tensionless Net-brane (red line). On the right, there are multiple branches, but for simplicity, only one is shown. The subsystem $A$ consists of $CN=\overset{(1)}{L} $ and multi $NB=\overset{(p)}{L} $, where $\overset{(1)}{L} >\overset{(2)}{L} =...=\overset{(p)}{L} $. In the left branch, the CW and EW are given by the region between the orange and blue curves, and the AdS boundary, respectively. We have EW$ \subset$CW for $T=0$ and $p\ge 3$. The figure on the right demonstrates the strategy for minimizing the RT surface by adjusting the joint point on the Net-brane.} 
 \label{EWCW3p}
\end{figure}

Now, we are ready to argue that ``EW \( \supseteq \) CW" leads to $T>0$ for AdS/NCFT with $p\ge 3$. As illustrated in Fig. \ref{EWCW3p} (left), we consider the AdS/NCFT with a tensionless Net-brane (red line) and $p$ branches. Note that there are multiple branches on the right side; for simplicity, we depict only one. We consider the subsystem \( A \) consisting of \( CN = \overset{(1)}{L}  \) and multiple \( NB = \overset{(p)}{L}  \), where \( \overset{(1)}{L}  > \overset{(2)}{L}  = ... = \overset{(p)}{L}  \). In the left branch, the CW and EW regions are given by the areas between the orange and blue curves, and the AdS boundary, respectively. Note that the CW is identical to that of AdS/ICFT, as shown in Fig. \ref{EWCW2p} (left), since they share the same domain of dependence \( D(A) \) and bulk geometry in the left branch. Below, we will explain why \( EW \subset CW \) for $T=0$ when \( p \ge 3 \).

Recall that EW = CW for \( T = 0 \) and \( p = 2 \). In other words, the RT surface corresponds to the orange curve in Fig. \ref{EWCW3p} (left) for \( p = 2 \). As shown in Fig. \ref{EWCW3p} (right), let us slightly deform the RT surface by pulling down the joint point (the orange point) of the RT surface and the Net-brane by \( \delta z_b \). Since the RT surface is extremal, its area remains invariant at the linear order of \( \delta z_b \):
\begin{align}\label{sect 3: RT1}
\delta QO+\delta PO=\big(QO'-QO\big)+\big(PO'-PO\big)=0,  \ \text{for} \ p=2,
\end{align}
where \( \delta QO > 0 \) and \( \delta PO < 0 \), as illustrated in Fig. \ref{EWCW3p} (right). Now, returning to the AdS/NCFT with \( p \ge 3 \), the same deformation yields:
\begin{align}\label{sect 3: RT2}
\delta QO+(p-1)\delta PO=(p-2)\delta PO<0,  \ \text{for} \ p\ge 3.
\end{align} 
It means that pulling down the joint point (the orange point) can reduce the area of the surface, 
leading to EW \( \subset \) CW.
Furthermore, the greater the number of branches, the larger the reduction of the EW. 
Since the argument only relies on the local deformation of the extremal surface near the Net-brane, it also 
applies to the case of a Net-brane with arbitrary tension.

In summary, we have argued that EW$\subset$CW for $T = 0$ and $p \ge 3$; the larger $p$ is, the smaller EW is relative to CW. Previous discussions of AdS/ICFT suggest that decreasing the brane tension tends to shrink the EW, and increasing the number of edges further enhances this shrinking effect. Therefore, to satisfy ``EW \( \supseteq \) CW" for AdS/NCFT, we must have \( T > 0 \). Additionally, with larger \( p \), the lower bound of the tension also increases. These discussions align with the quantitative result (\ref{sect 3: EWCW bound}). Finally, let us comment on how to derive the quantitative result (\ref{sect 3: EWCW bound}). We find EW=CW for the symmetric subsystem $\overset{(1)}{L}=\overset{(2)}{L}=...=\overset{(p)}{L}$. In the opposite limit, when $\overset{(1)}{L}=1, \overset{(2)}{L}=...=\overset{(p)}{L}\to 0$, the requirement ``EW \( \supseteq \) CW" leads us to the causal bound (\ref{sect 3: EWCW bound}). For further details, please refer to the next subsection. 

\subsection{The causal bound}

In this subsection, we derive the causal bound on brane tension based on the wedge inclusion condition, specifically ``EW \( \supseteq \) CW." Without losing generality, we choose the subsystem on the \(m\)-th edge as \(\overset{(m)}{x} \leq \overset{(m)}{L}\), where \(\overset{(1)}{L} \geq \overset{(2)}{L} \geq \ldots \geq \overset{(p)}{L}\). We begin by examining the case where \(G_{N~m}=G_{N}\) and will subsequently generalize the results to accommodate arbitrary Newton's constants. To simplify our analysis, we focus on a specific time slice and denote the intersection points of the EW and CW with the Net-brane as \(z_b\) and \(z_c\), respectively. The condition ``EW \( \supseteq \) CW" requires that 
\begin{align}\label{sect 3: condition for intersections}
z_b \ge z_c.
\end{align}
Without loss of generality, we take the EW and CW in branch \(B_1\) to determine the intersection points \(z_b\) and \(z_c\).
The EW and CW in other branches will yield the same results since both the EW and CW are continuous across the Net-brane. 

As discussed in the previous subsection, the CW in branch $B_1$ is determined by $\overset{(1)}{L}$ and the smallest subregion length $\overset{(p)}{L}$. The boundary of CW is a circular arc with a diameter $\overset{(1)}{L}+\overset{(p)}{L}$, and its intersection point with the Net-brane can be determined as 
\begin{align}\label{sect 3: zc}
    z_{c}=\frac{\cos\theta}{2}\left( \sqrt{(\overset{(1)}{L}-\overset{(p)}{L})^{2}\sin^{2}\theta +4\overset{(1)}{L}\overset{(p)}{L}}-(\overset{(1)}{L}-\overset{(p)}{L})\sin\theta \right),
\end{align}
where $(\theta+\frac{\pi}{2})$ is the angle between edge $E_1$ and the Net-brane, which is related to the brane tension as $T=p\tanh\rho=p\sin\theta$.

Now, let us examine the intersection point \(z_b\) of EW with the Net-brane. According to \cite{Guo:2025sbm}, the holographic entanglement entropy (HEE) can be expressed as:
\begin{align}\label{sect 3: HEE for zb}
   S(z_b)=\sum_{m}^{p}\frac{1}{4G_{N}}\log \left( \frac{(\overset{(m)}{L}+z_b \tan\theta )^{2}+z_b^{2}}{z_b\epsilon} \right),
\end{align}
where we take the same Newton's constants across all branches, and $z_b$ is determined by minimizing the entropy. For cases of different Newton's constants in different branches, \(G_{N}\) should be replaced with \(G_{N~m}\) in the equation above.

\begin{figure}
    \centering
    \includegraphics[width=0.7\linewidth]{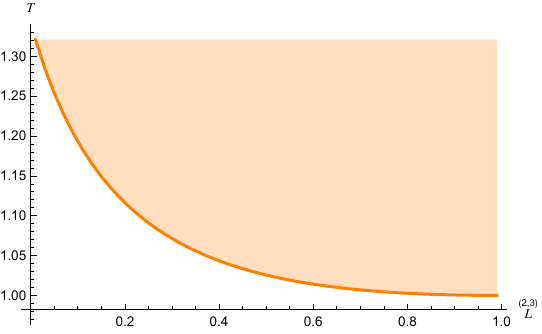}
    \caption{Tension bound for $p=3$, $G_{N~m}=G_{N}$. We fix \(\overset{(1)}{L}=1\) and every \(\overset{(2,3)}{L}\) give a bound for the tension $T$. The allowed range of the tension is shown by the orange region. It is clear that the strongest bound is obtained in the limit \(\overset{(2,3)}{L}\to 0\).}
    \label{fig: Tension bound example}
\end{figure}

Due to the minimal nature of $S(z_{b})$ and EW \( \supseteq \) CW on the Net-brane, namely $ z_{b}\geq z_{c}$, we require
\begin{align}\label{sect 3: EWCW dS}
	S'(z_{c})\le 0.
\end{align}
We fix the values of \( \overset{(1)}{L} \) and \( \overset{(p)}{L} \), under which \( z_c \) is fully determined by \( \theta \). As we decrease \( \overset{(q)}{L} \) (where \( q \neq 1, p \)) among the remaining branches, the corresponding \( z_b \) decreases accordingly. Hence, we find that, with \( \overset{(1)}{L} \) and \( \overset{(p)}{L} \) fixed, the case where \( \overset{(q)}{L} = \overset{(p)}{L} \) imposes the strongest constraint. Furthermore, this constraint becomes even stronger in the limit as \( \overset{(p)}{L} \) approaches zero. Therefore, this limit provides the most stringent constraint. For an illustrative example, see Fig.~\ref{fig: Tension bound example}.

In the limit $\overset{(2)}{L}=...=\overset{(p)}{L} \to 0$, from (\ref{sect 3: zc}) and (\ref{sect 3: HEE for zb}), we derive the following expression for $p = 2$:
\begin{align}\label{sect 3: EWCW dS2}
    p=2:~S'(z_{c})=\begin{cases}
        \frac{-2\tan \theta \sin ^{2} \theta }{G_{N}  (5-3 \cos (2 \theta ))\overset{(p)}{L}}+O(1),~\theta\in(0,\pi/2),\\
        0,~\theta=0,\\
        -\frac{\csc \theta \sec \theta \overset{(p)}{L}}{G_{N}\overset{(1)}{L}{}^{2}}+O(\overset{(p)}{L}{}^{2}),~\theta\in(-\pi /2,0).
    \end{cases}
\end{align}
For $p=2$, the requirement $S'(z_{c})\leq 0$ establishes a causal constraint $T=2\sin \theta \geq 0$ on the Net-brane tension.
For the AdS$_{3}$/NCFT$_{2}$ case with $p\geq 3$, the corresponding expansion is given by
\begin{align}\label{sect 3: EWCW dS1}
    p\geq 3:~S'(z_{c})=\begin{cases}
        \frac{\tan \theta  ((p+2) \cos (2 \theta )+p-6)}{4G_{N}  (5-3 \cos (2 \theta ))\overset{(p)}{L}}+O(1),~\theta\in(0,\pi/2),\\
        \frac{p-2}{4G_{N} (\overset{(1)}{L}\overset{(p)}{L})^{1/2} }+O(\overset{(p)}{L}{}^{1/2}),~\theta=0,\\
        -\frac{(p-2) \csc (\theta ) \sec (\theta )}{4G_{N} \overset{(1)}{L}}+O(\overset{(p)}{L}),~\theta\in(-\pi /2,0).
    \end{cases}
\end{align}
For the case where \(\theta > 0\) and $p\geq 3$, the requirement \(S'(z_c) \le 0\) establishes a causal constraint on the tension of the Net-brane:
\begin{align}\label{sect3: tension bound final}
    T=p\sin\theta\geq p\sqrt{\frac{p-2}{p+2}}. 
\end{align}
Conversely, for \(\theta \le 0\) and $p\geq 3$, 
we always have \(S'(z_c) > 0\), indicating that the condition EW\( \supseteq \)CW 
is violated for non-positive tension of the Net-brane.
These results about $p=2$ and $p\geq 3$ share the same form of (\ref{sect3: tension bound final}), and are consistent with the discussion in the preceding subsection.

To conclude this subsection, we will discuss the scenario involving different Newtonian constants across various branches. Since CW is determined by the geometry and is independent of the Newtonian constants, the interaction point \( z_c \) with the Net-brane is still given by equation (\ref{sect 3: zc}). The value of \( z_b \) is determined by minimizing the following entropy:
\begin{align}\label{sect 3: HEE for zb differ GN}
   S(z_b)=\sum_{m}^{p}\frac{1}{4G_{N~m}}\log \left( \frac{(\overset{(m)}{L}+z_b \tan\theta )^{2}+z_b^{2}}{z_b\epsilon} \right),
\end{align}
where \( G_{N~m} \) represents the Newton’s constant in branch \( B_m \). We choose the subregion in the branch with the largest Newtonian constant to have a finite size, while we set \( \overset{(q)}{L}\to 0\) 
for the subregions on the remaining branches. By following the above approach, we derive the causal bound for AdS\(_{3}\)/NCFT\(_{2}\) within the framework of Einstein gravity with arbitrary Newtonian constants:
\begin{align}\label{sect 3: tension bound final2}
	T=p\sin\theta\geq \max_{q}\left\{p\sqrt{\frac{p-\frac{2G_{N}}{G_{N~q}}}{p+\frac{2G_{N}}{G_{N~q}}}}\right\},~p\geq 2.
\end{align}

In summary, this section first presents a qualitative argument demonstrating that the condition ``EW \( \supseteq \) CW" requires a positive tension of the Net-brane in AdS$_3$/NCFT$_2$ for \( p \geq 3 \). We then quantitatively derive the causal bounds (\ref{sect 3: EWCW bound}) and (\ref{sect 3: tension bound final2}) for scenarios with both the same and different Newton's constants, respectively. Notably, the causal bound (\ref{sect 3: EWCW bound}) is stronger than the unitary bound (\ref{sect 3: reflectivity bound}) derived from the positivity of holographic reflectivity. It would be interesting to explore the constraint ``EW \( \supseteq \) CW" in higher dimensions; however, given the complexity of the calculations, we will leave this for future work.

\section{Holographic compact network}
\label{Holographic compact network}

In this section, we investigate the holographic duals of compact networks featuring both nodes and outer boundaries. Consequently, the bulk geometry encompasses both Net-branes and end-of-the-world (EOW) branes (see Fig. \ref{fig: geometry p=3} for an example). For clarity, we concentrate on Einstein gravity and a single-node network with $p$ finite edges, noting that our results readily generalize to general compact networks with multiple nodes and loops. Building on the discussion of connected RT surfaces in Section \ref{Holographic entanglement entropy}, 
we propose that the EOW branes meet at a common joint on the Net-brane, and we derive the corresponding joint condition from the variational principle. We further propose that the vacuum states of compact networks are dual to appropriately glued AdS solitons, and provide an explicit example in AdS$_3$/NCFT$_2$ 
with tensive Net-branes.

\subsection{Holographic setup}

\begin{figure}
    \centering
    \includegraphics[width=0.5\linewidth]{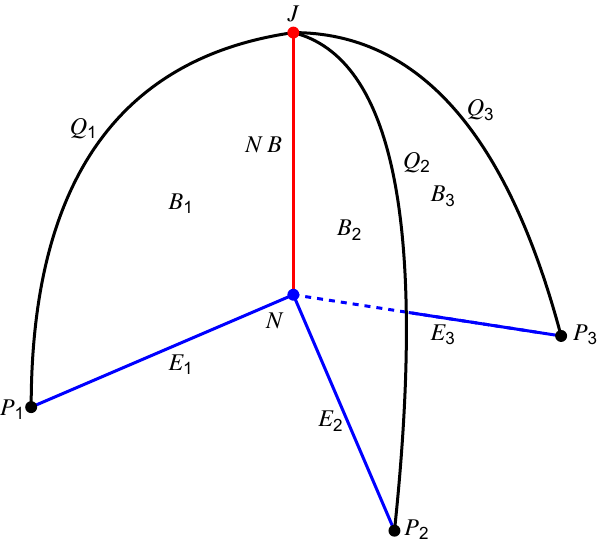}
    \caption{Geometry of the holographic compact network. In this representative example, we consider a positive tension Net-brane and a network with a single node connected by three edges with identical length. The blue lines indicate the edges on which the NCFT is defined. The red curve denotes the Net-brane, while the black curve represents the EOW brane. The blue, red, and black points denote the node \(N\), the joint \(J\), and the endpoints of the edges \(D_m = \partial E_m\), respectively.}
    \label{fig: geometry p=3}
\end{figure}

We begin by examining the geometry of a holographic compact network. As illustrated in Fig. \ref{fig: geometry p=3}, the node $N$ is dual to a Net-brane $NB$, while the outer boundaries $P_m$ are dual to EOW branes $Q_m$ in the bulk. The EOW branes all intersect at a common joint $J = NB \cap Q_m$ located on the Net-brane. Each edge $E_m$ is dual to a bulk branch $B_m$, which is bounded by both the Net-brane $NB$ and the EOW brane $Q_m$.

The action of the holographic compact network is given by
\begin{align}\label{sect4: action}
    I=&\sum_{m=1}^{p}\frac{1}{16\pi G_{N~m}}\left(\int_{B_{m}}d^{d+1}x\sqrt{|g|}\left(R+\frac{d(d-1)}{l_{m}^2}\right)+2\int_{Q_{m}}d^{d}y \sqrt{|h|}(-T_{m}+\overset{(m)}{K_Q})\right)\nonumber\\
    &+\frac{1}{8\pi G_{N}}\int_{NB}d^{d}y\sqrt{|h|}\left(-T+\sum_{m}^{p}\frac{G_{N}}{G_{N~m}}\overset{(m)}{K}\right)+I_{J}.
\end{align}
Here, $\overset{(m)}{K}$ and $\overset{(m)}{K_Q}$ denote the extrinsic curvatures on the Net-brane and the EOW brane $Q_m$, respectively. $T_m$ and $T$ represent the tensions of the EOW brane $Q_m$ and the Net-brane $NB$, respectively. Finally, $I_J$ refers to the Hayward term \cite{Hayward:1993my}, which is included to ensure a well-defined variation of the action. The explicit form of the Hayward term will be derived below.

Following the approach of Section \ref{Holographic Network}, 
we impose the continuity condition for the induced metrics 
\begin{align}\label{sect4: continuous condition}
    NB:~\overset{(m)}{h}_{ij}|_{NB}=h_{ij}|_{NB},
\end{align}
and the junction condition for the extrinsic curvatures on the Net-brane:
\begin{align}\label{sect4: junction condition}
    NB:~\sum_{m}^{p}\frac{G_{N}}{G_{N~m}}(\overset{(m)}{K}_{ij}-\overset{(m)}{K}h_{ij})=-Th_{ij}.
\end{align}
Similarly, we impose the Neumann boundary condition on the EOW brane \cite{Takayanagi:2011zk} \footnote{See \cite{Miao:2018qkc} and \cite{Chu:2021mvq} for discussions of Dirichlet boundary condition and conformal boundary condition.}
\begin{align}\label{sect4: NBC}
    Q_{m}:~\overset{(m)}{K_Q}{}_{ij}-\overset{(m)}{K_Q}h_{ij}=-T_{m}h_{ij},
\end{align}
where $h_{ij}$ denotes the induced metrics on corresponding branes.

\begin{figure}[!h]
    \centering
    \includegraphics[width=0.6\linewidth]{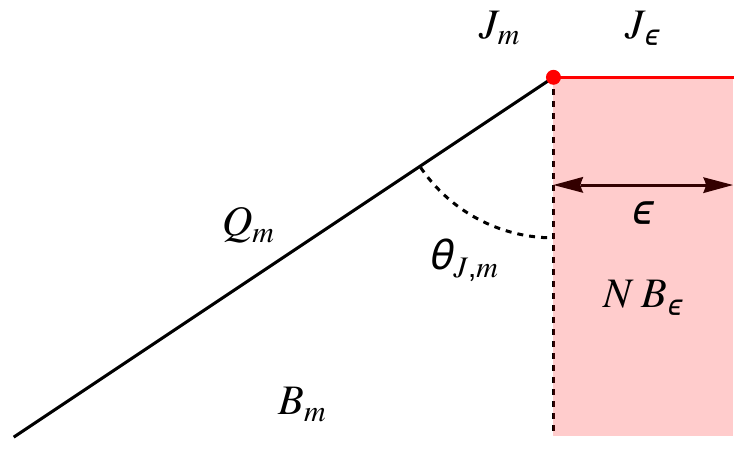}
    \caption{Local geometry of the regularized joint. The red region and the red line denote the regularized Net-brane $NB_{\epsilon}$ and the regularized joint $J_{\epsilon}$, respectively. The red point represents the secondary joint formed between the EOW brane $Q_m$ and the regularized joint $J_{\epsilon}$. After taking the limit $\epsilon \to 0$, we have $J_m \to J_{\epsilon} \to J$ and $NB_{\epsilon} \to NB$, where $\overset{(m)}{\theta}_{J}$ denotes the angle between the EOW brane $Q_m$ and the Net-brane.}
    \label{fig: joint smoothing}
\end{figure}

Next, we analyze the Hayward term and derive the joint condition for the EOW branes at the intersection $J$. We begin by regularizing both the Net-brane and the joint $J$ through the extension of the regulator $\epsilon$ into the branch $B_m$, as depicted in Fig.~\ref{fig: joint smoothing}. In this illustration, the red region corresponds to the regularized Net-brane $NB_{\epsilon}$, the red line to the regularized joint $J_{\epsilon}$, and the red point $J_m = Q_m \cap J_{\epsilon}$ marks the secondary joint formed by the intersection of the EOW brane $Q_m$ with the regularized joint $J_{\epsilon}$.

It is important to note that the regularized Net-brane $NB_{\epsilon}$ is part of the spacetime itself, rather than serving as a boundary like the EOW brane. Consequently, the Hayward term $(\pi - \overset{(m)}{\theta}_J)$, which corresponds to the angle $\overset{(m)}{\theta}_J$ between the EOW brane $Q_m$ and the Net-brane $NB$, should not be included. Instead, we introduce the Hayward term $(\pi - (\overset{(m)}{\theta}_J+\frac{\pi}{2}))$ for the angle $(\overset{(m)}{\theta}_J+\frac{\pi}{2})$ between the EOW brane $Q_m$ and the regularized joint $J_{\epsilon}$. Taking the limit $\epsilon \to 0$ and summing over all branches yields the Hayward term at the joint $J$
\begin{align}\label{sect4: Hayward-term}
I_{J}= \sum_{m=1}^{p}\frac{1}{8\pi G_{N,m}}\int_{J} d^{d-1}y \sqrt{|\sigma|}
\left(\frac{\pi}{2}-\overset{(m)}{\theta}_{J}\right),
\end{align}
where $\overset{(m)}{\theta}_{J}$ represent the angle between the EOW brane $Q_m$ and the Net-brane $NB$.

One can verify that the Hayward term above ensures a well-defined variation of the total action (\ref{sect4: action}). At the joint $J$, this leads to
\begin{align}\label{sect4: joint variation}
\delta I|_{J}=\sum_{m=1}^{p}\frac{ 1}{8\pi G_{N~m}}\int_{J}d^{d-1}y\sqrt{|\sigma|} \left( \frac{\pi}{2}-\overset{(m)}{\theta}_{J} \right)\frac{\sigma^{ab}}{2}\delta\sigma_{ab}
\end{align}
which leads to the following joint condition at the joint $J$:
\begin{align}\label{sect4: joint condition}
\sum_{m=1}^{p}\frac{1}{G_{N,m}}
\left(\frac{\pi}{2}-\overset{(m)}{\theta}_{J}\right)=0.
\end{align}
For detailed derivations, refer to Appendix \ref{app C}.

\subsection{Vacuum solutions}

In this subsection, we present the vacuum solution that satisfies all conditions—namely, the continuity condition (\ref{sect4: continuous condition}), the junction condition (\ref{sect4: junction condition}), the NBC (\ref{sect4: NBC}), and the joint condition (\ref{sect4: joint condition})—outlined in the preceding subsection. For clarity, we focus on a single-node network with $p$ edges in AdS$_3$/NCFT$_2$. 
Recall that the vacuum state of a strip is dual to an AdS soliton in AdS/BCFT \cite{Takayanagi:2011zk}. Similarly, the vacuum of a compact network is also dual to an AdS soliton in AdS/NCFT. Below, we provide further arguments to support this assertion.

Due to the symmetry of the vacuum and the geometry, the vacuum expectation value of the energy-momentum tensor takes the following form on edge $m$:
\begin{align}\label{sect 4.2: Tuv}
\langle \overset{(m)}{T}{}^{i}_{\ j} \rangle=\text{diag}\big( -\overset{(m)}{\rho}(x), \ \overset{(m)}{p}(x)\big),
\end{align}
where the energy density $\overset{(m)}{\rho}(x)$ and pressure $\overset{(m)}{p}(x)$ are independent of time. Since there is no Weyl anomaly in flat space, we find that
\begin{align}\label{sect 4.2: Tuv 1}
\langle \overset{(m)}{T}{}^{i}_{\ i} \rangle=0 \ \ \Rightarrow \ \overset{(m)}{\rho}(x)=\overset{(m)}{p}(x).
\end{align}  
Using energy-momentum conservation, we further derive:
\begin{align}\label{sect 4.2: Tuv 2}
\partial_{i}\langle T^{i}_{\ x} \rangle=\partial_{x} \overset{(m)}{p}(x)=0 \ \ \Rightarrow \ \overset{(m)}{\rho}(x)=\overset{(m)}{p}(x)\ \text{are constants}.
\end{align}  
In conclusion, the vacuum expectation value of the energy-momentum tensor is constant in a compact network. Calculations regarding the Casimir effect of free scalars support this conclusion \cite{Zhao:2025npv}. 

Next, we discuss the gravitational dual of a compact network. For each edge, two natural candidates yield constant holographic energy-momentum tensors: the BTZ black hole and the AdS soliton. The BTZ black hole corresponds to a positive constant energy density, while the AdS soliton corresponds to a negative constant energy density \cite{Miao:2024gcq}. The vacuum is defined as the quantum state with minimal energy; therefore, it is dual to an AdS soliton rather than a BTZ black hole.

In summary, we have demonstrated why the vacuum of a compact network is dual to an AdS soliton in the bulk. 
The bulk metric in each branch $B_{m}$ is given by:
\begin{align}\label{sect4: AdS soliton metric}
    ds^{2}=\frac{1}{z^{2}}\left( \frac{dz^{2}}{f_{m}(z)}+f_{m}(z)d\overset{(m)}{x}{}^{2}-dt^{2} \right),~f_{m}(z)=1-\frac{z^{2}}{z_{m}^{2}}.
\end{align}
Here, we set the AdS radius $l_{m}=1$ for all branches, and $z_m$ is a constant. The edge $E_m$ is defined by $0 \leq \overset{(m)}{x} \leq \hat{L}_m$, with the node located at $\overset{(m)}{x}=0$ on the AdS boundary $z=0$. Next, we specialize the continuity condition (\ref{sect4: continuous condition}), the junction condition (\ref{sect4: junction condition}), the NBC (\ref{sect4: NBC}), and the joint condition (\ref{sect4: joint condition}) to this setup.

It is important to note that the causal bound (\ref{sect 3: tension bound final2}) on the Net-brane tension obtained in Section \ref{Wedge inclusion condition} 
also applies to the setup discussed in this section. That is because, in the vicinity of the node, the AdS soliton geometry can be effectively approximated by the Poincaré AdS geometry. By selecting sufficiently small boundary subregions near the node, Poincaré AdS can accurately describe the local geometry. Consequently, the analysis from the previous section can be used to derive the same causal bound (\ref{sect 3: tension bound final2}). An open question remains as to whether a stronger constraint can be obtained when considering a larger subregion. However, since this falls outside the main focus of this section, we will leave it for future work. Note also that the tension of EOW brane can be negative \cite{Miyaji:2021ktr}, while the tension of Net-brane should be positive. 

\begin{figure}[t]
  \centering
  \begin{subfigure}{0.47\linewidth}
    \centering
    \includegraphics[width=\linewidth]{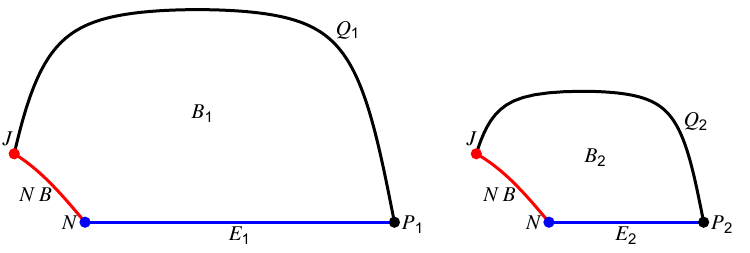}
    \caption{$L_{1}=2L_{2,3},~G_{N~1}=G_{N~2,3}$.}
    \label{fig: case a}
  \end{subfigure}
  \hfill
  \begin{subfigure}{0.47\linewidth}
    \centering
    \includegraphics[width=\linewidth]{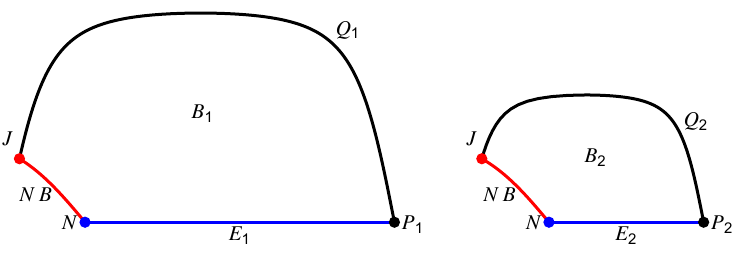}
    \caption{$L_{1}=2L_{2,3},~G_{N~1}=2G_{N~2,3}$.}
    \label{fig: case b}
  \end{subfigure}
  \caption{Two cases of solutions: Since branches $B_2$ and $B_3$ are symmetric, we only show branches $B_1$ and $B_2$. In these two figures, the blue line, red curve, and black curve represent the edge $E_{m}$, Net-brane $NB$, and EOW brane $Q_{m}$, respectively. The blue, red, and black dots correspond to the node $N$, joint $J$, and endpoint $P_{m}$ of the edge. Different branches are connected by the Net-brane.}
  \label{fig: two cases}
\end{figure}

In analogy with the AdS/BCFT construction, applying the NBC (\ref{sect4: NBC}), we derive the embedding function of the EOW brane \cite{Takayanagi:2011zk}
\begin{align}\label{sect4: EOW brane}
Q_m:~
\frac{d\overset{(m)}{x}}{dz}
= \frac{\pm}{f_m(z)}\frac{T_{m}}{\sqrt{f_{m}(z)-T_{m}^{2}}}.
\end{align}
Note that we should select the ``+" sign at the boundary \( P_m \), and choose the ``-" sign after the EOW brane passes the turning point. The turning point is defined as \( \overset{(m)}{x}{}' \to \pm \infty \), which results in \( z_{Q_m} = z_m \sqrt{1 - T_m^2} \). Refer to Fig. \ref{fig: two cases} for the EOW brane configuration with a turning point.
Next, we assume the embedding function of the Net-brane in each branch $B_{m}$ takes the form:
\begin{align}\label{sect4: NB embedding function}
NB:~\overset{(m)}{x}=X_{m}(z).
\end{align}
By substituting (\ref{sect4: NB embedding function}) into (\ref{sect4: AdS soliton metric}), we obtain the induced metric on the Net-brane:
\begin{align}
ds^{2}_{NB}=\frac{1}{z^{2}}\left( \frac{f_{m}(z)^{2}X_{m}'(z)^{2}+1}{f_{m}(z)}dz^{2}-dt^{2} \right).
\end{align}
Imposing the continuity of the induced metric (\ref{sect4: continuous condition}) across different branches yields
\begin{align}\label{sect4: continuous condition 2}
\frac{f_m(z)^{2} X_m'(z)^{2} + 1}{f_m(z)}=\frac{f_q(z)^{2} X_q'(z)^{2} + 1}{f_q(z)}\equiv F(z),
\end{align}
where $F(z)$ is a function of $z$ independent of the branch index $m$. Using (\ref{sect4: AdS soliton metric}), (\ref{sect4: NB embedding function}), and (\ref{sect4: continuous condition 2}), starting from the node $N$, the junction condition (\ref{sect4: junction condition}) imposes the following constraint:
\begin{align}\label{sect4: junction condition 2}
\sum_{m=1}^{p} \frac{G_{N}}{G_{N~m}}f_{m}(z)X_{m}'(z)=-T\sqrt{F(z)}.
\end{align}
Note that $\overset{(m)}{\theta}_J$ in the joint condition (\ref{sect4: joint condition}) is defined as
\begin{align}
\overset{(m)}{\theta}_{J}=\pi-\arccos{(\hat{n}_{Q_{m}}\cdot \overset{(m)}{\hat{n}}_{NB})}|_{J}.
\end{align}
where $\hat{n}_{Q_{m}}$ and $\overset{(m)}{\hat{n}}_{NB}$ represent the outward-pointing unit normal vectors in the bulk $B_{m}$ that are orthogonal to the EOW brane $Q_{m}$ and the Net-brane, respectively. Note that, in the case of $l_{m}=1$ and the tension bound (\ref{sect 3: tension bound final2}), $\overset{(m)}{\theta}_{J}<\pi$.

We now turn to a more concrete configuration, where the Net-brane tension $T$ satisfies the causal bound \eqref{sect 3: tension bound final2}. Initially, we consider a simplified scenario in which all edges have the same negative EOW brane tension \( T_m = T_{\text{EOW}} < 0 \) and share an identical length \( \hat{L}_m = \hat{L} \). In this setup, the Newton constant is allowed to vary across different edges. An exact solution can be derived in a straightforward analytic form:
\begin{align}\label{sect4: sym case EOW}
    Q_{m}:&~\overset{(m)}{x}=\begin{cases}
    &\hat{L}- z_{m}\left(\pi-\arctan{\left( \frac{z}{z_{m}\sqrt{f_{m}(z)/T_{\text{EOW}}^{2}-1}} \right)}\right),~z_{J}\leq z\leq z_{Q_{m}},\\
    &\hat{L}-z_{m}\arctan{\left( \frac{z}{z_{m}\sqrt{f_{m}(z)/T_{\text{EOW}}^{2}-1}} \right)},~z_{Q_{m}}\geq z\geq 0,
    \end{cases}\\
    NB:&~\overset{(m)}{x}=-z_{m}\arctan{\left( \frac{z}{z_{m}\sqrt{p^2f_{m}(z)/T^{2}-1}} \right)},~z_{J}\geq z\geq 0\label{sect4: sym case NB}\\
    J:  &~\overset{(m)}{\theta}_{J} = \frac{\pi}{2}, \label{sect4: sym case J}
\end{align}
where these branches share the identical configuration with \( z_{m} = z_{1} \) and \( f_{m}(z_{Q_{m}}) = T_{\text{EOW}}^{2} \). The length \( \hat{L} \) is given by
\begin{align}\label{sect4: sym case hat L}
	\hat{L}= z_{m}\left(\pi-\arctan{\left( \frac{z_{J}}{z_{m}\sqrt{f_{m}(z_{J})/T_{\text{EOW}}^{2}-1}} \right)}-\arctan{\left( \frac{z_{J}}{z_{m}\sqrt{p^2f_{m}(z_{J})/T^{2}-1}} \right)} \right).
\end{align}

The geometry of this configuration is illustrated in Fig.~\ref{fig: geometry p=3}. Although the Newton constants may differ among different branches, $G_{N\,m} \neq G_{N\,q}$,
the joint condition (\ref{sect4: joint condition}) remains satisfied due to $\overset{(m)}{\theta}_{J} = \frac{\pi}{2}$. 
These two types of branes, referred to as the Net-brane and the EOW brane, intersect at the joint located at $z_{J}=z_{m}\sqrt{1-T_{\text{EOW}}^{2}-T^{2}/p^{2}}$, 
which imposes an upper bound on the tension of the Net-brane in the length-symmetric case
\begin{align}\label{sect4: sym case upper bound of T}
T^{2}\leq p^{2}(1-T_{\text{EOW}}^{2}).
\end{align}
This upper bound, combined with the lower bound specified in (\ref{sect 3: tension bound final2}) for the Net-brane, establishes a constraint on the tension of the EOW brane:
\begin{align}\label{sect4: sym case upper bound of TEOW}
T_{\text{EOW}}^2 \le 1- \max_{q}\left\{\frac{p-\frac{2G_{N}}{G_{N~q}}}{p+\frac{2G_{N}}{G_{N~q}}}\right\}.
\end{align}
Only when the tensions of the Net-brane and the EOW brane satisfy these constraints can a symmetric configuration exist with \( T_{\text{EOW}} \leq 0 \). The scenario in which \( T_{\text{EOW}} \geq 0 \) is more complicated and will be addressed in future studies. It is not surprising that there are additional constraints on the tensions in the gravity dual of compact networks as compared to open networks. In fact, the gravity dual of the strip in AdS/BCFT also imposes further constraints on brane tension \cite{Miyaji:2021ktr, Miao:2024gcq}. 

Next, we examine a less symmetric case where two edges are symmetric; specifically, $\hat{L}_{2}=\hat{L}_{3}=1$ and $G_{N~2}=G_{N~3}$. The remaining edge, however, differs from these two, $\hat{L}_{1}=2\hat{L}_{2,3}$. Additionally, the tension of the Net-brane is set to $T=5/2$, which satisfies the causal bound (\ref{sect 3: tension bound final2}) in two scenarios. Meanwhile, the tensions of the EOW branes, $Q_{m}$, are fixed at $T_{1,2,3}=-1/3$. Depending on the source of asymmetry, we distinguish two cases:
\begin{itemize}
  \item[(a)] the remaining edge has the same Newton constant but a different edge length, namely
  $\hat{L}_{1}=2 \hat{L}_{2,3}$ and $G_{N\,1}=G_{N\,2,3}$;
  \item[(b)] the remaining edge has both a different Newton constant and a different edge length, namely
  $\hat{L}_{1}=2 \hat{L}_{2,3}$ and $G_{N\,1}=2G_{N\,2,3}$.
\end{itemize}
From (\ref{sect4: continuous condition 2}) and (\ref{sect4: junction condition 2}), we derive
\begin{align}
	&F(z)=\frac{\Big[\Big(\frac{G_{N}}{G_{N~1}}\Big)^{2}-\Big(\frac{2G_{N}}{G_{N~2}}\Big)^{2}\Big]\Big[\Big(\frac{G_{N}}{G_{N~1}}\Big)^{2}f_{1}-\Big(\frac{2G_{N}}{G_{N~2}}\Big)^{2}f_{2}\Big]}{\Big[\Big(\frac{G_{N}}{G_{N~1}}\Big)^{2}f_{1}-\Big(\frac{2G_{N}}{G_{N~2}}\Big)^{2}f_{2}\Big]^{2}+T^{2}\Big[T^{2}-2\Big(\frac{2G_{N}}{G_{N~2}}\Big)^{2}f_{2}-2\Big(\frac{G_{N}}{G_{N~1}}\Big)^{2}f_{1}\Big]}\nonumber\\
	&+\frac{2\frac{G_{N}}{G_{N~1}}\frac{2G_{N}}{G_{N~2}}\sqrt{T^{2}-\Big(\Big(\frac{G_{N}}{G_{N~1}}\Big)^{2}-\Big(\frac{2G_{N}}{G_{N~2}}\Big)^{2}\Big)(f_{1}-f_{2})}}{\Big[\Big(\frac{G_{N}}{G_{N~1}}\Big)^{2}f_{1}-\Big(\frac{2G_{N}}{G_{N~2}}\Big)^{2}f_{2}\Big]^{2}+T^{2}\Big[T^{2}-2\Big(\frac{2G_{N}}{G_{N~2}}\Big)^{2}f_{2}-2\Big(\frac{G_{N}}{G_{N~1}}\Big)^{2}f_{1}\Big]},
\end{align}
where we omit $(z)$ of the functions $f_{m}(z)$ for the sake of brevity. Next, with (\ref{sect4: continuous condition 2}), we derive the embedding functions of the Net-brane in each branch $B_{m}$ as
\begin{align}
	\frac{d\overset{(m)}{x}}{dz}=X_{m}'(z)=\frac{-1}{f_{m}(z)}\sqrt{f_{m}(z)F(z)-1}.
\end{align}

From (\ref{sect4: EOW brane}) and the two equations mentioned above, we can derive the edge length as follows: 
\begin{align}\label{sect4: edge length} 
\hat{L}_m = \int_0^{z_J} X_{m}'(z) \, dz + \int_{z_J}^0 \overset{(m)}{x}{}_{\text{EOW}}'(z) \, dz, 
\end{align} 
where \( z_J \) denotes the joint location, and \( X_{m}(z) \) and \( \overset{(m)}{x}_{\text{EOW}}(z) \) represent the embedding functions of the Net-brane and the EOW brane, respectively.  Given the edge length \( \hat{L}_{1}=2\hat{L}_{2,3}=2 \), Newton's constants \( G_{N\ m} \), and the brane tension \( T_m = -1/3, \, T =  5/2\), one can numerically solve (\ref{sect4: edge length}) to determine the solution parameters \( z_J \) and \( z_m \). These two illustrative examples are shown in Fig. \ref{fig: two cases}.

Finally, we evaluate the free energy of holographic compact networks. The free energy can be evaluated from the renormalized Euclidean on-shell action, which differs from the Lorentzian action by a negative sign. We have:
\begin{align}\label{sect 4: Euclidean action}
    I_{\text{Euclidean}}&=\beta F=- I_{\text{Lorentzian}}\nonumber\\
    &=-I-\sum_{m=1}^{3}\frac{1}{8\pi G_{N~m}}\left[ \int_{E_{m}}d^{2}y\sqrt{|h|}(K-1)\right.\nonumber\\
    &\left.-\int_{P_m} dy \sqrt{|\sigma|}
    (\overset{(m)}{\theta}_{P}-\overset{(m)}{\theta}_{P,0})-\int_{N} dy \sqrt{|\sigma|}
    (\overset{(m)}{\theta}_{N}-\overset{(m)}{\theta}_{N,0}) \right],
\end{align}
where \(\beta\) represents the period of Euclidean time, and \(F\) denotes the free energy. The term \(\overset{(m)}{\theta}_{P}\) represents the angle between the EOW brane \(Q_m\) and the AdS boundary \(E_m\) at their intersection point \(P_m\). Meanwhile, \(\overset{(m)}{\theta}_{N}\) denotes the angle between the Net-brane and the AdS boundary \(E_m\) at the node \(N\). Additionally, \(\overset{(m)}{\theta}_{\#,0}\) refers to the angle \(\overset{(m)}{\theta}_{\#}(z=0)\). After performing some calculations, we derive the free energy
\begin{align}\label{sect 4: free energy NCFT}
F_{\text{NCFT}}=W_{\text{Casimir}}=\sum_{m}^{p}\frac{-1}{16\pi G_{N~m}} \frac{\hat{L}_{m}}{z_{m}^{2}},
\end{align}
which is equal to the Casimir energy \(W_{\text{Casimir}}\). Here, \(T_{tt} = \frac{-1}{16\pi G_{N~m}} \frac{1}{z_{m}^{2}}\) represents the holographic energy density \cite{Miao:2024gcq}, and \(\hat{L}_{m}\) is the length for edge \(E_m\). This result aligns with the fact that the free energy equals the internal energy at zero temperature. As discussed after equation (\ref{sect4: edge length}), we can numerically determine \(z_{m}\) and thus calculate the free energy for given network lengths, the brane tensions, and values of Newton's constants.

The free energy of \( p \) isolated holographic strips is given by the expression \cite{Takayanagi:2011zk}:
\begin{align}\label{sect 4: free energy BCFT}
F_{\text{BCFT}}=\sum_{m}^{p}\frac{-1}{16\pi G_{N~m}} \frac{\pi^{2}}{ \hat{L}_{m}}. 
\end{align}
We define the binding energy of the network as the energy difference between the network and its isolated constituent units (the strips):
\begin{eqnarray} \label{sect 4: network binding energy}
 F_{\text{bind}}=F_{\text{NCFT}}-F_{\text{BCFT}} 
 \end{eqnarray}
From (\ref{sect 4: free energy NCFT}) (\ref{sect 4: free energy BCFT}) and (\ref{sect 4: network binding energy}), we derive
\begin{align}\label{sect 4: bind energy}
 F_{\text{bind}}=\sum_{m}^{p}\frac{-1}{16\pi G_{N~m}}\left( \frac{\hat{L}_{m}}{z_{m}^{2}}-\frac{\pi^{2}}{ \hat{L}_{m}} \right) \ge 0,
\end{align}
where we have used the inequality \( \hat{L}_{m} \le \pi z_m \). It is important to note that we have glued parts of holographic strips together to construct the gravity dual of compact networks. For each strip, the strip width is given by \( \pi z_m \) \cite{Takayanagi:2011zk}, indicating that \( \hat{L}_{m} \leq \pi z_m \) for a part. To verify the inequality (\ref{sect 4: bind energy}), we consider the two cases discussed earlier. In case (a), we find that \( F_{\text{bind}} \approx 13.63 > 0 \), while in case (b), \( F_{\text{bind}} \approx 24.33 > 0 \). Here, we have set \( 16\pi G_{N~1} = 1 \). These holographic results suggest that the network's binding energy is non-negative, indicating that energy must be expended to construct a network from individual components. It is intriguing to investigate whether the conclusion \( F_{\text{bind}} \geq 0 \) holds for general NCFTs. We will leave general studies and tests for future work.

\section{Traversable Parallel Universe}
\label{Traversable Parallel Universe}
This section shows that the AdS/NCFT is a natural realization of the parallel universe. Each branch in the bulk can be viewed as its own universe, potentially exhibiting different geometries and physical laws than other universes. These universes can be consistently connected through the junction conditions on the Net-brane. Notably, it is possible to travel from one universe to another without violating fundamental physical principles, such as causality and unitarity. It is important to emphasize that our concept of parallel universes differs from both the many-worlds interpretation in quantum mechanics \cite{Everett,Tegmark:2007wh} and the multiverse proposed in eternal inflation \cite{Eternalinflation}, as these are typically non-traversable between different realms.

Let us discuss how different the parallel universe can be in principle. 

\begin{itemize}

 \item  The coupling constants of interactions can vary across different universes. We have provided examples in Section \ref{Holographic Network}
 regarding gravity and Maxwell's theory. Now, let us discuss the scalar field with \( \Phi^4 \) interactions:
 \begin{align}\label{sect5: scalar action} 
I = \sum_{m=1}^p \int_{B_m} d^{d+1}x \sqrt{|g|} \left(-\frac{1}{2} \nabla \Phi \nabla \Phi -\frac{1}{2} M_m^2 \Phi^2+ \lambda_m \Phi^4 \right),
\end{align} 
where the mass \( M_m \) and the coupling constant \( \lambda_m \) can differ across various universes. Since the junction condition is derived from the kinetic energy term, it remains independent of the parameters \( M_m \) and \( \lambda_m \). We have two types of junction conditions \cite{Pang:2025flq}:
\begin{align}\label{sect5: scalar JCI}
& \text{JC I: }\ \ \ \sum_{m}\partial_{\hat{n}}\overset{(m)}{\Phi}|_{NB}=0, \ \ \ \overset{(m)}{ \Phi}|_{NB}=\overset{(n)}{\Phi}|_{NB}, \\
& \text{JC II: }\ \ \ \sum_{m}\overset{(m)}{\Phi}|_{NB}=0, \ \ \ \partial_{\hat{n}}\overset{(m)}{ \Phi}|_{NB}=\partial_{\hat{n}}\overset{(n)}{\Phi}|_{NB}, \label{sect5: scalar JCII}
\end{align}
which correspond to NBC and DBC for BCFTs, respectively. 

 \item The number of fundamental fields can vary across different universes. For instance, we have the standard model with four fundamental interactions in our universe, but could have no strong or weak interactions in other universes. In such cases, we apply junction conditions for the fields that appear in all universes, while implementing reflecting boundary conditions for the field that exists only in a single universe. This prevents it from propagating to other universes.

To illustrate this concept, let's consider scalar fields. Assume we have a scalar field that exists in all universes, as shown in action (\ref{sect5: scalar action}), alongside another scalar field that resides solely in universe I. The action for this scalar field is given by:
  \begin{align}\label{sect5: scalar action 1} 
I = \int_{B_1} d^{d+1}x \sqrt{|g|} \left(-\frac{1}{2} \nabla \Phi_1 \nabla \Phi_1 -\frac{1}{2} M^2 \Phi_1^2+ \lambda \Phi_1^2 \Phi^2 \right),
\end{align} 
which includes interactions between the two scalar fields. We impose JCs (\ref{sect5: scalar JCI},\ref{sect5: scalar JCII}) for $\Phi$, while applying either NBC or DBC for $\Phi_1$
\begin{align}\label{sect5: scalar NBC}
& \text{NBC: }\ \ \ \partial_{\hat{n}} \Phi_1 |_{NB}=0, \\
& \text{DBC: }\ \ \ \Phi_1|_{NB}=0. \label{sect5: scalar DBC}
\end{align}
One can verify that the energy conservation (\ref{sect2: matter Tij JC}) holds on the Net-brane under the given junction and boundary conditions.

Additionally, gravity may exist in some universes but not in others. Further examples can be found in Section \ref{Gravity-free universe}.

  \item It is interesting to explore whether fundamental constants such as \(\hbar\), \(k_B\), and \(c\) can vary across different universes. In this discussion, we will focus on the speed of light \(c\), leaving the examination of other fundamental constants for future work. In a \((1+1)\)-dimensional spacetime, it is possible for different universes to have different speeds of light without violating causality. To see this, let us consider a similar scenario involving a network composed of strings in our real world. The transverse micro-displacement of these strings can be described by a free scalar with the speed of sound given by \(v = \sqrt{T/\lambda}\), where \(T\) represents the string tension and \(\lambda\) represents the linear mass density. Of course, we can join the strings with different $ v = \sqrt{T/\lambda}$ in the real world, and nothing goes wrong. The case of higher dimensions is more subtle. If we require that the induced speeds of light from each branch to the Net-brane be uniform, then all the universes must have the same speed of light. This issue warrants further investigation, and we will address it in future research.
  
 \item  The glue between various universes is not arbitrary. The entire network of spacetime must adhere to fundamental physical principles, such as unitarity and causality. We observed these constraints in Section \ref{KK modes and stability}, 
 where the ghost-free (unitarity) and tachyon-free (causality) conditions imposed restrictions on the Newton's constants, Gauss-Bonnet couplings, the AdS radius, and the brane tension. See (\ref{sect2: ghost-free condition}) for KK modes. Additionally, we identify further constraints on the tension of the Net-brane in Section \ref{Wedge inclusion condition}. 
 Please see the unitary bound (\ref{sect 3: reflectivity bound}) and the causal bound (\ref{sect 3: EWCW bound}) in AdS$_3$/NCFT$_2$.
 
 \item The parallel universes in our setup are traversable in terms of probability. For instance, consider $p$ flat universes connected at a flat junction. Under the JC I (\ref{sect5: scalar JCI}) and JC II (\ref{sect5: scalar JCII}), when a free scalar wave travels from universe I to the junction, it has a probability of $(p-2)^2/p^2$ of being reflected and a probability of $(2/p)^2$ of being transmitted to universe II or other universes \cite{Guo:2025sbm}. In other words, when we try to travel from universe I to universe II, we at most have a probability $(2/p) ^2$ to succeed under the JC I and JC II.

 \end{itemize}

\subsection{Threefold universe}

The geometries of spacetime can differ across various universes. This section explores a simplified model of a parallel universe that links flat, de Sitter (dS), and anti-de Sitter (AdS) spacetimes in $1+3$ dimensions. Since flat, dS, and AdS spacetimes are the only three maximally symmetric spaces, this model holds a unique significance among all types of parallel universes. For simplicity, we refer to it as the ``threefold universe." 

The metric for each universe is given by
\begin{align}\label{sect5: threefold universe}
ds^2=-\frac{(1+\kappa \frac{r^2}{L^2})}{(1+\kappa \frac{r_{NB}^2}{L^2})}dt^2+\frac{dr^2}{1+\kappa  \frac{r^2}{L^2}}+r^2 d\Omega^2,\ \ 0\le r \le r_{NB}, 
 \end{align}
 where \(\kappa = 0, -1, 1\) corresponds to flat, de Sitter (dS), and Anti-de Sitter (AdS) spacetimes with a radius \(L\). For our analysis, we set \(L = 1\). The term \(r_{NB}\) denotes the location of the Net-brane. We have rescaled time such that the induced metric on the Net-brane is given by:
 \begin{align}\label{sect5: metric NB}
ds_{NB}^2=-dt^2+r_{NB}^2 d\Omega^2.  
 \end{align}
We assume that the threefold universes obey Einstein's gravity, with the same Newton's constant but different cosmological constants. The JC (\ref{JC Einstein gravity}) requires the matter stress tensor on the Net-brane:
 \begin{align}\label{sect5: Tij NB} 
 \mathcal{T}^{i}_{\ j}=\sum_{m=1}^3 \frac{1}{8\pi G_N}\Big(\overset{(m)}{K}{}^{i}_{\ j}-\overset{(m)}{K} h^{i}_{\ j}\Big)|_{NB}=\text{diag}\Big(-\rho, p,p\Big),
\end{align} 
 where the energy density and pressure are given by
  \begin{align}\label{sect5: energy density} 
&8\pi G_N \rho=\frac{2 \left(\sqrt{1-r^2}+\sqrt{r^2+1}+1\right)}{r}|_{r=r_{NB}} >0, \\
&8\pi G_N p=-\frac{ \left(\sqrt{1-r^4}+2 \left(\sqrt{1-r^2}-\sqrt{r^2+1}\right) r^2+\sqrt{1-r^2}+\sqrt{r^2+1}\right)}{r \sqrt{1-r^4}}|_{r=r_{NB}}. \label{sect5: pressure} 
\end{align}
It is important to note that we have \(\rho > 0\) and \(\rho + p > 0\):
   \begin{align}\label{sect5: null energy condition} 
8\pi G_N (\rho+p)=\frac{ \left(\sqrt{1-r^4}+\sqrt{1-r^2}+\sqrt{r^2+1}\right)}{r \sqrt{1-r^4}}|_{r=r_{NB}}>0.
\end{align} 
Thus, the matter fields on the brane are well-defined since they satisfy both the weak and null energy conditions. One can realize the brane stress tensor (\ref{sect5: Tij NB}) exactly by considering a tension term and a free scalar field, given by \(\phi = c_1 t\) on the Net-brane:
 \begin{align}\label{sect5: Tij NB 1}
 \mathcal{T}_{ij}=\frac{-T}{8\pi G_{N}} h_{ij}+\partial_i \phi \partial_j \phi-\frac{1}{2}(\nabla\phi)^2h_{ij},
\end{align} 
with the parameters defined as follows:
 \begin{align}\label{sect5: threefold T} 
&T=\frac{4 \left(\sqrt{1-r^2}-\sqrt{r^2+1}\right) r^2+3 \left(\sqrt{1-r^4}+\sqrt{1-r^2}+\sqrt{r^2+1}\right)}{2 r \sqrt{1-r^4}}|_{r=r_{NB}},\\
&c_1^2=\frac{\sqrt{1-r^4}+\sqrt{1-r^2}+\sqrt{r^2+1}}{8\pi G_{N}r \sqrt{1-r^4}}|_{r=r_{NB}}>0. \label{sect5: threefold cc} 
\end{align} 

In summary, we have constructed a toy model of a threefold universe using physically well-defined brane matters. Further investigation of this model will be conducted in a separate paper.

\subsection{Gravity-free universe}
\label{Gravity-free universe}
This subsection presents examples of how the gravitational universe interacts with non-gravitational universes. These toy models help to study the black hole information paradox, in which gravity in the black hole region is linked to non-gravitational baths in the radiation region. For simplicity, we will focus on just two universes with \( p = 2 \); extending the discussion to \( p > 2 \) is straightforward.

In the first model, we consider Einstein gravity in only universe I, while matter fields exist in both universes I and II. The action is given by
\begin{align}\label{sect5: gravity model I} 
I =\sum_{m=1}^2 \int_{B_m} d^{d+1}x \sqrt{|g|} \left(\frac{\epsilon_m}{16\pi G_N} \left(R-2\Lambda\right)-\frac{1}{2} \nabla \Phi \nabla \Phi -\frac{1}{4} \mathcal{F}^2 \right),
\end{align} 
where $\epsilon_1=1, \epsilon_2=0$, indicating that gravity is confined to universe I. We focus on massless matter fields, as they dominate Hawking radiation. We impose DBC for the induced metric on the Net-brane
 \begin{align}\label{sect5: model I gravity DBC} 
\text{DBC}: \ \delta h_{ij}|_{NB}=0,
\end{align} 
ensuring that gravity remains restricted to universe I. We do not choose Neumann boundary conditions (NBC) because they typically lead to fluctuating gravity on the Net-brane, which is inconsistent with the fixed spacetime background of the non-gravitational universe II. Unlike gravity, we employ the JCs (\ref{sect5: scalar JCI})(\ref{sect5: scalar JCII}) 
for the scalar and the following JCs for Maxwell fields
\begin{align}\label{sect5: vector JCI}
& \text{JC I: }\ \ \ \sum_{m}\overset{(m)}{\mathcal{F}}_{\hat{n}i}|_{NB}=0, \ \ \ \overset{(m)}{\mathcal{F}}_{ij}|_{NB}=\overset{(n)}{\mathcal{F}}_{ij}|_{NB}, \\
& \text{JC II: }\ \ \ \sum_{m}\overset{(m)}{\mathcal{F}}_{ij}|_{NB}|_{NB}=0, \ \ \ \overset{(m)}{\mathcal{F}}_{\hat{n}i}|_{NB}=\overset{(n)}{\mathcal{F}}_{\hat{n}i}|_{NB}, \label{sect5: vector JCII}
\end{align}
which allows matter fields to propagate freely between the two universes.

In the second model, we consider Einstein gravity in universe I and non-dynamical gravity in universe II. For example, in four dimensions, the Gauss-Bonnet gravity is non-dynamical. The action of this model is given by
\begin{align}\label{sect5: gravity model II} 
I =\sum_{m=1}^2 \frac{1}{16\pi G_N}\int_{B_m} d^{4}x \sqrt{|g|}  \Big(\epsilon_m\left(R-2\Lambda\right)+(1-\epsilon_m) \mathcal{L}_{\text{GB}} \Big),
\end{align} 
where $\epsilon_1=1, \epsilon_2=0$ and $\mathcal{L}_{\text{GB}} =R_{\mu\nu\alpha\beta}R^{\mu\nu\alpha\beta}-4R_{\mu\nu}R^{\mu\nu}+R^2$.  For this model, we can choose either DBC (\ref{sect5: model I gravity DBC}) or NBC for the induced metric on the Net-brane
 \begin{align}\label{sect5: model II gravity NBC} 
\text{NBC}:\  \Big(K_{ij}-(K-T) h_{ij}\Big)|_{NB}=0,
\end{align} 
where $K_{ij}$ is the extrinsic curvature defined from universe I to the Net-brane. Notably, the Brown-York stress tensor for Gauss-Bonnet gravity vanishes in four bulk dimensions. Since gravity is non-dynamical and topological in universe II, the metric in this universe and the induced metric on the Net-brane are allowed to fluctuate. Therefore, it is appropriate to impose the NBC (\ref{sect5: model II gravity NBC})  on the Net-brane.   

Let us discuss the black hole solution to the first model with DBC (\ref{sect5: model I gravity DBC}), 
which is closely related to the black hole information paradox. For simplicity, we will focus on the case of a negative cosmological constant. The metric of universe I is given by
\begin{align}\label{sect5: model I metric 1} 
ds_{\text{I}}^2=\frac{\frac{dz^2}{h(z)}-h(z) dt^2+\sum_{a=1}^{d-1} dy_a^2}{z^2}, \ \ z\ge \epsilon,
\end{align} 
where \( h(z) = 1 - \frac{z^d}{z_h^d} \), and \( z = \epsilon \) is the location of the Net-brane. We have set the AdS radius to \(L = 1\). The metric of universe II can be represented either by a flat metric
\begin{align}\label{sect5: model I metric 2} 
ds_{\text{II}}^2=\frac{dz^2-h(\epsilon) dt^2+\sum_{a=1}^{d-1} dy_a^2}{\epsilon^2}, \ \ z\ge \epsilon,
\end{align} 
or a AdS metric exactly as (\ref{sect5: model I metric 1}). There is considerable freedom in choosing the metric for universe II, as long as it produces the required induced metric on the Net-brane.

In summary, we have coupled massless gravity in universe I with a non-gravitational bath in universe II. Some comments are in order. {\bf 1}. In double holography, such as AdS/BCFT or braneworld scenarios, one can also couple dynamical gravity on the AdS brane to a non-gravitational bath in flat space (the AdS boundary). However, in this context, the gravity on the brane is massive. This situation leads to the so-called  ``massive island" problem \cite{Geng:2020qvw,Geng:2021hlu}, which argues that all proposed resolutions of the black hole information paradox based on double holography apply only to massive gravity models, rather than to long-range Einstein gravity. {\bf 2}. Through the constructions presented in this section, we have successfully coupled massless gravity in AdS with a non-gravitational bath in flat space. Whether an entanglement island can emerge in the gravitational universe I is an intriguing and significant question; however, it falls outside the scope of this paper and will be addressed in future work. {\bf 3}. In previous studies, the non-gravitational bath has been modeled using strongly coupled CFTs, where calculations are feasible mainly for low-dimensional SYK models or holographic CFTs with gravitational duals. In lower dimensions, true gravity does not exist. In higher-dimensional holographic CFTs, we encounter the ``massive island" problem in the double holography framework. The advantage of our model is that it considers non-gravitational baths as free theories, making the calculations of entanglement entropy more accessible.

\section{Conclusions and Discussions}
\label{Conclusions and Discussions}
This paper examines a holographic network that features different conformal field theories (CFTs) at its edges. We focus on Gauss-Bonnet (GB) gravity with varying parameters across different bulk branches and explore various aspects of AdS/NCFT. We develop a new method, holographic Noether's theorem, to derive conservation laws at network nodes from the junction conditions on the Net-brane. This approach is much simpler than the Gauss-law method used in previous work \cite{Guo:2025sbm}. Additionally, we analyze the linear stability of the gravitational Kaluza-Klein (KK) modes on the Net-brane and obtain constraints on the GB couplings that are stronger than those found in AdS/CFT without Net-branes. Next, we study holographic entanglement entropy (HEE) and derive the connecting conditions of the Ryu-Takayanagi (RT) surfaces on the Net-brane. We explore various proposals for network entropy and confirm that type I and type II network entropies satisfy the holographic g-theorem in general dimensions, whereas type III network entropy remains non-negative. It suggests that type I and type II network entropies serve as effective measures of node degrees of freedom. In contrast, type III network entropy can characterize the information of internal edges. We also discuss the correlation functions of stress tensors and examine examples involving free scalars, as well as a holographic network with a tensionless Net-brane. 
We find that a tensionless Net-brane results in negative reflectivity at the node, suggesting that \( T=0 \) corresponds to a non-unitary parameter.

We examine the wedge inclusion condition in AdS/NCFT, which stipulates that the entanglement wedge (EW) must encompass the causal wedge (CW), expressed as ``EW$\supseteq$CW."  This relationship establishes a lower bound on the tension of Net-branes, which is stronger than the unitary bound derived from the positivity of holographic reflectivity \cite{Liu:2025khw}. Our findings indicate that the tension of Net-branes must be positive, and that an increase in the number of edges results in a larger lower bound on this tension. For the sake of simplicity, we focus on AdS$_3$/NCFT$_2$ in this paper. It would be interesting to extend these discussions further into higher dimensions.

We carefully study the holographic compact network, characterized by edges with outer boundaries. Due to the non-zero Casimir effect \cite{Zhao:2025npv}, the vacuum state of a compact network corresponds to appropriately glued AdS solitons instead of the Poincaré AdS. We derive the joint conditions for the EOW branes on the Net-brane, and present some vacuum solutions in AdS$_3$/NCFT$_2$. For simplicity, we focus on the scenario in which the Casimir force is attractive at all edges in this paper. According to \cite{Zhao:2025npv}, it is also possible for the Casimir force to be repulsive at certain edges, given a specific range of parameters. We plan to address the gravity dual associated with this nontrivial case in our future work.

We explore traversable parallel universes within the framework of AdS/NCFTs. Each branch in the bulk can be considered a separate universe, allowing for probabilistic travel between them. For instance, when attempting to send a single photon from Universe I to Universe II, we cannot guarantee a 100\% success rate. Instead, the photon has specific probabilities of being reflected or transmitted to other universes.
Traversability is a key distinction between our model of parallel universes and the many-worlds interpretation of quantum mechanics \cite{Everett, Tegmark:2007wh}, as well as the multiverse suggested by eternal inflation, where these universes are generally non-traversable. Notably, parallel universes with different geometries and physical laws can be consistently connected at the junction. The junction conditions ensure the conservation of energy and current but do not constrain other details of the theory. Causality and unitarity impose specific constraints on the parameters of the theory, but still allow for considerable freedom in connecting different universes consistently. We examine various possible traversable parallel universes and construct two toy models. The first model is known as the threefold universe, which combines flat, de Sitter (dS), and anti-de Sitter (AdS) universes. Remarkably, this model satisfies all energy conditions, indicating that it is physically well-defined. The second toy model connects universes with and without gravity, demonstrating that we can consistently couple massless gravity with a non-gravitational environment. This toy model has potential applications for addressing the black hole information paradox \cite{Penington:2019npb, Almheiri:2019psf, Almheiri:2020cfm}, and we will leave its study for future work. Finally, it is important to note that our models of traversable parallel universes differ from traversable wormholes, as they satisfy the null energy condition.

In the future, it will be intriguing to apply our AdS/NCFT framework to investigate real physical systems, such as certain aspects of neuronal networks in the brain and the quantum effects in chips. Recent studies have examined energy transport along intersecting lines \cite{Liu:2025khw} and the GHZ state in non-manifolds \cite{Jiang:2025iet}. It would be interesting to generalize these findings to broader networks.

\acknowledgments

Miao thank Cheng Peng and Ping Gao for the valuable comments and discussions. Miao acknowledges the support from the National Natural Science Foundation of China (NSFC) grant (No.12275366). 
The authors thank Yukawa Institute for Theoretical Physics at Kyoto University, where this work was improved during ``YITP-IAS workshop: Interfaces $\&$ Symmetry"  (YITP-I-25-04).


\appendix

\section{Stability of KK modes}\label{app A}

This appendix derives the ghost-free and tachyon-free conditions (\ref{sect2: ghost-free condition}) for the gravitational KK modes on the Net-brane. 

Let us begin with the orthogonal relation of the KK modes. Following the Sturm–Liouville theory \cite{Arfken:SturmLiouville}, we rewrite the linearized Einstein equation (\ref{sect2: EOMMBCmassiveH}) for the bulk branch \( B_{m} \) in the following form:
\begin{align}\label{app A: EOM bulk M}
    \frac{d}{dr}\left( l_{m}^{d}\cosh^{d}\left(\frac{r}{l_{m}}\right)\overset{(m)}{H}{}'_{M}(r) \right)+l_{m}^{d-2}\cosh^{d-2}\left( \frac{r}{l_{m}} \right)M^{2}\overset{(m)}{H}_{M}(r)=0.
\end{align}
where the prime $'$ denotes differentiation with respect to $r$. By combining this equation for different KK modes, we construct the following integral:
\begin{align}\label{app A: ortho construct 1}
    \sum_{m}^{p}&A_{m}\int_{-\infty}^{\rho_{m}}dr\left\{\left[\frac{d}{dr}\left( l_{m}^{d}\cosh^{d}{\left( \frac{r}{l_{m}} \right)}\overset{(m)}{H}{}'_{M} \right)+l_{m}^{d-2}\cosh^{d-2}{\left( \frac{r}{l_{m}} \right)}M^{2}\overset{(m)}{H}_{M}\right]\overset{(m)}{H}_{M'}\right.\nonumber\\
    &\left.-\left[\frac{d}{dr}\left( l_{m}^{d}\cosh^{d}{\left( \frac{r}{l_{m}} \right)}\overset{(m)}{H}{}'_{M'} \right)+l_{m}^{d-2}\cosh^{d-2}{\left( \frac{r}{l_{m}} \right)}M'^{2}\overset{(m)}{H}_{M'}\right]\overset{(m)}{H}_{M}\right\}=0,
\end{align}
where $\overset{(m)}{H}_{M}=\overset{(m)}{H}_{M}(r)$. By performing integration by parts and the DBC on the AdS boundary $\overset{(m)}{H}_{M}(-\infty)=0$, we derive:
\begin{align}\label{app A: ortho construct 2}
    &\sum_{m}^{p}A_{m}l_{m}^{d}\cosh^{d}{\left( \frac{\rho_{m}}{l_{m}} \right)}\left( \overset{(m)}{H}{}_{M}'(\rho_{m})\overset{(m)}{H}{}_{M'}(\rho_{m}) - \overset{(m)}{H}{}_{M'}'(\rho_{m})\overset{(m)}{H}{}_{M}(\rho_{m})\right)\nonumber\\
    =&(M'^2-M^2)\sum_{m}^{p}A_{m}\int_{-\infty}^{\rho_{m}}dr~l_{m}^{d-2}\cosh^{d-2}{\left( \frac{r}{l_{m}} \right)}\overset{(m)}{H}{}_{M}(r)\overset{(m)}{H}{}_{M'}(r).
\end{align}
Recall that the parameters \( A_m \) and \( C_m \) are defined in (\ref{sect2: AmBm}). As previously discussed in Section \ref{KK modes and stability}, \( A_m > 0 \). 
By applying the junction condition on the Net-brane:
\begin{align}\label{app A: junction condition}
    \sum_{m}^{p}(A_{m}\overset{(m)}{H}{}_{M}'-C_{m}M^{2}\overset{(m)}{H}_{M})|_{NB}=0,
\end{align}
we can rewrite (\ref{app A: ortho construct 2}) into the following form:
\begin{align}\label{app A: ortho construct 3}
    (M^{2}-M'^{2})\cdot\langle H_{M},H_{M'} \rangle=0,
\end{align}
where the inner product is defined as
\begin{align}\label{app A: orthogonal condition}
    \langle H_{M},H_{M'} \rangle=c_{M}\delta_{M,M'}=\sum_{m}^{p}&\left[A_{m}\int_{-\infty}^{\rho_{m}}dr~l_{m}^{d-2}\cosh^{d-2}{\left( \frac{r}{l_{m}} \right)}\overset{(m)}{H}{}_{M}(r)\overset{(m)}{H}{}_{M'}(r)\right.\nonumber\\
    &\left.+C_{m}l_{m}^{d}\cosh^{d}{\left( \frac{\rho_{m}}{l_{m}} \right)}\overset{(m)}{H}{}_{M}(\rho_{m})\overset{(m)}{H}{}_{M'}(\rho_{m})\right].
\end{align}
Note that (\ref{app A: ortho construct 3}) implies that \( \langle H_{M}, H_{M'} \rangle = 0 \) for \( M \neq M' \). Therefore, \( \langle H_{M}, H_{M'} \rangle \) defined in (\ref{app A: orthogonal condition}) represents the orthogonal relation for the KK modes.

Now we are ready to discuss the ghost-free condition, which requires non-negative inner products for all the modes:
\begin{align}\label{app A: positive inner product}
 c_M=\langle H_{M},H_{M} \rangle \ge 0. 
\end{align}
Following the approach of \cite{Miao:2023mui}, we define a step function \(\Pi_0\) that is non-zero only on the Net-brane. In each bulk branch \(B_m\), we have:
\begin{align}\label{app A: characteristic function}
\Pi_0(r)|_{B_m}=\overset{(m)}{\Pi}_{0}(\overset{(m)}{r})=\begin{cases}
        1,~\overset{(m)}{r}=\rho_{m},\\
        0,~\overset{(m)}{r}< \rho_{m}.
    \end{cases}
\end{align}
Similarly, we label the eigenfunctions of  KK modes by $H_M(r)$:
\begin{align}\label{app A: eigenfunctions}
H_M(r)|_{B_m}=\overset{(m)}{H}_M(\overset{(m)}{r}),
\end{align}
where $\overset{(m)}{H}_M(\overset{(m)}{r})$ is given by (\ref{sect2: Htwocase}) with $M$ obeying the spectrum constraint (\ref{sect2: spectrum constraint}). 
By expanding \(\Pi_0(r)\) in terms of the eigenfunctions \(H_M(r)\) and utilizing the orthogonality condition (\ref{app A: orthogonal condition}), we arrive at:
\begin{align} \label{app A: ghost-free proof 1}
 \Pi_{0}(r)=\sum_{M}\frac{\langle \Pi_{0},H_{M} \rangle}{\langle H_{M},H_{M} \rangle}H_{M}(r)&=\sum_M \frac{H_M(r)}{c_M}\sum_m^p C_m l_{m}^{d}\cosh^{d}{\left( \frac{\rho_{m}}{l_{m}} \right)} \overset{(m)}{H}_{M}(\rho_{m})\nonumber\\
 &= \sum_M \frac{H_M(r)}{c_M} \left(l_{1}^{d}\cosh^{d}{\left( \frac{\rho_{1}}{l_{1}} \right)} \overset{(1)}{H}_{M}(\rho_{1})\right)\sum_m^p C_m,
\end{align}
where we have used the continuity condition (\ref{sect 2: induced metric condition}) to derive the last line. Without loss of generality, we evaluate (\ref{app A: ghost-free proof 1}) at $r=\overset{(1)}{r}= \rho_1$. By using the definition of the step function (\ref{app A: characteristic function}), we derive the spectrum identity:
\begin{align}\label{app A: spectrum identity}
    \sum_{M}\frac{\overset{(1)}{H}{}_{M}(\rho_{1})^{2}}{c_{M}}=\frac{\left(l_{1}^{d}\cosh^{d}{\left( \frac{\rho_{1}}{l_{1}} \right)}\right)^{-1}}{\sum_{m}^{p}C_{m}},
\end{align}
where the index $1$ can be replaced with any edge index $n$. 

If the spectrum includes complex \( M^2 \), these must appear in complex-conjugate pairs. This requirement arises because the equation of motion (\ref{app A: EOM bulk M}), the boundary condition \(\overset{(m)}{H}(-\infty)=0\), and the junction condition (\ref{app A: junction condition}) are all real. Complex-conjugate pairs consist of a growing mode and a decaying mode, leading to instability. Therefore, for the stability of KK modes, a real spectrum is necessary. Additionally, we require that all inner products be non-negative, meaning \( c_M = \langle H_{M}, H_{M} \rangle \ge 0 \). These two conditions ensure that the left-hand side of equation (\ref{app A: spectrum identity}) is positive and provide a necessary ghost-free condition for the theory parameter \( C_m \) (\ref{sect2: AmBm}):
\begin{align}\label{app A: ghost-free condition 1}
 \sum_m^p C_m \ge 0.
\end{align}

Now, let us prove that if the condition (\ref{app A: ghost-free condition 1}) holds, the mass spectrum must be real and free of ghosts. Assuming there are complex mass pairs with $M\ne M^*$, we have
\begin{align} \label{app A: real mass 1}
    \langle H_{M}, H_{M^{*}} \rangle = c_{M}\delta_{M,M^{*}} = 0.
\end{align}
However, the pair of mutually complex-conjugate modes give rise to a positive inner product:
\begin{align}\label{app A: real mass 2}
 \langle H_{M}, H_{M^{*}} \rangle&=\sum_{m=1}^{p} 
A_{m}\int_{-\infty}^{\rho_{m}}dr\, 
l_{m}^{d-2}\cosh^{d-2}\!\left(\frac{r}{l_{m}}\right)
|\overset{(m)}{H}_{M}(r)|^{2}\nonumber\\
&\ +l_{1}^{d}\cosh^{d}\!\left(\frac{\rho_{1}}{l_{1}}\right)
|\overset{(1)}{H}_{M}(\rho_{1})|^{2} \sum_{m=1}^{p}C_{m}
 > 0,
\end{align}
where we have applied (\ref{app A: ghost-free condition 1}) and \( A_m > 0 \). This contradiction between the two equations indicates that there are no complex masses in the spectrum. For real values of \( M^2 \), we can directly verify that (\ref{app A: ghost-free condition 1}) is a sufficient condition for a positive inner product as specified in (\ref{app A: orthogonal condition}). In summary, we have demonstrated that (\ref{app A: ghost-free condition 1}) is both a necessary and sufficient condition for ensuring a real and ghost-free mass spectrum.

Next, we will demonstrate that the spectrum is automatically free of tachyons under the condition outlined in (\ref{app A: ghost-free condition 1}). To do this, we construct the following positive-definite integral:
\begin{align}
\sum_{m=1}^{p}
A_{m}\int_{-\infty}^{\rho_{m}}dr\,
l_{m}^{d}\cosh^{d}\!\left(\frac{r}{l_{m}}\right)
 \overset{(m)}{H}{}_{M}'(r) ^{2} > 0 .
\label{app A: construct integral 1}
\end{align}
By performing integration by parts and substituting the EOM~(\ref{app A: EOM bulk M}) and junction condition~(\ref{app A: junction condition}), we obtain
\begin{align}\label{app A: construct integral 2}
& M^{2}
\sum_{m=1}^{p}\Big[
A_{m}\int_{-\infty}^{\rho_{m}}dr\,
l_{m}^{d-2}\cosh^{d-2}\!\left(\frac{r}{l_{m}}\right)
\overset{(m)}{H}_{M}(r)^{2}
+
C_{m}l_{m}^{d}\cosh^{d}\!\left(\frac{\rho_{m}}{l_{m}}\right)
\overset{(m)}{H}_{M}(\rho_{m})^{2}
\Big]\nonumber\\
&= M^{2} c_{M} > 0 ,
\end{align}
where $c_{M}>0$ under the ghost-free condition (\ref{app A: ghost-free condition 1}). Thus, the KK spectrum is strictly positive, i.e., $M^{2}>0$. In Einstein gravity, the spectrum identity (\ref{app A: spectrum identity})  is no longer valid. Nevertheless, the KK spectrum remains real and positive, as can be demonstrated using (\ref{app A: real mass 2}) and (\ref{app A: construct integral 2}) in the Einstein gravity with $C_{m}=0$.

In summary, we have established the ghost-free and tachyon-free condition (\ref{app A: ghost-free condition 1}) for gravitational KK modes in this appendix.

\section{Junction condition for RT surfaces} \label{app B}

In this appendix, we derive the junction condition for RT surfaces in AdS/NCFT for Gauss-Bonnet gravity. We label the embedding function of the RT surface $\Gamma_m$ and its intersection to the Net-brane $\gamma=\Gamma_m\cap NB$ as $\overset{(m)}{x}{}^{\mu}(\overset{(m)}{\xi}{}^{\alpha})$ and $\overset{(m)}{x}{}^{\mu}(s^{a})$, respectively. Subsequently, we derive the induced metrics on \(\Gamma_m\) and \(\gamma\) as follows:
\begin{align}\label{app B: RT metric}
h_{\Gamma_m \ \alpha\beta}=\frac{\partial \overset{(m)}{x}{}^{\mu}}{\partial \overset{(m)}{\xi}{}^{\alpha}} \frac{\partial \overset{(m)}{x}{}^{\nu}}{\partial \overset{(m)}{\xi}{}^{\beta}} g_{\mu\nu}, \ \ h_{\gamma \ ab}=\frac{\partial \overset{(m)}{x}{}^{\mu}}{\partial s^{a}} \frac{\partial \overset{(m)}{x}{}^{\nu}}{\partial s^{b}} g_{\mu\nu}.
\end{align}

Taking variations of the entropy functional (\ref{sect2: HEEGB}), we obtain
\begin{align} \label{app B: dS}
\delta S &=\sum_m^p \frac{1}{4 G_{N\ m}} \int_{\Gamma_m} d^{d-1}\xi\sqrt{h_{\Gamma_m}}\left(\frac{1+\hat{\lambda}_m \mathcal{R}_m}{2}h_{\Gamma_m}^{\alpha\beta}-\hat{\lambda}_m  \mathcal{R}_m^{\alpha\beta}\right) \delta h_{\Gamma_m \ \alpha\beta} \nonumber\\
&+  \sum_m^p \frac{1}{ 4G_{N\ m}} \int_{\gamma} d^{d-2}s \sqrt{h_{\gamma}} \hat{\lambda}_m \left(\mathcal{K}_m h_{\gamma}^{ab}-\mathcal{K}_m^{ab}\right) \delta h_{\gamma \ ab},
\end{align}
where 
\begin{align}\label{app B: hat lambda}
\hat{\lambda}_m=\frac{2L_m^2 \lambda_{m}}{(d-2)(d-3)}.
\end{align}
Substituting the metric variations
\begin{align}\label{app B: metric variations 1}
&\delta h_{\Gamma_m \ \alpha\beta}=\partial_{\alpha} \delta \overset{(m)}{x}{}^{\mu} \partial_{\beta} \overset{(m)}{x}{}^{\nu} g_{\mu\nu}+\partial_{\alpha}  \overset{(m)}{x}{}^{\mu} \partial_{\beta} \delta \overset{(m)}{x}{}^{\nu} g_{\mu\nu}+ \partial_{\alpha} \overset{(m)}{x}{}^{\mu} \partial_{\beta} \overset{(m)}{x}{}^{\nu}\partial_{\rho}g_{\mu\nu} \delta \overset{(m)}{x}{}^{\rho},\\
&\delta h_{\gamma \ ab}=\partial_{a} \delta \overset{(m)}{x}{}^{\mu} \partial_{b} \overset{(m)}{x}{}^{\nu} g_{\mu\nu}+\partial_{a}  \overset{(m)}{x}{}^{\mu} \partial_{b} \delta \overset{(m)}{x}{}^{\nu} g_{\mu\nu}+ \partial_{a} \overset{(m)}{x}{}^{\mu} \partial_{b} \overset{(m)}{x}{}^{\nu}\partial_{\rho}g_{\mu\nu} \delta \overset{(m)}{x}{}^{\rho}, \label{app B: metric variations 2}
\end{align}
into (\ref{app B: dS}) and integrating by parts, we obtain
\begin{align}\label{app B: variation RT area}
\delta S&=\sum_m^p \frac{-1}{4 G_{N\ m}} \int_{\Gamma_m}d^{d-1}\xi \sqrt{h_{\Gamma_m}} \left( \overset{(m)}{K}_{\Gamma\ \mu} +\hat{\lambda}_m\left( \mathcal{R}_m  \overset{(m)}{K}_{\Gamma\ \mu}-2 \mathcal{R}_m^{\ \alpha\beta} \overset{(m)}{K}_{\Gamma\ \mu \alpha\beta} \right) \right) \delta \overset{(m)}{x}{}^{\mu}\nonumber\\
&+\sum_m^p \frac{1}{ 4G_{N\ m}} \int_{\gamma} d^{d-2}s\sqrt{h_{\gamma}} \overset{(m)}{n}_{\alpha} \left( (1+\hat{\lambda}_m \mathcal{R}_m )h_{\Gamma_{m}}^{\alpha\beta}-2\hat{\lambda}_m \mathcal{R}_m^{\alpha\beta} \right) \partial_{\beta} \overset{(m)}{x}{}^{\nu}\delta \overset{(m)}{x}{}^{\mu}g_{\mu\nu}\nonumber\\
&+\sum_m^p \frac{-2\hat{\lambda}_m}{ 4G_{N\ m}} \int_{\gamma} d^{d-2}s\sqrt{h_{\gamma}} \left( \hat{D}_a (\mathcal{K}_m h_{\gamma}^{ab}-\mathcal{K}_m^{ab})\partial_b \overset{(m)}{x}{}^{\nu} g_{\mu\nu} +  (\mathcal{K}_m h_{\gamma}^{ab}-\mathcal{K}_m^{ab})  \overset{(m)}{K}_{\gamma\ \mu ab}\right) \delta \overset{(m)}{x}{}^{\mu},
\end{align}
 where $\overset{(m)}{n}{}^{\alpha}$ and $\mathcal{K}_m$ represent the normal vectors and extrinsic curvatures defined from the RT surface $\Gamma_m$ to the intersection $\gamma = \Gamma_m \cap NB$. The covariant derivative on $\gamma$ is denoted as $\hat{D}_a$. The extrinsic curvatures $\overset{(m)}{K}_{\Gamma}$ and $\overset{(m)}{K}_{\gamma}$ are defined from the bulk branch $B_m$ to the surfaces $\Gamma_m$ and $\gamma$, respectively. They are expressed as follows:
\begin{align}\label{app B: KGamma}
&\overset{(m)}{K}_{\Gamma}{}^{\mu}_{\alpha\beta}= \partial_{\alpha}\partial_{\beta} \overset{(m)}{x}{}^{\mu}-\overset{(m)}{(\Gamma_{\Gamma})}{}^{\gamma}_{\alpha\beta} \partial_{\gamma} \overset{(m)}{x}{}^{\mu}+\overset{(m)}{(\Gamma_{B})}{}^{\mu}_{\sigma\rho}\partial_{\alpha} \overset{(m)}{x}{}^{\sigma}\partial_{\beta} \overset{(m)}{x}{}^{\rho},\\ \label{app B: Kgamma}
&\overset{(m)}{K}_{\gamma}{}^{\mu}_{ab}= \partial_{a}\partial_{b} \overset{(m)}{x}{}^{\mu}-\overset{(m)}{(\Gamma_{\gamma})}{}^{c}_{ab} \partial_{c} \overset{(m)}{x}{}^{\mu}+\overset{(m)}{(\Gamma_{B})}{}^{\mu}_{\sigma\rho}\partial_{a} \overset{(m)}{x}{}^{\sigma}\partial_{b} \overset{(m)}{x}{}^{\rho}. 
\end{align}
In this context, $\overset{(m)}{(\Gamma_{B})}{}^{\mu}_{\sigma\rho}$, $\overset{(m)}{(\Gamma_{\Gamma})}{}^{\gamma}_{\alpha\beta}$, and $\overset{(m)}{(\Gamma_{\gamma})}{}^{\gamma}_{ab}$ represent the Levi-Civita connections in bulk branch $B_m$, on the RT surface $\Gamma_m$, and at the intersection $\gamma$, respectively.

From the bulk term of (\ref{app B: variation RT area}), we derive an extremal condition for the RT surfaces
\begin{align}\label{app B: extremal surface}
 \overset{(m)}{K}_{\Gamma\ \mu} +\hat{\lambda}_m\left( \mathcal{R}_m  \overset{(m)}{K}_{\Gamma\ \mu}-2 \mathcal{R}_m^{\ \alpha\beta} \overset{(m)}{K}_{\Gamma\ \mu \alpha\beta} \right)=0.
\end{align}
Let us denote the embedding function of $\gamma$ into the Net-brane as $y^i(s^a)$. Since \( \delta  \overset{(m)}{x}{}^{\mu}|_{\gamma} = \left( \frac{\partial  \overset{(m)}{x}{}^{\mu}}{\partial y^i(s^a)} \right) \delta y^i(s^a)|_{\gamma} \) can vary along the Net-brane, the boundary term of (\ref{app B: variation RT area}) provides us with the junction condition (\ref{sect2: JC RT}) for the RT surfaces on $\gamma$. It is expressed as follows:
 \begin{align}\label{app B: JC RT}
&\sum_{m=1}^p \frac{1}{G_{N\ m}} \Big[\overset{(m)}{n}_{\alpha} \left( (1+\hat{\lambda}_m \mathcal{R}_m )h_{\Gamma_{m}}^{\alpha\beta}-2\hat{\lambda}_m \mathcal{R}_m^{\alpha\beta} \right) \partial_{\beta} \overset{(m)}{x}{}^{\nu}g_{\mu\nu}\nonumber\\
&-2\hat{\lambda}_m \left( \hat{D}_a (\mathcal{K}_m h_{\gamma}^{ab}-\mathcal{K}_m^{ab})\partial_b \overset{(m)}{x}{}^{\nu} g_{\mu\nu} +  (\mathcal{K}_m h_{\gamma}^{ab}-\mathcal{K}_m^{ab})  \overset{(m)}{K}_{\gamma\ \mu ab}\right) \Big] \frac{\partial \overset{(m)}{x}{}^{\mu}}{\partial y^i}|_{\gamma}=0.
\end{align}

\section{Joint condition for EOW branes} \label{app C}

In this appendix, we derive the joint condition from the total bulk action~(\ref{sect4: action}). Without loss of generality, we focus on the case in which the angle between the EOW brane $Q_{m}$ and the Net-brane satisfies $\overset{(m)}{\theta}_{J}<\pi$. Using the same method, one can show that the resulting joint condition remains valid when $\overset{(m)}{\theta}_{J}>\pi$.
Before proceeding with the derivation, we first fix the notation. We denote by $\overset{(m)}{n}_{NB}$ and $\overset{(m)}{n}_{Q}$ the outward-pointing unit normal vectors in the bulk branch $B_{m}$ that are orthogonal to the Net-brane and the EOW brane $Q_{m}$, respectively. Correspondingly, $\overset{(m)}{l}_{NB}$ and $\overset{(m)}{l}_{Q}$ denote the outward-pointing unit normal vectors intrinsic to the Net-brane and the EOW brane that are orthogonal to the joint $J$. 

Let us first examine the variation of the joint action (\ref{sect4: Hayward-term}), which results in:
\begin{align}\label{app C: dIJ}
\delta I_{J}=\sum_{m=1}^{p}\frac{1}{8\pi G_{N~m}}\int_{J}d^{d-1}y\sqrt{|\sigma|}\left[-\delta\overset{(m)}{\theta}_{J}-\frac{1}{2}\sigma_{ab}\left( \frac{\pi}{2}-\overset{(m)}{\theta}_{J} \right)\delta\sigma^{ab} \right],
\end{align}
where $\overset{(m)}{\theta}_{J}=\pi-\arccos{(\overset{(m)}{n}{}_{Q}\cdot\overset{(m)}{n}_{NB})}$. The actions of the Net-brane and the EOW brane also contribute to the variation at the joint. To understand this, let us consider the variation of the extrinsic curvature:
\begin{align}
    \delta \overset{(m)}{K}_{\#}
    = -\frac{1}{2} \overset{(m)}{n}{}_\#^{\mu}\bigl(\nabla^{\nu}\delta g_{\mu\nu}
    - g^{\lambda\nu}\nabla_{\mu}\delta g_{\lambda\nu}\bigr)
    + \frac{1}{2} \overset{(m)}{K}_{\#~\mu\nu}\delta g^{\mu\nu}
    - \overset{(m)}{D}{}_{\#~\mu}\!\left(\frac{1}{2} \overset{(m)}{h}{}_{\#}^{\mu\lambda} \overset{(m)}{n}{}_{\#}^{\nu}\delta g_{\lambda\nu}\right),
\end{align}
where the index $\#$ represent either the EOW brane $Q_{m}$ ($\overset{(m)}{K}_{Q}$) or Net-brane $NB$ ($\overset{(m)}{K}_{NB}=\overset{(m)}{K}|_{NB}$).
Besides, $\overset{(m)}{D}{}_{\#~\mu}$ and $\overset{(m)}{h}{}_{\#}^{\mu\nu}=g^{\mu\nu}-\overset{(m)}{n}{}_{\#}^{\mu}\overset{(m)}{n}{}_{\#}^{\nu}$ denote the covariant derivative and the projector on corresponding brane $\#$, respectively.
The last term gives rise to a boundary contribution at the joint $J=Q_{m}\cap NB$. It contributes to the joint variation as follows: 
\begin{align}\label{app C: dIother}
    &\delta (I-I_{J})|_{J}\nonumber\\
    =&\sum_{m=1}^{p}\frac{-1}{16\pi G_{N~m}}\left(\int_{\partial Q_{m}}d^{d-2}y\sqrt{|\sigma|}\overset{(m)}{l}{}_{Q}^{\mu}\overset{(m)}{n}{}_{Q}^{\nu}\delta g_{\mu\nu}+\int_{\partial NB} d^{d-2}y\sqrt{|\sigma|} \overset{(m)}{l}{}_{NB}^{\mu}\overset{(m)}{n}{}_{NB}^{\nu}\delta g_{\mu\nu} \right)\nonumber\\
    =&\sum_{m=1}^{p}\frac{1}{8\pi G_{N~m}}\int_{J}d^{d-2}y\sqrt{|\sigma|}\frac{\overset{(m)}{n}{}_{Q~\mu}\overset{(m)}{n}{}_{NB~\nu}\delta g^{\mu\nu}+\overset{(m)}{n}{}_{NB}^{\mu}\delta\overset{(m)}{n}{}_{Q~\mu}+\overset{(m)}{n}{}_{Q}^{\mu}\delta\overset{(m)}{n}{}_{NB~\mu}}{\sqrt{1-(\overset{(m)}{n}{}_{Q}\cdot\overset{(m)}{n}_{NB})^{2}}}\nonumber\\
    =&\sum_{m=1}^{p}\frac{-1}{8\pi G_{N~m}}\int_{J}d^{d-2}y\sqrt{|\sigma|}~\delta\left( \arccos{(\overset{(m)}{n}{}_{Q}\cdot\overset{(m)}{n}_{NB})}\right),\nonumber\\
    =&\sum_{m=1}^{p}\frac{ 1}{8\pi G_{N~m}}\int_{J}d^{d-1}y\sqrt{|\sigma|}\delta \overset{(m)}{\theta}_{J},
\end{align}
where $\overset{(m)}{l}_{Q~\mu},\overset{(m)}{l}_{NB~\mu}$ denote the outer normal vectors to the joint $J$ on the EOW brane $Q_{m}$ and Net-brane from bulk branch $B_{m}$, respectively. Besides, on above variation, we have constructed $\overset{(m)}{l}_{Q~\mu}$ by projecting $\overset{(m)}{n}_{NB~\mu}$ onto $Q_{m}$ and than normalizing it, while the $\overset{(m)}{l}_{NB~\mu}$ can be constructed with the same method: 
\begin{align}
    \overset{(m)}{l}_{Q~\mu}=\frac{\overset{(m)}{h}{}_{Q~\mu}^{~\nu}\overset{(m)}{n}{}_{NB~\nu}}{\sqrt{\overset{(m)}{h}{}_{Q}^{~\rho\sigma}\overset{(m)}{n}_{NB~\rho}\overset{(m)}{n}_{NB~\sigma}}},~\overset{(m)}{l}_{NB~\mu}=\frac{\overset{(m)}{h}{}_{NB~\mu}^{~\nu}\overset{(m)}{n}_{Q~\nu}}{\sqrt{\overset{(m)}{h}{}_{NB}^{~\rho\sigma}\overset{(m)}{n}_{Q~\rho}\overset{(m)}{n}_{Q~\sigma}}}.
\end{align}

From (\ref{app C: dIJ}) and (\ref{app C: dIother}), we obtain the variation of total bulk action at joint:
\begin{align}
    \delta I|_{J}=\sum_{m=1}^{p}\frac{1}{8\pi G_{N~m}}\int_{J}d^{d-1}y\sqrt{|\sigma|}\left[-\frac{1}{2}\sigma_{ab}\left( \frac{\pi}{2}-\overset{(m)}{\theta}_{J} \right)\delta\sigma^{ab} \right],
\end{align}
which leads to the joint condition at the joint:
\begin{align}\label{app C: joint condition}
    \sum_{m=1}^{p}\frac{1}{G_{N\,m}}
    \left(\frac{\pi}{2}-\overset{(m)}{\theta}_{J}\right)=0.
\end{align}




\end{document}